\documentclass[fleqn,usenatbib]{mnras}

\usepackage{newtxtext,newtxmath}

\usepackage[T1]{fontenc}

\DeclareRobustCommand{\VAN}[3]{#2}
\let\VANthebibliography\thebibliography
\def\thebibliography{\DeclareRobustCommand{\VAN}[3]{##3}\VANthebibliography}

\usepackage{graphicx}	
\usepackage{amsmath}	

\title[Photometric bulges in robust decompositions]{Robust galaxy image decompositions with Differential Evolution optimisation and the problem of classical bulges in and beyond the nearby Universe}

\author[Dimitri A. Gadotti]{
Dimitri A. Gadotti,$^{1}$\thanks{E-mail: dimitri.a.gadotti@durham.ac.uk}
\\
$^{1}$Centre for Extragalactic Astronomy, Department of Physics, Durham University, South Road, Durham DH1 3LE, UK\\
}

\date{Accepted XXX. Received YYY; in original form ZZZ}

\pubyear{2024}

\begin{document}
\label{firstpage}
\pagerange{\pageref{firstpage}--\pageref{lastpage}}
\maketitle

\begin{abstract}
Deconstructing galaxies through two-dimensional decompositions has been shown to be a powerful technique to derive the physical properties of stellar structures in galaxies. However, most studies employ fitting algorithms that are prone to be trapped in local minima, or involve subjective choices. Furthermore, when applied on samples beyond the nearby Universe, results on the fraction of classical bulges in disc galaxies do not agree with studies on nearby galaxies. The latter studies point to a small fraction of classical bulges, possibly challenging our merger-driven picture of galaxy formation. Therefore, understanding the discrepancy between observations in and beyond the nearby Universe is of paramount importance. In this paper, I use a sample of 16 nearby galaxies drawn from the TIMER project, which previously have been shown to {\em not} host classical bulges, and perform decompositions applying different methodologies and employing the original images as well as artificially redshifted images. I show that the Differential Evolution algorithm is able to provide accurate measurements of structural properties with little subjective intervention, correctly indicating the presence of nuclear discs (not classical bulges). However, I also show that when the physical spatial resolution is not adequate, a systematic overestimation of the photometric bulge S\'ersic index leads to the false conclusion of the presence of classical bulges. I discuss how this may be the root cause of the discrepancy mentioned above, and point out how this issue may be a problem even with data from facilities such as Euclid, HST and JWST.
\end{abstract}

\begin{keywords}
galaxies: bar -- galaxies: bulges -- galaxies: formation -- galaxies: evolution -- galaxies: structure -- methods: data analysis
\end{keywords}



\section{Introduction}

The observation of disc galaxies in the nearby Universe has for a long time revealed a complex plethora of different stellar structures in these galaxies, and this has become more and more evident as observational facilities become more powerful, particularly with increased spatial resolution. To a certain degree, this has even been pushed beyond the nearby Universe with the Hubble Space Telescope (HST) and the James Webb Space Telescope (JWST), for example. The structural properties of the different stellar structures in disc galaxies hold evidence of the formation and evolution process in galaxies that have shaped them. Therefore, galaxy images have been used to study the structure in galaxies, particularly through photometric decompositions \citep[see, e.g.,][amongst many other studies]{deJ96c,LauSalBut05,AllDriGra06,BenDzaFre07,HauMcIBar07,DurSulBut08,WeiJogKho09,Gad09b,SimMenPat11,SalLauLai15,GaoHo17,KruLinBam18,Haussler2022,Chugunov2024}, employing bespoke software such as {\sc gim2d} \citep{Sim98,Simard2002}, {\sc galfit} \citep{PenHoImp02,PenHoImp10}, {\sc budda} \citep{deSGaddos04,Gad08}, {\sc gasp2d} \citep{MenAguCor08}, {\sc imfit} \citep{Erwin2015}, and {\sc profit} \citep{Robotham2017}.

These decompositions consist in fitting the 2D surface brightness distribution in galaxy images with a number of mathematical models, each aiming to represent different structural components, such as the disc, a photometric bulge\footnote{The photometric bulge is defined as the excess above the inward extrapolation of the disc radial surface brightness profile, after accounting for the profiles of additional components, such as a bar, when they are accounted for in the fit.}, and a bar. The term `photometric bulge' is a useful way to highlight the lack of knowledge on the physical nature of the component; that is, only identifying a central brightness excess in the galaxy 2D surface brightness distribution does not inform whether the component is a classical bulge (i.e., a central, pressure-supported spheroid), a rotation-supported nuclear disc (which used to be associated to the term `pseudo-bulge'), a box/peanut (i.e., the vertically thickened part of a bar, which also used to be associated to the term `pseudo-bulge'), or even other possibilities, such as a nuclear star cluster. Further analysis is required to distinguish between the different possibilities \citep[see, e.g.,][]{Ath05b,FisDro16,Gadotti2020}, including the central brightness or mass concentration of the component, often parameterised with the S\'ersic index \citep{Ser68}, or a direct measurement of the stellar kinematics in the component. In the absence of the latter, it is common practice\footnote{Although I note further below in the paper that this practice suffers from important limitations.} to classify a photometric bulge as a classical bulge if the S\'ersic index $n$ is larger than two. If $n$ is less than two, then the S\'ersic function is close to the exponential function that generally well describes galaxy discs, and thus the photometric bulge is understood to be a nuclear disc. It is important to note that exponential (low $n$) photometric bulges are not found only at the late-type end of the Hubble sequence. In fact, \citet{Aguerri2005}, for example, found exponential photometric bulges in lenticular (S0) galaxies. Such realisation can be traced back, albeit from a different perspective, to the proposition by \cite{van76}, who suggested that S0 galaxies should form a new branch of the Hubble sequence, parallel to the branch of spirals, and likewise in increasing order of `disc-to-bulge' ratio \citep[see also][]{KorBen12}.

The optimisation of the models fitted in 2D decompositions is commonly done via the Levenberg-Marquardt (LM) algorithm \citep{Levenberg1944,Marquardt1963}, e.g., with {\sc galfit} and {\sc gasp2d}. The LM algorithm is fast, but due to its gradient-search nature, it can be trapped in local minima of the statistic being minimised, particularly in the complicated topology of the multi-dimensional parameter space that characterises galaxy image decompositions. Such local minima often do not correspond to the best possible fit. {\sc gim2d} uses the Metropolis algorithm, which is much slower than LM, but is less prone to be trapped in local minima of the parameter space \citep{Metropolis1953}. This algorithm also employs Monte-Carlo sampling techniques to provide confidence levels for each parameter fitted. {\sc budda} uses the Nelder-Mead (NM) simplex algorithm \citep{Nelder1965}, which is also less prone to be trapped in local minima. {\sc profit} is not tied to a specific optimiser but has access to different algorithms, including gradient-search algorithms and more sophisticated options. In {\sc imfit}, the user can choose between different algorithms including the LM and NM algorithms, as well as two other algorithms that will be explored further below. To counteract the possibility of being trapped in spurious local minima, it is common practice to carefully choose the initial values of the parameters being fitted from which these algorithms start their optimisations of the fit. In addition, users can adjust these values after inspection of the first fitted models in an iterative process. Both procedures, however, have the drawback of adding subjectivity to the fits.

Many studies have explored the structure of galaxies through such decompositions, often employing the codes mentioned above, or with similar bespoke codes. A number of such studies have discussed the prevalence of classical bulges as a constraint to cosmological models of galaxy formation and evolution, in particular the importance of mergers. The reasoning is that, if mergers lead to the formation of classical bulges (but not necessarily to the formation of nuclear discs), then if the fraction of classical bulges in disc galaxies, or the fraction of the stellar mass content in disc galaxies that are in classical bulges, is sufficiently low, then this {\em could be} a challenge to the hierarchical picture of galaxy formation, where mergers are important \citep[see, e.g.,][]{Bittner2020,Peebles2020,Peebles2022,Fraser-McKelvie2024}. Because these studies employ different samples (in different environments) and selection functions, different ways to compute stellar mass and classify classical bulges, and different structural models and approaches to account for the Point Spread Function (PSF), it is unclear whether there is an agreement. Nevertheless, while studies employing samples of nearby galaxies often find a low prevalence of classical bulges \citep[particularly in barred galaxies;][but see \citealt{KimGadShe14}]{LauSalBut04b,WeiJogKho09,Kormendy2010,Fisher2011,SalLauLai15,Mendez-Abreu2017,KruLinBam18}, it appears that studies employing samples at higher redshifts (and comparably lower physical spatial resolution) find a more balanced distribution, or even favour the prevalence of classical bulges \citep[see, e.g.,][]{AllDriGra06,DriAllLis07,Gad09b,SimMenPat11}. Cosmic variance could perhaps explain this discrepancy, and it should be noted that the latter studies did take precautions to minimise the unwanted effects of low spatial resolution. However, more in depth analyses of the effects of low physical spatial resolution are in order. For example, an important question to address is whether nuclear discs are robustly retrieved as exponential (low $n$) photometric bulges in and beyond the local Universe.

In this paper, I use {\sc imfit} to test three optimisation algorithms in decompositions of a sample of nearby disc galaxies known -- from direct observations of the stellar kinematics -- to {\em not} host massive classical bulges, but instead host prominent nuclear discs. This allows me to test the accuracy of algorithms for which subjective choices are minimised. I then artificially degrade the spatial resolution of their images to test the effects of the lower physical spatial resolution on the derived S\'ersic index of the photometric bulge and its corresponding photometric classification. In Sect.\,\ref{sec:sampledata}, I describe the sample and the imaging data employed, while in Sect.\,\ref{sec:nearby} I present the decompositions of the original galaxy images, employing the three optimisation algorithms, and in Sect.\,\ref{sec:beyond} the corresponding decompositions of the artificially ``redshifted'' images. I discuss the results in Sect.\,\ref{sec:discuss}, in the context of the problem of classical bulges mentioned above, and in the context of improving the robustness of galaxy image decompositions. Finally, in Sect.\,\ref{sec:conc}, I summarise the main conclusions from this work. I assume a Hubble constant of ${\rm H_0}=67.8\,\rm{km}\,\rm{s}^{-1}\,\rm{Mpc}^{-1}$ and $\Omega_{\rm m}=0.308$, in a Universe with flat topology \citep[see][]{AdeAghArn15}.

\section{Sample and Data}
\label{sec:sampledata}

\begin{figure*}
\includegraphics[trim=0.2cm 0.2cm 0.2cm 0.2cm,clip=true,width=\columnwidth]{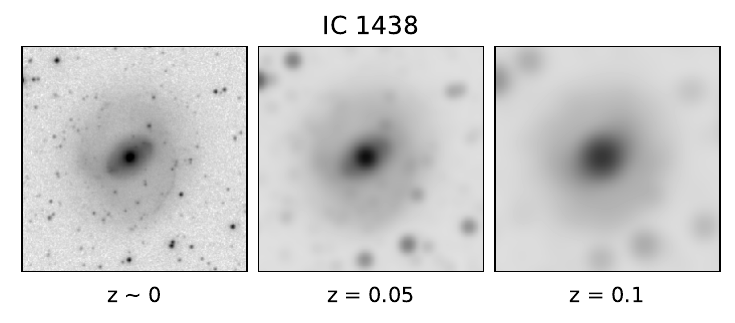}
\includegraphics[trim=0.2cm 0.2cm 0.2cm 0.2cm,clip=true,width=\columnwidth]{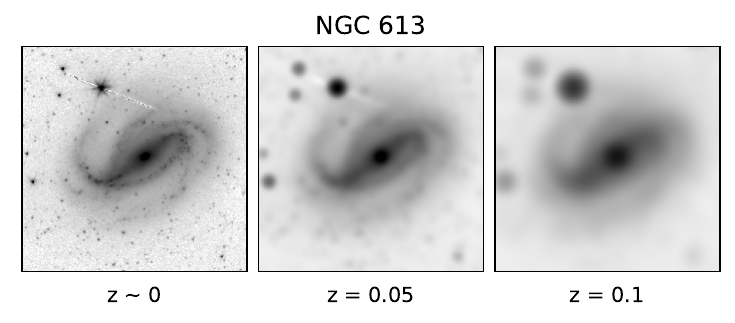}
\includegraphics[trim=0.2cm 0.2cm 0.2cm 0.2cm,clip=true,width=\columnwidth]{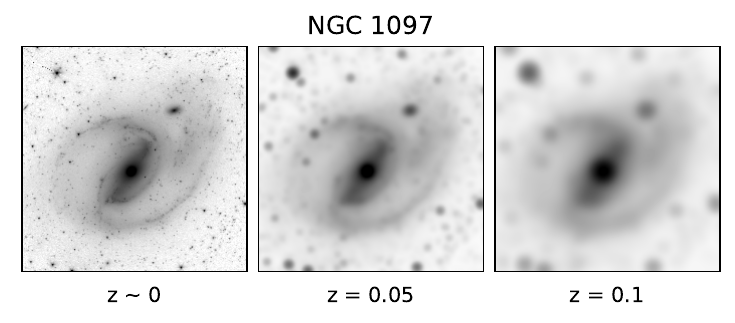}
\includegraphics[trim=0.2cm 0.2cm 0.2cm 0.2cm,clip=true,width=\columnwidth]{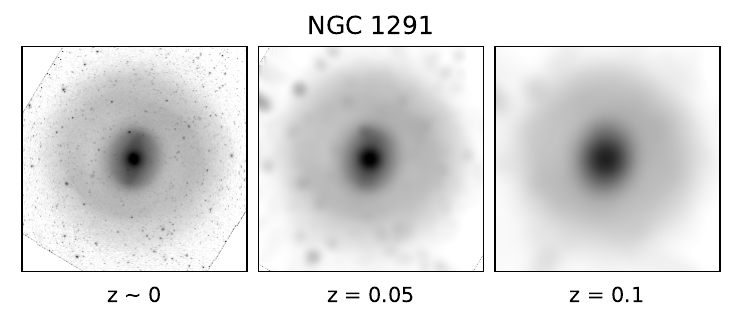}
\includegraphics[trim=0.2cm 0.2cm 0.2cm 0.2cm,clip=true,width=\columnwidth]{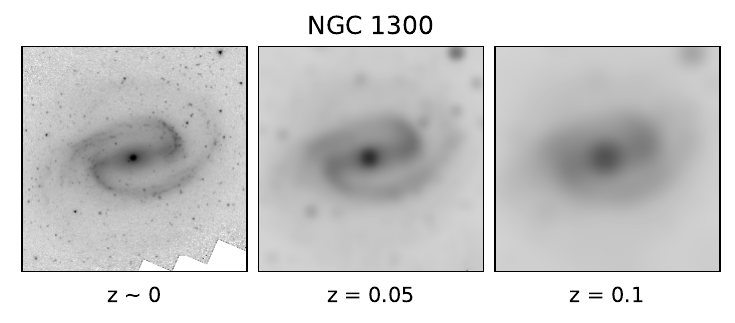}
\includegraphics[trim=0.2cm 0.2cm 0.2cm 0.2cm,clip=true,width=\columnwidth]{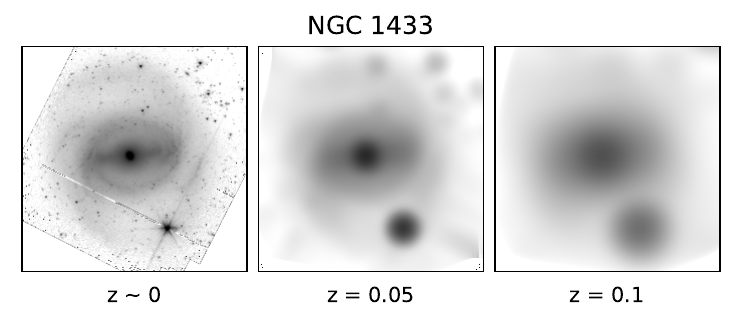}
\includegraphics[trim=0.2cm 0.2cm 0.2cm 0.2cm,clip=true,width=\columnwidth]{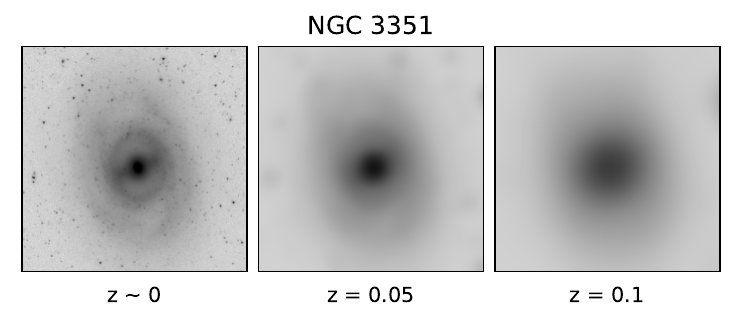}
\includegraphics[trim=0.2cm 0.2cm 0.2cm 0.2cm,clip=true,width=\columnwidth]{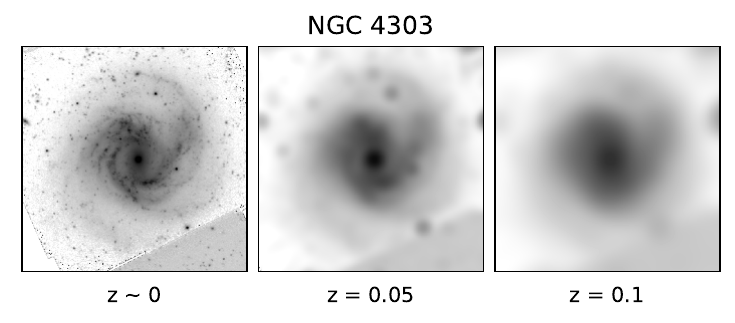}
\includegraphics[trim=0.2cm 0.2cm 0.2cm 0.2cm,clip=true,width=\columnwidth]{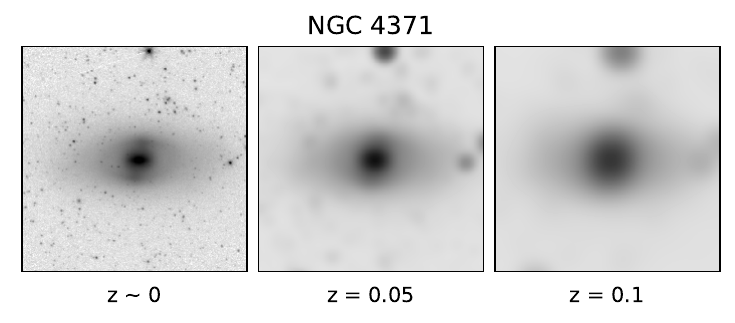}
\includegraphics[trim=0.2cm 0.2cm 0.2cm 0.2cm,clip=true,width=\columnwidth]{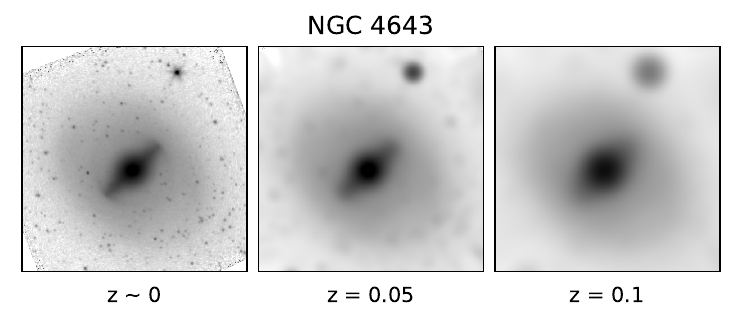}
\includegraphics[trim=0.2cm 0.2cm 0.2cm 0.2cm,clip=true,width=\columnwidth]{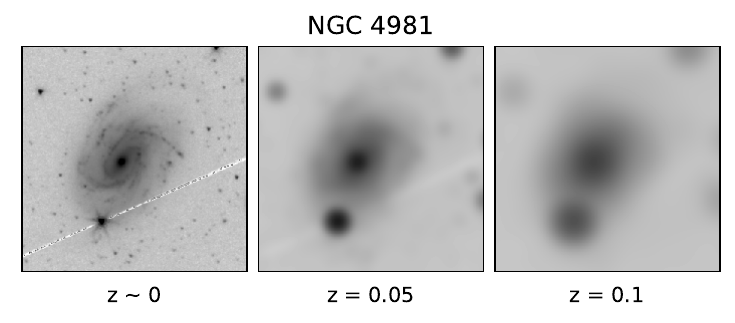}
\includegraphics[trim=0.2cm 0.2cm 0.2cm 0.2cm,clip=true,width=\columnwidth]{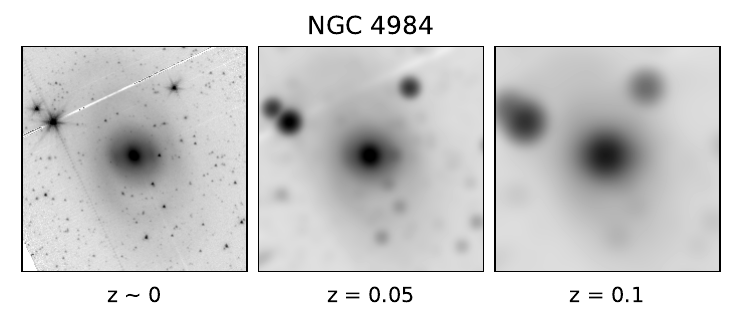}
\caption{Images used as input to the decompositions. For each galaxy, the left panel shows the original, unaltered S$^4$G image (with an average PSF FWHM of $\approx170$\,pc), whereas the central and right panels show the artificially redshifted images, in which the PSF FWHM is $\approx1.7$\,kpc and $\approx3.4$\,kpc, respectively.}
\label{fig:input}
\end{figure*}

\begin{figure*}
\includegraphics[trim=0.2cm 0.2cm 0.2cm 0.2cm,clip=true,width=\columnwidth]{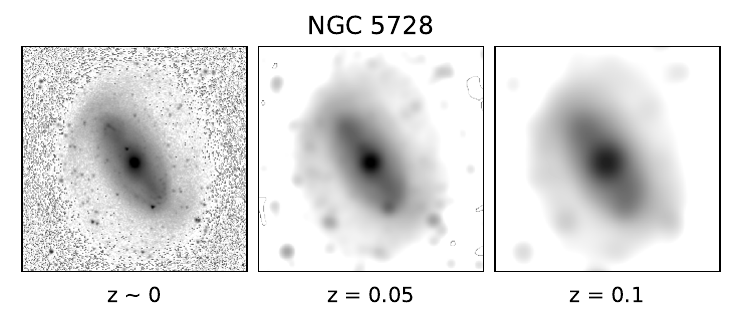}
\includegraphics[trim=0.2cm 0.2cm 0.2cm 0.2cm,clip=true,width=\columnwidth]{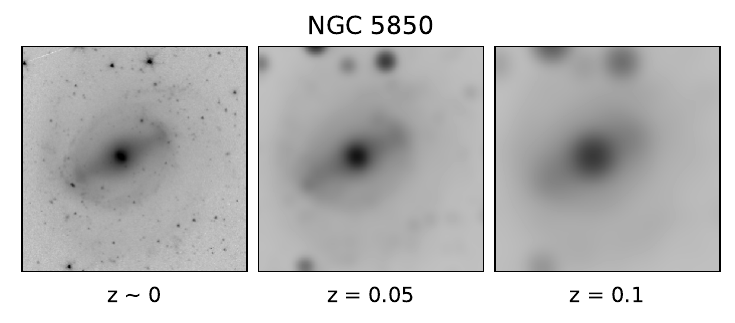}
\includegraphics[trim=0.2cm 0.2cm 0.2cm 0.2cm,clip=true,width=\columnwidth]{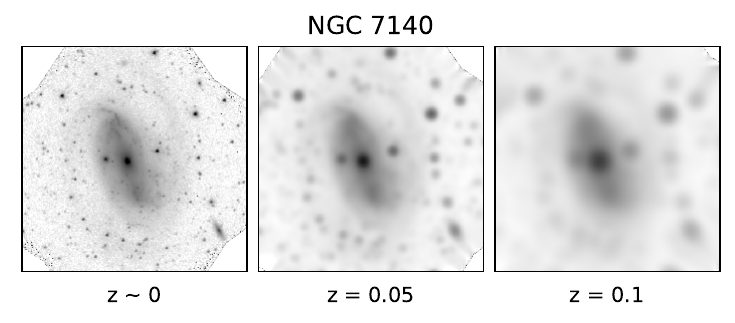}
\includegraphics[trim=0.2cm 0.2cm 0.2cm 0.2cm,clip=true,width=\columnwidth]{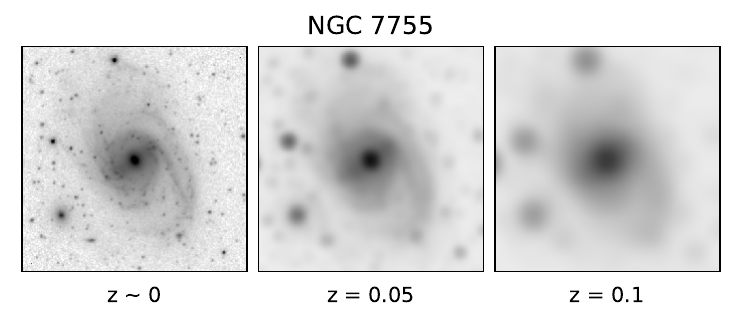}
\addtocounter{figure}{-1}
\caption{continued.}
\end{figure*}

Since one of the main goals of this work is to verify if nuclear discs are robustly retrieved in photometric decompositions as exponential photometric bulges both in and beyond the nearby Universe, the best possible sample is one that contains galaxies with kinematically confirmed nuclear discs. For this reason, the parent sample consists of the 18 galaxies from the TIMER sample \citep{GadSanFal19} for which the presence of a nuclear disc is unequivocal in the stellar kinematics maps presented in \citet{Gadotti2020}. To this parent sample I added another galaxy (NGC\,1291), whose signatures of a nuclear disc are not as clear in the kinematics maps due to the very low inclination of the galaxy ($i\approx11^\circ$). However, given the proximity of the galaxy ($d\approx9$\,Mpc), the physical spatial resolution of available images ($\approx40$\,pc per arcsecond) is such that the nuclear disc is visually very clear \citep[e.g.,][]{ButSheAth15}. In addition, the obvious presence of a nuclear bar is also an indication of the presence of a nuclear disc (see Fig.\,\ref{fig:ellfits}).

Details on the TIMER sample selection can be found in \citet{GadSanFal19}, but the main relevant selection criteria imply that all galaxies in this study are barred, more massive than $10^{10}$\,M$_\odot$, are at distances below 40\,Mpc, have inclinations below $\approx60^\circ$ and angular diameters above $1'$. With all galaxies being barred, this sample is also suitable thus to identify conditions in which the presence of a bar can no longer be established in image decompositions for galaxies beyond the local Universe with images with relatively low physical spatial resolution. The stellar kinematics analysis performed by the TIMER team and presented in \citet{Gadotti2020} reveals that none of these galaxies host a central, massive and kinematically hot spheroid (i.e., a classical bulge). However, NGC\,1291 is a strong candidate to host a so-called `composite bulge', wherein a small classical bulge is embedded within the nuclear disc \citep[see, e.g.,][]{ErwSagFab15,deLSanMen19}.

To perform the image decompositions, I used images from the Spitzer Survey of Stellar Structure in Galaxies \citep[S$^4$G;][]{shereghin10}. These are deep 3.6\,$\mu$m images from IRAC onboard the Spitzer Space Telescope, which are available for all galaxies, ensuring a homogenous analysis and high signal-to-noise that is particularly beneficial to derive the structural parameters of the main galaxy disc. The choice for a mid-infrared wavelength range stems from the need to minimise effects from dust absorption, which can be significant in the central region, and to obtain models that are more representative of the bulk of the stellar population. However, it should be noted that emission from heated dust can be important in this wavelength range in regions of pronounced star formation \citep[e.g.,][]{QueMeiSch15}. In addition, the PSF Full Width at Half Maximum (FWMH) of these images is $\approx1.7''$ (with a pixel size of $0.75''$), meaning that the angular spatial resolution is relatively poor. Nevertheless, this is not an issue in this study, given the short distances to these galaxies and their large angular sizes. Indeed, the physical spatial resolution in the sample (corresponding to the PSF FWHM of $1.7''$) ranges from $\approx70\,{\mathrm{pc}}$ to $\approx300\,{\mathrm{pc}}$, with a typical value of the order of $170\,{\mathrm{pc}}$.

To study the effects of the poorer spatial resolution typical of studies of samples at higher redshifts, the original S$^4$G images were blurred by convolving them with a Gaussian function. Each image was thus artificially redshifted to two different distances corresponding to $z\approx0.05$ and to $z\approx0.1$ by convolving each image with a different Gaussian function with the correct value for $\sigma$. Therefore, the physical spatial resolution of all images artificially redshifted to $z\approx0.05$ and to $z\approx0.1$ correspond, respectively, to $\approx1.7$\,kpc and $\approx3.4$\,kpc. There are two important points that need to be considered here. Firstly, it is obvious that with facilities such as HST or JWST one can obtain images of galaxies at $z\approx0.05$ and $z\approx0.1$ with much better physical spatial resolution, namely, $\sim150\,{\mathrm{pc}}$ and $\sim300\,{\mathrm{pc}}$ (e.g., with HST at $1.6\mu m$), respectively, and thus comparable to the resolution in the original S$^4$G images. The results from this study are thus suitable to be compared with imaging data similar to that of the Sloan Digital Sky Survey (SDSS), with a PSF FWHM of $\approx1.5''$, or the Dark Energy Camera Legacy Survey (DECaLS), with a PSF FWHM of $\approx1.2''$. Secondly, I am neglecting effects such as band-shifting and cosmological surface brightness dimming, which are likely to be secondary compared to the effects of poorer spatial resolution in photometric decompositions of galaxies at $z\lesssim0.1$.

In the course of this work, three galaxies were removed from the analysis: NGC\,5236 and NGC\,5248 were removed because in these galaxies the bar is relatively inconspicuous and difficult to fit properly in the photometric decompositions; and NGC\,7552 was removed due to saturated central pixels in the IRAC image. The total sample thus consists of 16 galaxies. Figure \ref{fig:input} shows the IRAC 3.6\,$\mu$m images of all galaxies in the sample studied here. The left panels show the original S$^4$G image while the middle and right panels show the artificially redshifted images as indicated. A comparison between the three panels in this figure for each galaxy shows the extent to which the poorer spatial resolution alters how one can discern different morphological features in the galaxies. This will be discussed at length further below. In addition, Table \ref{tab:sample} shows some fundamental parameters of these galaxies, including stellar mass, central stellar velocity dispersion, the peak value of $v/\sigma$ in the nuclear disc, and the radius at which this peak is located, defined in \citet{Gadotti2020} as the kinematic radius r$_{\rm k}$ of the nuclear disc (with $v$ and $\sigma$ here being, respectively, the mean stellar velocity and velocity dispersion). Note that all galaxies in this paper show strong signatures of the presence of a nuclear disc, including evidence of near-circular orbits.

\begin{table*}
\centering
\caption{Some fundamental properties of the sample studied here, as presented in \citet{MunSheGil13,MunSheReg15} and \citet{Gadotti2020}. Column (1) gives the galaxy designation, and column (2) shows the morphological classification by \citet{ButSheAth15}. The inclination of the galaxy main disc relative to the plane of the sky is given in column (3), and derived using $\cos i = b/a$, where $a$ and $b$ are, respectively, the semi-major and semi-minor axes of the 25.5\,AB\,mag\,arcsec$^{-2}$ isophote at 3.6\,$\mu$m. Column (4) shows the stellar mass derived within S$^4$G, and column (5) shows the galaxy distance, calculated as the mean of all redshift-independent measurements presented in the NASA Extragalactic Database (NED; \url{http://ned.ipac.caltech.edu/}). In column (6), I show the TIMER measurements of the central velocity dispersion, as measured within an aperture of $r_{\rm e,gal}/8$, with the corresponding errors shown in column (7), and $r_{\rm e,gal}$ being the effective (half-light) radius of the galaxy. Finally, columns (8) and (9) show, respectively, the peak value of $v/\sigma$ in the nuclear disc, and the radius at which this peak is located, defined as the kinematic radius r$_{\rm k}$ of the nuclear disc ($v$ and $\sigma$ here are the mean stellar velocity and velocity dispersion, respectively).}
\label{tab:sample}
\begin{tabular}{lcccccccc}
\hline
Galaxy & Type & $i$ & $M_\star$ & $d$ & $\sigma_{r_{\rm e}/8}$ & err($\sigma$) & $v/\sigma$ & r$_{\rm k}$\\
\omit & \omit & $^\circ$ & 10$^{10}$ M$_\odot$ & Mpc & km s$^{-1}$ & km s$^{-1}$ & \omit & kpc \\
(1) & (2) & (3) & (4) & (5) & (6) & (7) & (8) & (9)\\
\hline
IC 1438  & (R$_1$)SAB$_{\mathrm{a}}$(r$'$\underline{l},nl)0/a         & 24 & 3.1  & 33.8 & 101 & 2  & 2.57 & 0.60\\
NGC 613  & SB(\underline{r}s,bl,nr)b                                  & 39 & 12.2 & 25.1 & 125 & 3  & 2.35 & 0.59\\
NGC 1097 & (R$'$)SB(rs,bl,nr)ab pec                                   & 51 & 17.4 & 20.0 & 196 & 3  & 2.82 & 1.07\\
NGC 1291 & (R)SAB(l,bl,nb)0$^+$                                       & 11 & 5.8  & 8.6  & 168 & 7  & -- & -- \\
NGC 1300 & (R$'$)SB(s,bl,nrl)b                                        & 26 & 3.8  & 18.0 & 100 & 8  & 2.98 & 0.33\\
NGC 1433 & (R$'_1$)SB(r,p,nrl,nb)a                                    & 34 & 2.0  & 10.0 & 95 & 13  & 2.00 & 0.38\\
NGC 3351 & (R$'$)SB(r,bl,nr)a                                         & 42 & 3.1  & 10.1 & 98 & 8  & 2.57 & 0.24\\
NGC 4303 & SAB(rs,nl)b\underline{c}                                   & 34 & 7.2  & 16.5 & 79 & 8  & 1.36 & 0.21\\
NGC 4371 & (L)SB$_{\mathrm{a}}$(r,bl,nr)0$^{0/+}$                     & 59 & 3.2  & 16.8 & 132 & 12 & 2.02 & 0.95\\
NGC 4643 & (L)SB(\underline{r}s,bl,nl)0$^{0/+}$                       & 44 & 10.7 & 25.7 & 133 & 3  & 1.31 & 0.50\\
NGC 4981 & SA\underline{B}(s,nl)\underline{b}c                        & 54 & 2.8  & 24.7 & 95 & 3  & 0.99 & 0.14\\
NGC 4984 & (R$'$R)SAB$_{\mathrm{a}}$(l,bl,nl)0/a                      & 53 & 4.9  & 21.3 & 113 & 3  & 2.49 & 0.49\\
NGC 5728 & (R$_1$)SB(\underline{r}$'$l,bl,nr,nb)0/a                    & 44 & 7.1  & 30.6 & 160 & 7  & 2.84 & 0.63\\
NGC 5850 & (R$'$)SB(r,bl,nr,nb)\underline{a}b                         & 39 & 6.0  & 23.1 & 123 & 3  & 1.13 & 0.80\\
NGC 7140 & (R$'$)SA\underline{B}$_{\mathrm{x}}$(rs,nrl)a\underline{b} & 51 & 5.1  & 37.4 & 98 & 3  & 1.45 & 0.63\\
NGC 7755 & (R$'$)SAB(rs,nrl)\underline{b}c                            & 52 & 4.0  & 31.5 & 114 & 3  & 2.20 & 0.47\\
\hline
\end{tabular}
\end{table*}

\section{Galaxy Image Decompositions in the Nearby Universe}
\label{sec:nearby}

All photometric decompositions were done using {\sc imfit} v1.8 \citep{Erwin2015} and all used the masks provided by S$^4$G to reject, e.g., foreground stars and background objects. The pixel units were converted from MJy/sr to DN/s by dividing each image by the corresponding {\tt FLUXCONV} value in the header. {\sc imfit} is fed with information on the exposure time, gain, readout noise and the subtracted background value\footnote{The background value is taken as {\tt SKYDKMED}/{\tt FLUXCONV}, with {\tt SKYDKMED} also found in the header.} by editing the {\sc imfit} configuration file with the corresponding values taken from each image header. This information is necessary so that {\sc imfit} correctly computes the statistic being used to find the best fit model.

To account for the effects of the PSF, {\sc imift} is fed with a square PSF image that is modelled as a \citet{Mof69} function \citep[see][]{TruAguCep01b} with a FWHM of 2.28 pixels and $\beta=2.63$. These are the average values obtained from fits to a number of selected foreground stars in the fields of the galaxies in the parent sample. Since the IRAC PSF is not optimally sampled, the fits to the original S$^4$G images also employ a fivefold oversampled PSF. The use of an oversampled PSF will be shown below to be important to derive accurate structural parameters. Before comparing a model to the actual image being fitted, {\sc imfit} convolves the model with the PSF image, which in this work measures 40 times the PSF FWHM on a side. When an oversampled PSF image is used, {\sc imfit} convolves the model with the oversampled PSF image -- which here is chosen to be 20 times the PSF FWHM on a side -- within a square region also centred on the galaxy centre and measuring 20 times the PSF FWHM on a side. Beyond this oversampled PSF region, the convolution is done with the regular PSF, as long as its image is larger than the oversampled PSF region. These numbers were derived after performing a number of tests and verifying the behaviour of the derived S\'ersic index of the photometric bulge component. However, they are in most cases overkill: these tests showed that, in most cases, the size of the PSF images and oversampled PSF region can be a factor two or more smaller and convergence on the S\'ersic index is already obtained.

The galaxy centre is defined as the brightest pixel in the central region, and all fitted structural components are centred on the galaxy centre. All fits presented in this paper include three photometric/structural components, unless otherwise noted, with light/mass radial profiles as follows: a S\'ersic function to describe the centre-most component (i.e., the photometric bulge), a generalised S\'ersic function to describe the bar, and an exponential function to describe the disc. In addition, a flat sky function (described by a single free parameter) is used to account for any residual background not removed during the background subtraction. Typical values of this residual background amount to only a few per cent of the originally subtracted background.

The photometric bulge is thus built as concentric and homologous ellipses with position angle PA$_{\mathrm b}$ and ellipticity $\epsilon_{\mathrm b}$, where the latter is defined as $1-b/a$, with $a$ being the semi-major axis of the ellipse, and $b$ its semi-minor axis. The intensity along each ellipse thus follows:
\begin{equation}
I_{\mathrm b}(r) = I_e \exp\left( -b_n \left[ \left( \frac{r}{r_e}\right)^{1/n} -1 \right] \right),
\end{equation}

\noindent where $n$ is the S\'ersic index, and $b_n$ is computed such that $r_e$ is the effective, or half-light, radius (i.e., the radius containing half of the luminosity of the component), and $I_e$ is the intensity at $r_e$ ($r$ is always the distance from the centre along the ellipse semi-major axis).

The bar follows an equivalent radial profile:
\begin{equation}
I_{\mathrm{bar}}(r) = I_{e,{\mathrm{bar}}} \exp\left( -b_{n,{\mathrm{bar}}} \left[ \left( \frac{r}{r_{e,{\mathrm{bar}}}}\right)^{1/n_{\mathrm{bar}}} -1 \right] \right),
\end{equation}

\noindent where, however, the concentric and homologous ellipses are generalised ellipses, with position angle PA$_{\mathrm{bar}}$ and ellipticity $\epsilon_{\mathrm{bar}}$, and which are described as:
\begin{equation}
\left( \frac{|x|}{a}\right)^{c_0+2} + \left( \frac{|y|}{b}\right)^{c_0+2} = 1,
\end{equation}

\noindent where $|x|$ and $|y|$ are the distances from the ellipse centre in a coordinate system where the major axis of the ellipse is along the $x$ axis and the ellipse is centred at the origin. In this formulation, the ellipses can be discy (if $c_0 < 0$) or boxy (if $c_0 > 0$), and this allows to better represent the light/mass distribution within bars, which are known to be better described by boxy ellipses \citep[][see also \citealt{Gad11}]{AthMorWoz90}. For regular ellipses, $c_0 = 0$. Note that although the photometric bulge and the bar both follow a S\'ersic radial profile, the two descriptions are far from being degenerate, as long as the bar is well resolved and sufficiently conspicuous. Given that the bar is normally substantially more eccentric than the photometric bulge (unless the galaxy is nearly at an edge-on projection), at a different position angle, and that $n_{\mathrm{bar}}$ is often less than $n$, it is rather straightforward to constrain separately the structural parameters describing the bar and the photometric bulge.

Finally, the disc is also described by concentric and homologous ellipses with position angle PA$_{\mathrm d}$ and ellipticity $\epsilon_{\mathrm d}$, and with the intensity along each ellipse following:
\begin{equation}
I_{\mathrm d}(r) = I_0 \exp (-r/h),
\end{equation}

\noindent where $I_0$ is the central intensity of the disc and $h$ its scale length.

Therefore, a typical fit here has 18 free parameters, which at first sight may cast doubts on their validity. However, one should also note that, considering the original S$^4$G images used here for example, the number of pixels employed in the fits is in the range of $10^5$ to $10^6$, which translates into a considerable amount of information on the structure of each galaxy. In any case, at different points further below, this paper discusses statistical approaches to enhance the empirical value of galaxy image decompositions and the corresponding structural analysis.

\subsection{Supervised Nelder-Mead fits}

The first set of decompositions presented here follows the same approach as the typical fits done with {\sc galfit} or {\sc budda}, for example, in that an initial guess for each free parameter needs to be provided, a choice that may influence the final result. However, while most studies in the literature employ the Levenberg-Marquardt (LM) algorithm to minimise the $\chi^2$ value of the fit, here I use the Nelder-Mead (NM) simplex algorithm available within {\sc imfit} (and the algorithm employed in {\sc budda}), which is less likely to become trapped in local minima in the multidimensional space of the statistic considered in the fit (typically $\chi^2$). In addition, instead of using $\chi^2$ as a goodness-of-fit parameter, I employ the Cash statistic $C$, which is more accurate, particularly in regimes of low signal-to-noise ratio \citep[see][]{Erwin2015}.

In this set of decompositions, each galaxy was treated individually. Before the first {\sc imfit} run, initial guesses for each free parameter were estimated either by visual inspection of the galaxy image or after running ellipse fits to the image isophotes to create radial profiles of their mean intensity, ellipticity, position angle and the Fourier b4 parameter, which indicates when isophotes are boxy (b4 $<0$) or discy (b4 $>0$; see \citealt{Jed87}, \citealt{Bender1987}, as well as \citealt{Kormendy2009} and references therein). Such radial profiles (shown for each galaxy in the sample in Appendix \ref{app:ellipse}) are crucial tools in that they help one to estimate initial guesses for all free parameters in the {\sc imfit} runs. Furthermore, after the first decomposition, the results for each galaxy were inspected individually, i.e., a comparison is done between the galaxy, model and residual images (an image obtained by subtracting the model for the fitted image) in order to ascertain if the fit is satisfactory or if it can be improved. In addition, the numerical values of each fitted parameter are also inspected to verify if any of those are unphysical or absurd when compared to the galaxy image or isophotal radial profiles. After this first inspection, the initial guesses are adjusted if necessary and {\sc imfit} is run again with the updated initial guesses. The procedure is repeated until no further significant improvement is obtained. Given this iterative process, these are called supervised fits.

\begin{figure}
\includegraphics[trim=0.2cm 0.2cm 0.2cm 0.2cm,clip=true,width=\columnwidth]{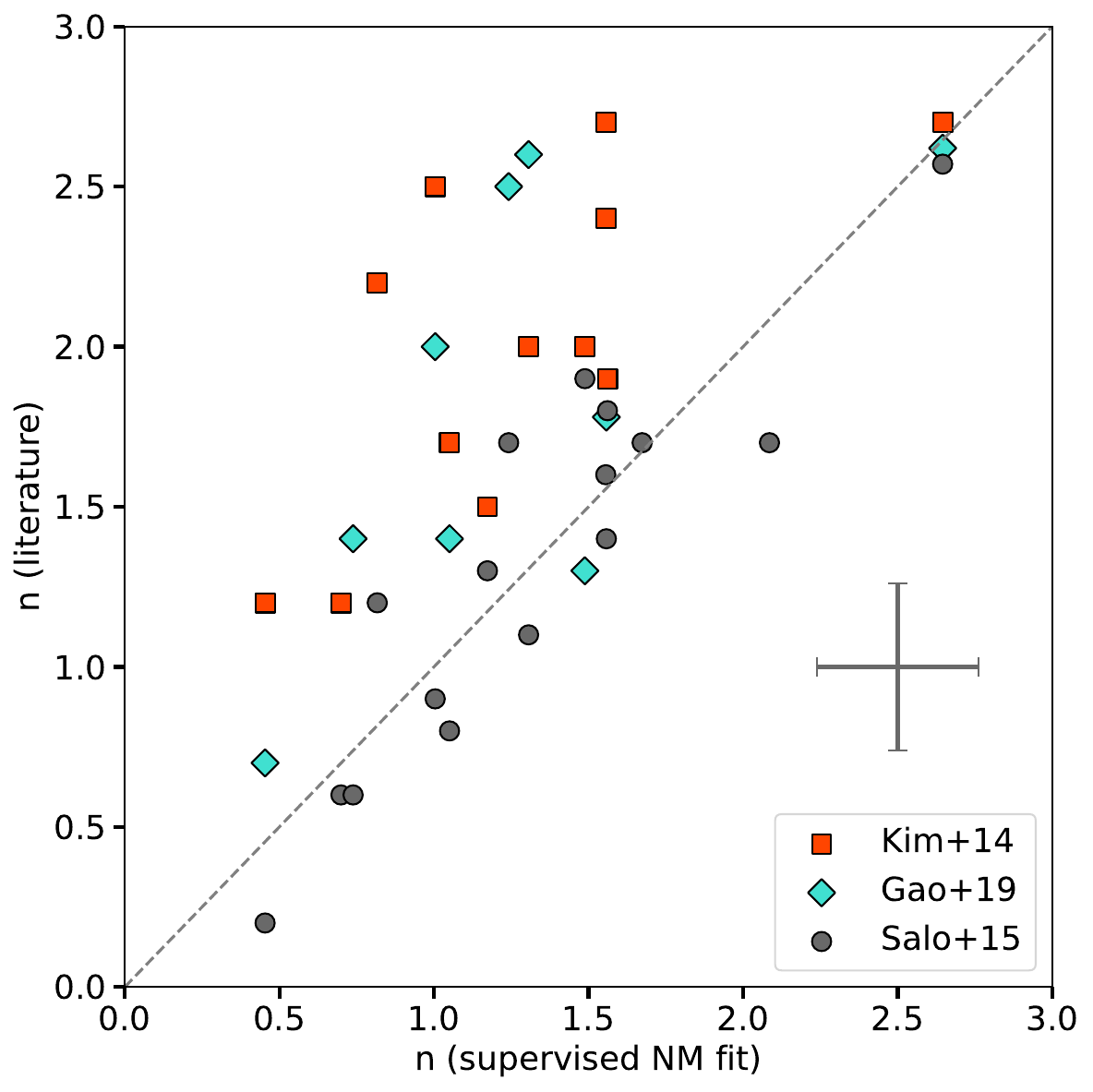}
\caption{Photometric bulge S\'ersic index $n$ obtained for the galaxies in our sample in three previous studies plotted against the values derived in this work via the supervised Nelder-Mead fits. Also plotted on the bottom right corner are fiducial errors bars derived by \citet{KimGadShe14} using {\sc budda}. The relative average error on $n$ found by Kim et al. corresponds to 13\%, which, for the galaxies studied here, implies a mean 1$\sigma$ error of 0.26.}
\label{fig:Compn_lit}
\end{figure}

Given the focus of this work on the properties of the photometric bulge component, it is instructive to compare the S\'ersic index $n$ of this component obtained with the supervised Nelder-Mead fits with results from the literature on the same galaxies. I make this comparison using three such studies: \citealt{KimGadShe14}, \citealt{SalLauLai15} and \citealt{Gao2019}. These works are particularly helpful in this context because they too concern supervised fits in which a bar component is also included in the model, and, in particular, the works by Kim et al. and Salo et al. have employed the same S$^4$G images I use here. While Salo et al. and Gao et al. used {\sc galfit}, Kim et al. used {\sc budda} to produce their fits. This comparison is done in Fig.\,\ref{fig:Compn_lit}, which shows a striking agreement with the results from \citet{SalLauLai15}, while the results from \citet{Gao2019} and particularly \citet{KimGadShe14} appear to yield systematically higher values for $n$. In Table \ref{tab:local}, I quantify these comparisons by computing the average values of $\Delta_n$ ($\left<\Delta_n\right>$) where $\Delta_n$ is defined as $n$ from the literature minus $n$ from the supervised Nelder-Mead fits. The table also shows the corresponding standard deviations. To interpret these results it is important to assess what is the typical uncertainty in the value of $n$ obtained in these fits. For that, I consider the relative average {\em statistical} error on $n$ found by Kim et al. (13\%)\footnote{\cite{KimGadShe14} have estimated uncertainties for each parameter using {\sc budda}, which yields robust {\em statistical} uncertainty estimates. The procedure consists in gradually varying the model parameter from its best-fit value, while keeping all other parameters fixed, and in computing the updated value for $\chi^2$ until it corresponds to a 1$\sigma$ variation.}, which, considering the fits in Kim et al. for the sample studied here, with a mean value of $n=2.0$, corresponds to $\sigma=0.26$ (see their Sect.\,3.2; note that the systematic uncertainty can often be significantly larger). These numbers confirm the very encouraging and excellent agreement with the results from Salo et al. (indeed with $\left<\Delta_n\right>=0.00$!), and with low scatter (in fact, the scatter is identical to the fiducial uncertainty derived by Kim et al.). On the other hand, these numbers also show the systematic offset by more than 2$\sigma$ in the results from both Kim et al. and Gao et al., and their large scatter of about twice the scatter in the comparison to Salo et al.

\begin{table}
\centering
\caption{Statistical comparison between the resulting photometric bulge S\'ersic index $n$ for different fits of the same galaxies. $\Delta n$ is defined as $n$ as derived in the literature or with the unsupervised Differential Evolution (DE) or Markov Chain Monte Carlo (MCMC) fits {\em minus} the corresponding value as derived with the supervised Nelder-Mead (NM) fits. The first row shows the average values of $\Delta n$, whereas the second row shows the standard deviation. All fits employ the original S$^4$G images, except those in \citet{Gao2019}, which employ images from the Carnegie-Irvine Galaxy Survey \citep[][CGS]{Ho2011}.}
\label{tab:local}
\begin{tabular}{lccccc}
\hline
Parameter & Kim+2014 & Gao+19 & Salo+15 & DE & MCMC\\
\hline
$\left<\Delta_n\right>$ & 0.72 & 0.53 & 0.00 & 0.03 & 0.04\\
SD ($\Delta n$) & 0.43 & 0.55 & 0.26 & 0.30 & 0.30\\
\hline
\end{tabular}
\end{table}

\begin{figure}
\includegraphics[trim=0.2cm 0.2cm 0.2cm 0.2cm,clip=true,width=\columnwidth]{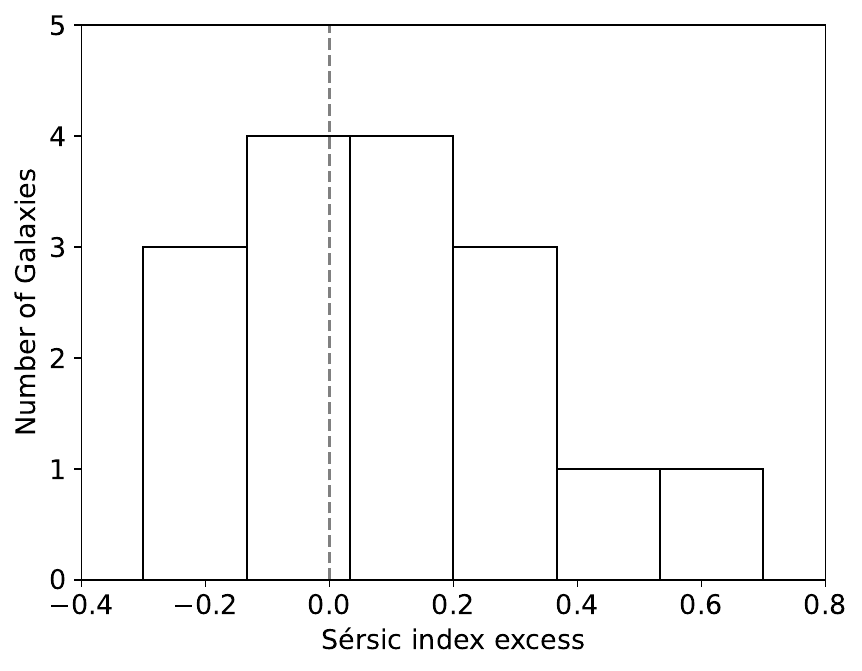}
\caption{Distribution of the difference between the photometric bulge S\'ersic indices obtained with and without employing an oversampled PSF. While the number of galaxies is small -- and a more statistically robust analysis is required to confirm the observed systematic offset -- the results shown here indicate that the bulge S\'ersic index can be significantly overestimated if the PSF treatment does not include an oversampled PSF model.}
\label{fig:nexcess}
\end{figure}

To fully understand what causes these systematic offsets is beyond the scope of this study, but because a major ingredient in deriving accurate values for $n$ is the PSF treatment, it is worth pointing out that while Salo et al. use an oversampled PSF in their decompositions, both Kim et al. and Gao et al. do not. To test the effects of using an oversampled PSF in the Nelder-Mead fits, I re-ran all those fits again without the oversample PSF and computed the S\'ersic index excess, i.e., the value of $n$ when no oversample PSF is employed minus $n$ when an oversampled PSF is included in the fit. Figure \ref{fig:nexcess} shows the distribution of the S\'ersic index excess. While a similar exercise with a larger number of galaxies is necessary to corroborate the observed trend, the results indicate that $n$ can be overestimated when the fit does not include an oversampled PSF. This may have caused or contributed to the systematic offsets found here in Fig.\,\ref{fig:Compn_lit} and Table \ref{tab:local}.

\subsection{Unsupervised Differential Evolution Fits}
\label{subsec:de}

\begin{figure}
\includegraphics[trim=0.2cm 0.2cm 0.2cm 0.2cm,clip=true,width=\columnwidth]{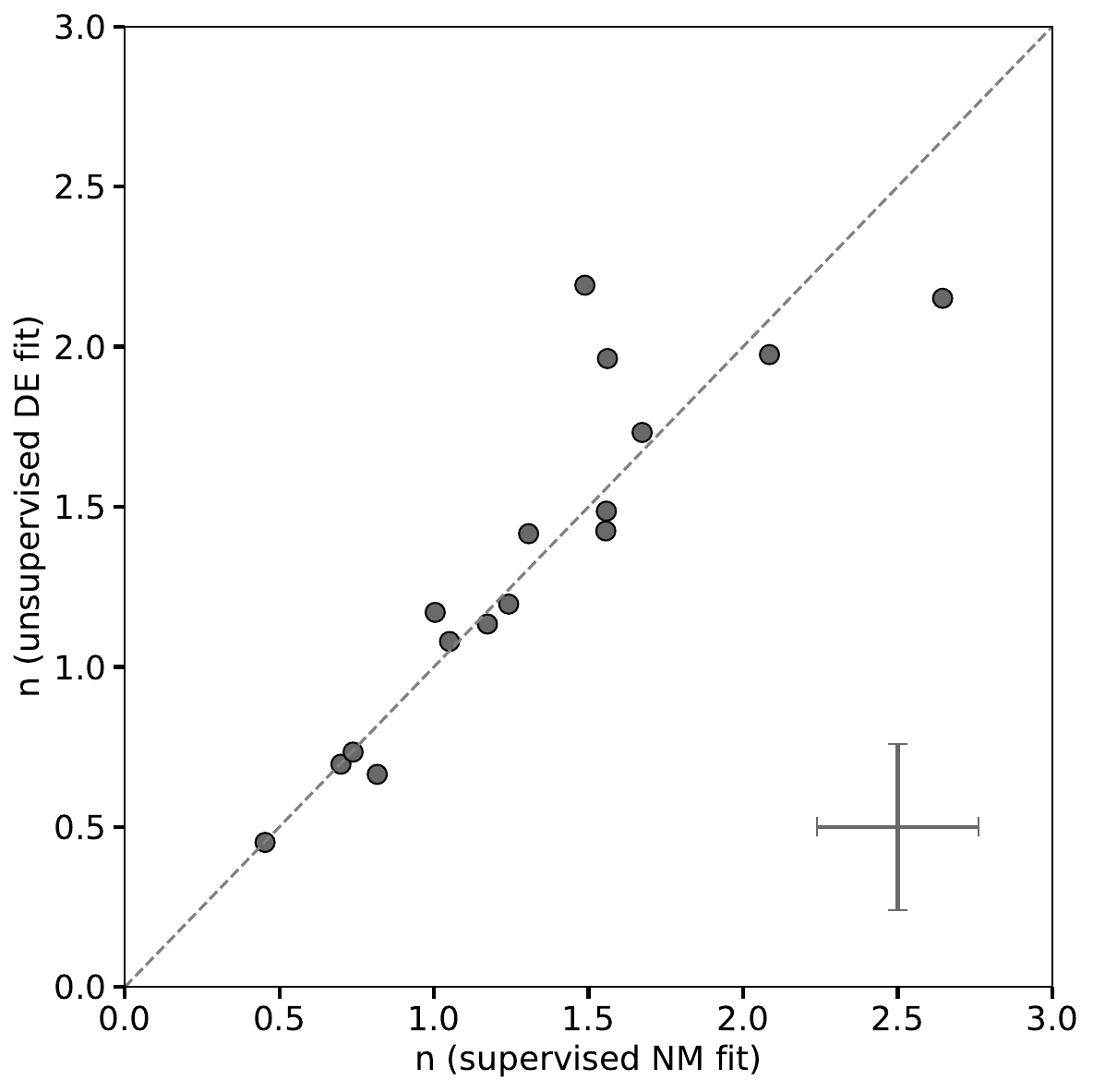}
\caption{Photometric bulge S\'ersic index $n$ obtained employing the Differential Evolution (DE) algorithm in unsupervised fits plotted against the values derived via the supervised Nelder-Mead (NM) fits. Also plotted on the bottom right corner are fiducial errors bars derived by \citet{KimGadShe14} using {\sc budda}. The relative average error on $n$ found by Kim et al. corresponds to 13\%, which, for the galaxies studied here, implies a mean 1$\sigma$ error of 0.26. This figure concerns only the fits to the original S$^4$G images and not the redshifted images.}
\label{fig:Compn_NMDE}
\end{figure}

\begin{figure*}
\includegraphics[trim=0.2cm 0.2cm 0.2cm 0.2cm,clip=true,width=\columnwidth]{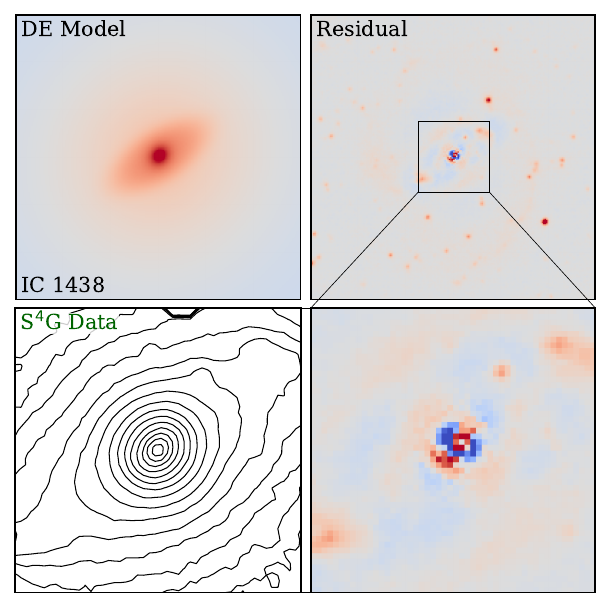}
\includegraphics[trim=0.2cm 0.2cm 0.2cm 0.2cm,clip=true,width=\columnwidth]{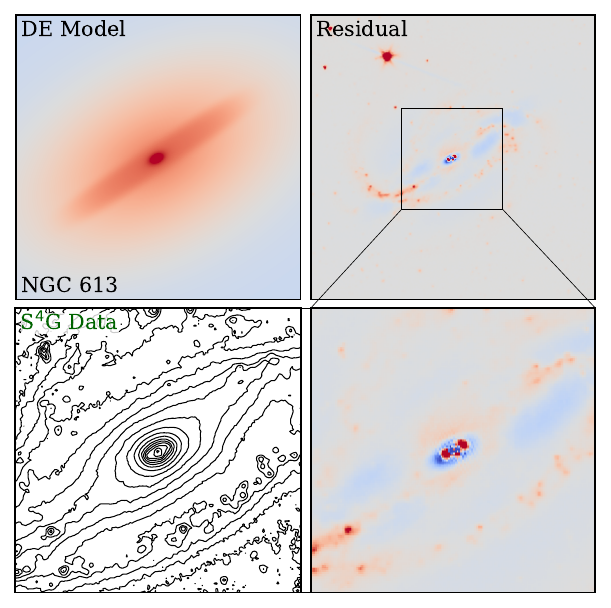}
\caption{Results from the unsupervised DE fits for each galaxy, as indicated. {\it Top left:} Full model; {\it top right}: residual image after subtracting the full model from the original image; {\it bottom left:} isophotal contours in zoomed-in original image; and {\it bottom right:} zoomed-in, central region of residual image. The black square in the top right panel indicates the zoomed-in region shown in the two bottom panels. In the residual image, red hues indicate regions where the model is fainter than the original image (e.g., where structures such as nuclear rings, nuclear bars and nuclear spiral arms dominate, as they are not included in the model). Conversely, blue shades indicate regions where the model is brighter than the original image. The residual image is shown with an augmented contrast to enhance the differences between the original image and the full model.}
\label{fig:modres}
\end{figure*}

\begin{figure*}
\includegraphics[trim=0.2cm 0.2cm 0.2cm 0.2cm,clip=true,width=\columnwidth]{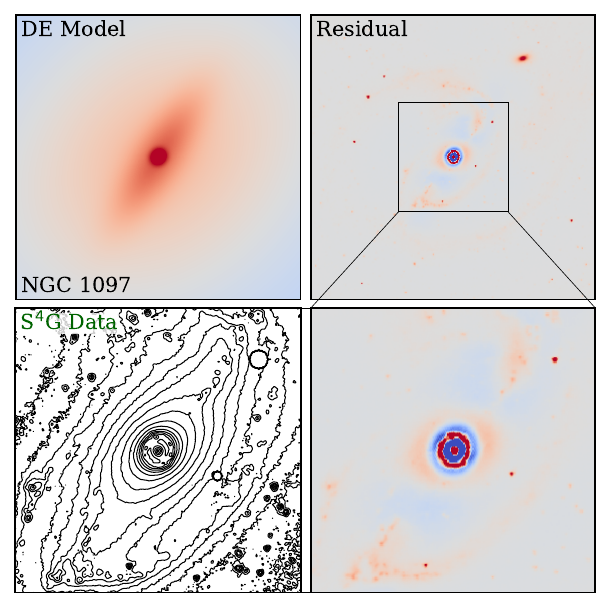}
\includegraphics[trim=0.2cm 0.2cm 0.2cm 0.2cm,clip=true,width=\columnwidth]{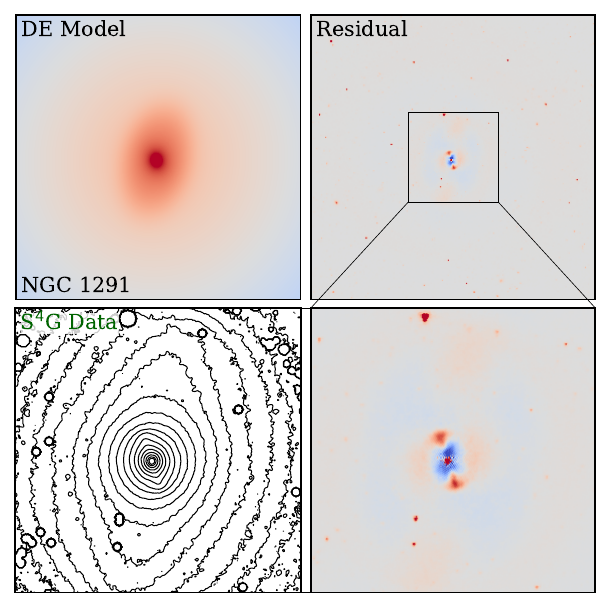}
\addtocounter{figure}{-1}
\caption{continued.}
\end{figure*}

\begin{figure*}
\includegraphics[trim=0.2cm 0.2cm 0.2cm 0.2cm,clip=true,width=\columnwidth]{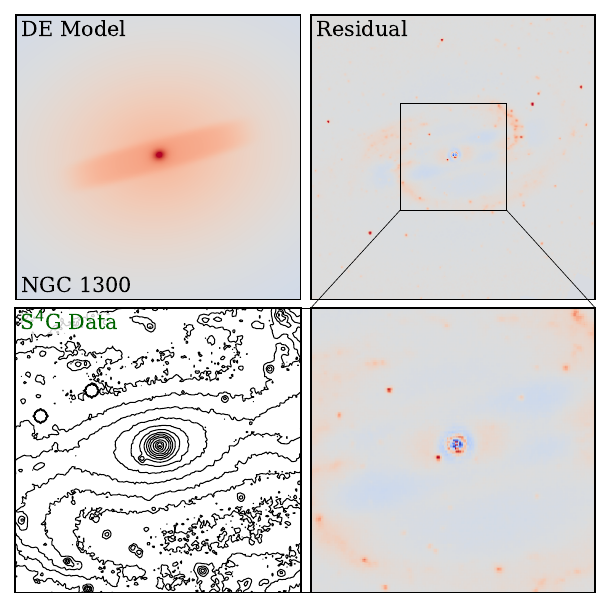}
\includegraphics[trim=0.2cm 0.2cm 0.2cm 0.2cm,clip=true,width=\columnwidth]{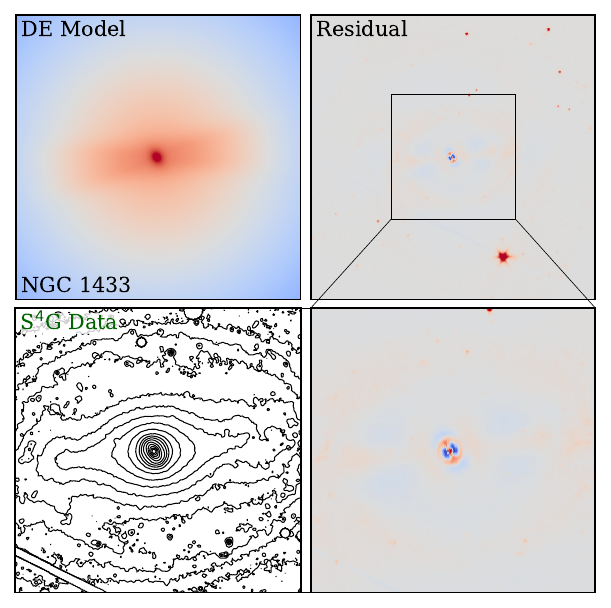}
\addtocounter{figure}{-1}
\caption{continued.}
\end{figure*}

\begin{figure*}
\includegraphics[trim=0.2cm 0.2cm 0.2cm 0.2cm,clip=true,width=\columnwidth]{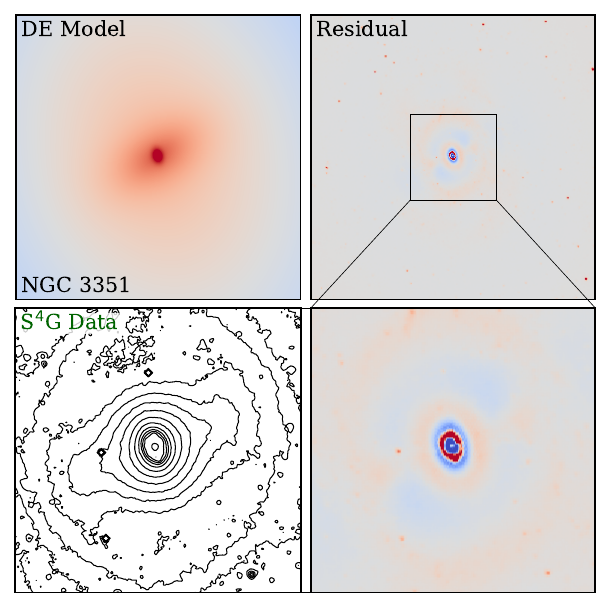}
\includegraphics[trim=0.2cm 0.2cm 0.2cm 0.2cm,clip=true,width=\columnwidth]{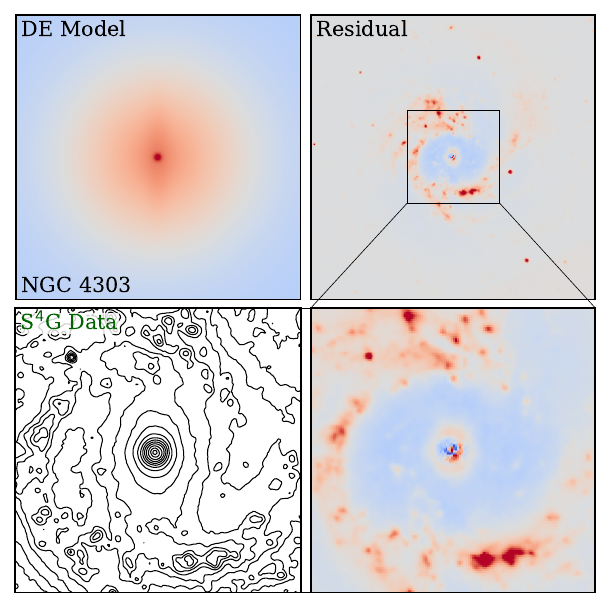}
\addtocounter{figure}{-1}
\caption{continued.}
\end{figure*}

\begin{figure*}
\includegraphics[trim=0.2cm 0.2cm 0.2cm 0.2cm,clip=true,width=\columnwidth]{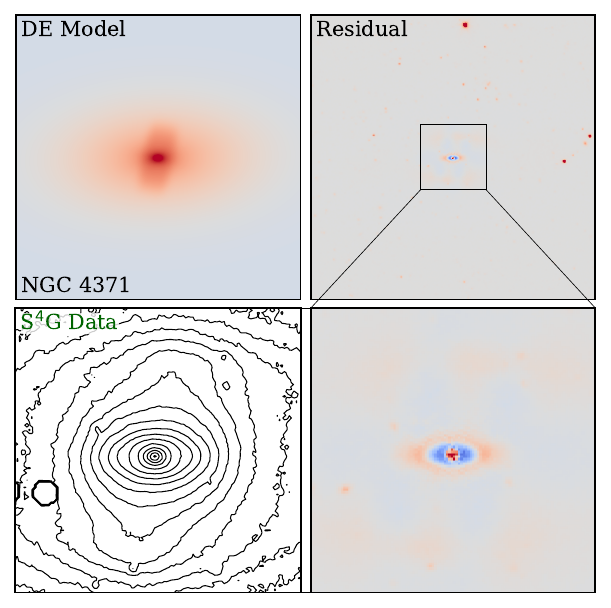}
\includegraphics[trim=0.2cm 0.2cm 0.2cm 0.2cm,clip=true,width=\columnwidth]{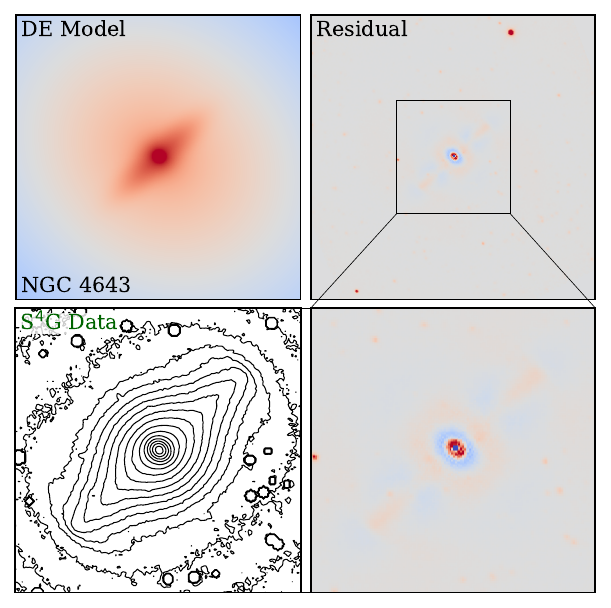}
\addtocounter{figure}{-1}
\caption{continued.}
\end{figure*}

\begin{figure*}
\includegraphics[trim=0.2cm 0.2cm 0.2cm 0.2cm,clip=true,width=\columnwidth]{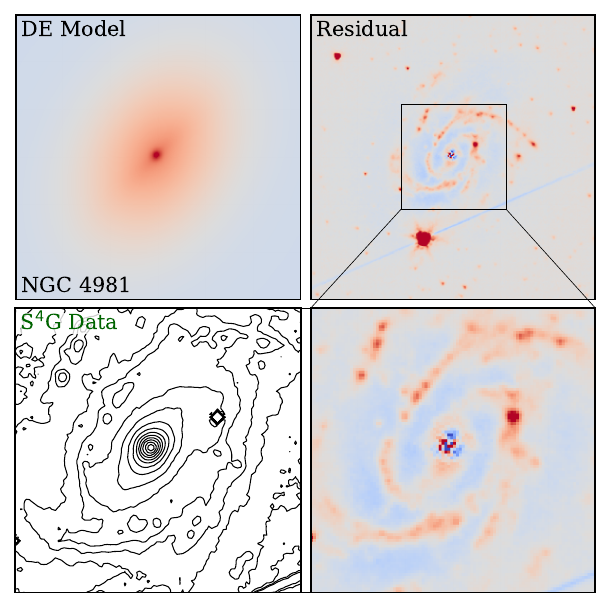}
\includegraphics[trim=0.2cm 0.2cm 0.2cm 0.2cm,clip=true,width=\columnwidth]{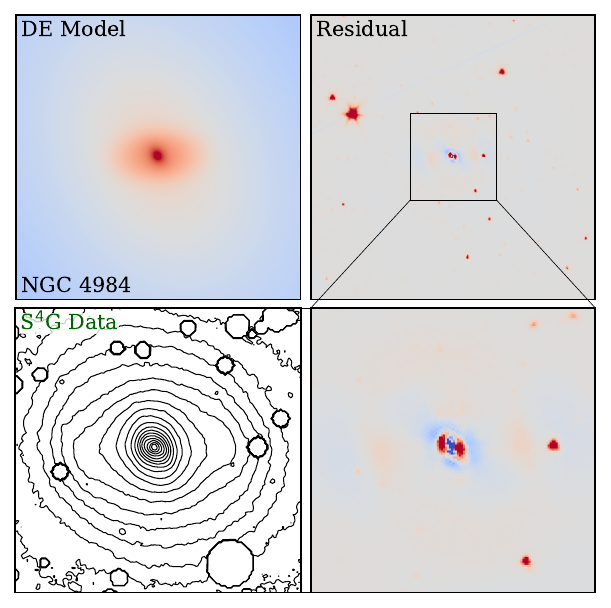}
\addtocounter{figure}{-1}
\caption{continued.}
\end{figure*}

\begin{figure*}
\includegraphics[trim=0.2cm 0.2cm 0.2cm 0.2cm,clip=true,width=\columnwidth]{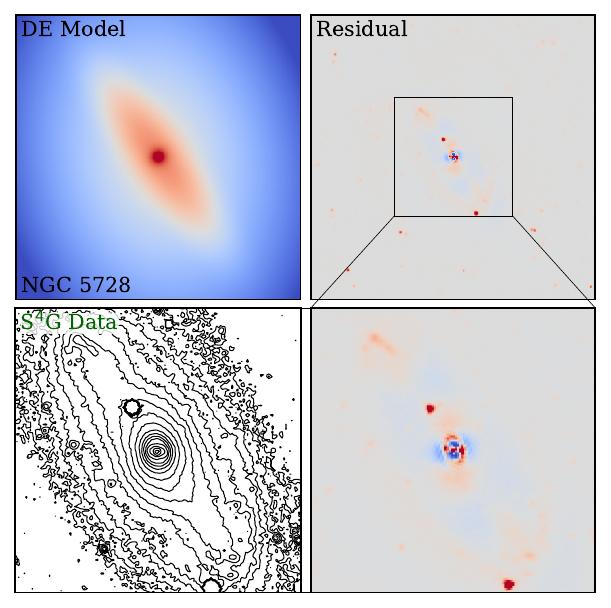}
\includegraphics[trim=0.2cm 0.2cm 0.2cm 0.2cm,clip=true,width=\columnwidth]{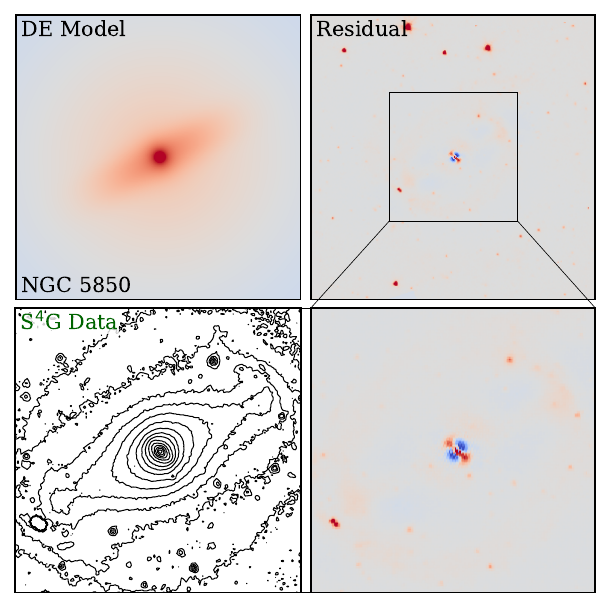}
\addtocounter{figure}{-1}
\caption{continued.}
\end{figure*}

\begin{figure*}
\includegraphics[trim=0.2cm 0.2cm 0.2cm 0.2cm,clip=true,width=\columnwidth]{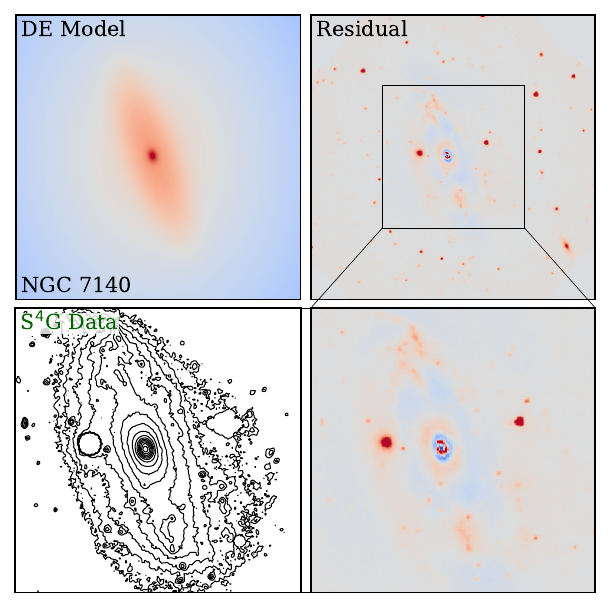}
\includegraphics[trim=0.2cm 0.2cm 0.2cm 0.2cm,clip=true,width=\columnwidth]{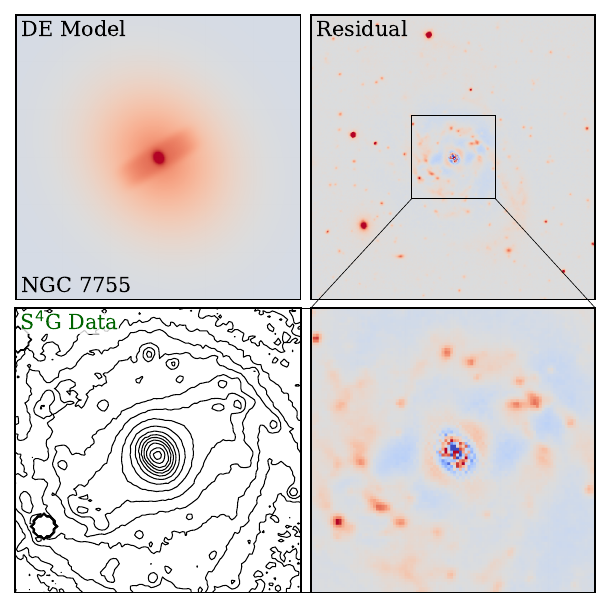}
\addtocounter{figure}{-1}
\caption{continued.}
\end{figure*}

The next set of fits presented here employing the original S$^4$G galaxy images makes use of the Differential Evolution (DE) algorithm \citep{Storn1997} available within {\sc imfit}, also using the Cash statistic. The goal is to verify if by using this algorithm one could remove altogether the need for an iterative, supervised procedure to produce accurate galaxy image decompositions. The DE algorithm does not require initial guesses but rather both a lower and an upper limit to all free parameters, and therefore it allows for a more objective approach to image decomposition. It is a genetic algorithm that is powerful against local minima in the model optimisation process, but that however comes at a cost: it is one order of magnitude slower than the NM algorithm employed above and two orders of magnitude slower than the LM algorithm. To generate the initial sampling of the multidimensional parameter space, I opted to use the Latin Hypercube sampling method rather than the default uniform sampling, since the former did result in more satisfactory fits in some test cases.

The lower and upper limits of the fitted parameters were determined with a principle in mind, namely, to be as least restrictive as possible but at the same time rejecting absurd models from the start. Thus, concerning the geometric parameters (position angle and ellipticity) of the different structural components, the lower and upper limits are such that they include values that, e.g., visually, are clearly beyond a possible range of correct values, but not overly off. The limits for the position angle then include all possibilities in the quadrant within which the position angle is identified visually or with the radial profile of position angle from the ellipse fits. Likewise, the limits for the ellipticity include values much rounder and much more eccentric than what a visual inspection or the ellipticity radial profile from the ellipse fits would allow for. To give an example, for the photometric bulge, bar and disc in NGC\,1300 the lower and upper limits for the position angle and ellipticity are, respectively: $80-160$, $0.0-0.3$, $90-160$, $0.5-0.9$, $90-180$ and $0.0-0.3$ (cf. Fig\,\ref{fig:ellfits}). A similar approach is followed to determine the lower and upper limits for intensities and scale parameters, including a careful visual inspection of the image, and stipulating ranges as wide as reasonably possible. One may argue that subjectivity still remains in this procedure in the choice for those limits, which is, to some extent, correct. However, there is much less scope for fine-tuning these limits as compared to choosing initial guesses, particularly following the principle described above. Importantly, the lower and upper limits for $n$ are the same for all galaxies, namely: $0.3-5.0$. The limits for $n_{\mathrm{bar}}$ and $c_0$ are also the same for all galaxies and are defined considering the results presented in \citet{Gad11}, which shows the distribution of these parameters obtained from photometric decompositions of nearly 300 barred galaxies. The limits are thus, for $n_{\mathrm{bar}}$ and $c_0$, respectively: $0.1-1.2$ and $0-2$. Finally, the allowed ranges for the galaxy centre and residual background correction and naturally much stricter. The galaxy centre is allowed to vary only $\pm$1 pixel from the brightest pixel, and the residual background correction is restricted to $\pm$7\% of the original subtracted background. These limits were chosen from initial tests, which showed that allowing for wider ranges is not necessary in this study.

Before defining the principle and limits mentioned above, I have tested much wider ranges, with the goal of probing the limits of the DE algorithm. For example, I tested with ranges for the position angle and ellipticity as wide as, respectively, $0-180$ and $0-1$. In addition, I also tested limits to the scale parameters covering the whole galaxy image. However, with these limits, the results would often be unsatisfactory. This shows that, naturally, the DE algorithm also has its limits.

The DE fits were ran only once, i.e., there was no adjustment of the ranges of any free parameter in an attempt to improve the fits. That was actually not necessary: these unsupervised DE fits proved to be very accurate when compared to the supervised NM fits. As an example, I show in Fig.\,\ref{fig:Compn_NMDE} the value of $n$ thus obtained as compared to the supervised NM fits. The agreement is striking, and the typical scatter smaller than the fiducial statistical uncertainty from \citet{KimGadShe14}. In fact, Table \ref{tab:local} shows that the average difference in $n$ is only 0.03, with a standard deviation of 0.30. Both values are very similar to the corresponding values in the comparison between the supervised NM fits and the results from the supervised fits of \citet{SalLauLai15}. It is worth stressing that the {\em typical} deviations from the one-to-one correlation line in Fig.\,\ref{fig:Compn_NMDE} are significantly smaller than the fiducial error bars (0.26), and the computed scatter of 0.30 in Table \ref{tab:local} is driven by a few more deviant cases. However, in the latter cases, a visual inspection shows that the unsupervised DE fits are actually more satisfactory than the supervised NM fits. This suggests that even with the iterative process, in these cases, the supervised NM fits got trapped in non-optimal local minima.

Figure \ref{fig:modres} shows for each galaxy the full model obtained via the unsupervised DE fits along with the corresponding residual image. The latter are also shown zoomed-in on the central region of the galaxy, for which the figure also shows isophotal contours. These zoomed-in panels allow one to see central structures not included in the models, such as nuclear rings (e.g., NGC\,1097) and nuclear bars (e.g., NGC\,5850). These structures add to the evidence for the presence of nuclear discs in these galaxies, and are also a reminder that even more accurate models can be obtained for these galaxies with even more complex fits including additional structural components. In fact, the presence of central bright residuals is generally a good diagnostic to indicate the presence of nuclear discs, since the latter often host sub-structures such as nuclear bars, nuclear rings or nuclear spiral arms, which may be difficult to model. Classical bulges, on the other hand, generally present a smoother distribution of stellar density and are well fitted with a single S\'ersic function, leading to faint residuals or no residuals at all \citep[see discussion on residuals in, e.g.,][]{LauSalBut05,Gad08,Erwin2015}. However, it should be noted that the absence of bright residuals does not imply the presence of a classical bulge, since some nuclear discs do not seem to host prominent sub-structures. Another way to improve the fits is to account for breaks in the main disc intensity radial profile, as in \citet{KimGadShe14}. However, these are generally more challenging fits and not pursued by most studies, in particular studies beyond the nearby Universe. I thus chose not to overly complicate the fits presented here, so that the results from this paper are more generally comparable with other results in the literature. In Appendix \ref{app:params}, I present the complete set of structural parameters derived for the photometric bulge, disc and bar with these DE fits for the entire sample.

\subsection{Unsupervised Markov Chain Monte Carlo Fits}
\label{subsec:mcmc}

\begin{figure}
\includegraphics[trim=0.2cm 0.2cm 0.2cm 0.2cm,clip=true,width=\columnwidth]{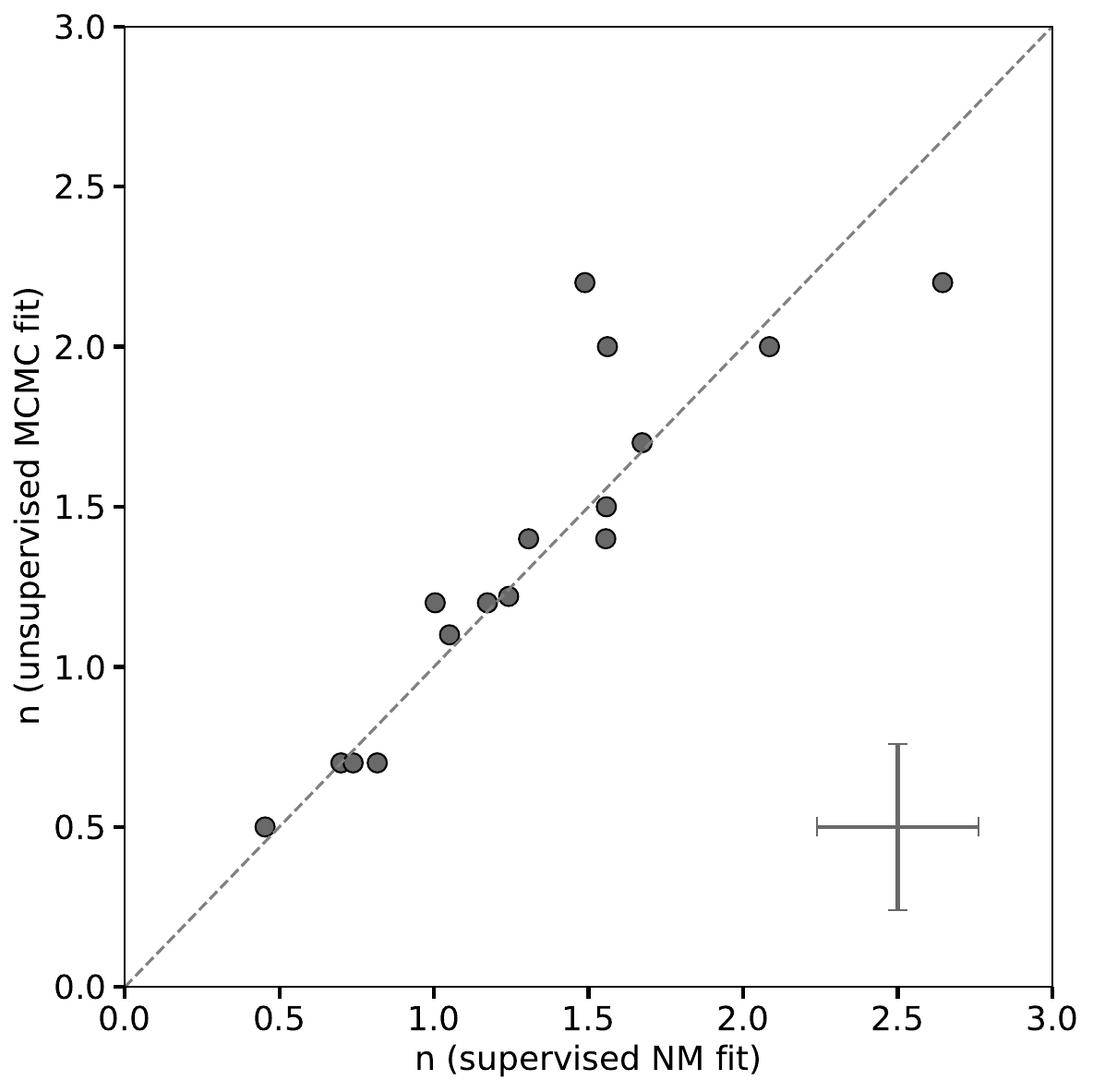}
\caption{Bulge S\'ersic index $n$ obtained employing the Markov Chain Monte Carlo (MCMC) algorithm in unsupervised fits plotted against the values derived via the supervised Nelder-Mead (NM) fits. Also plotted on the bottom right corner are fiducial errors bars derived by \citet{KimGadShe14} using {\sc budda}. The relative average error on $n$ found by Kim et al. corresponds to 13\%, which, for the galaxies studied here, implies a mean 1$\sigma$ error of 0.26. This figure concerns only the fits to the original S$^4$G images and not the redshifted images.}
\label{fig:Compn_NMMCMC}
\end{figure}

\begin{figure*}
\includegraphics[trim=0.2cm 0.2cm 0.2cm 0.2cm,clip=true,width=1.35\columnwidth]{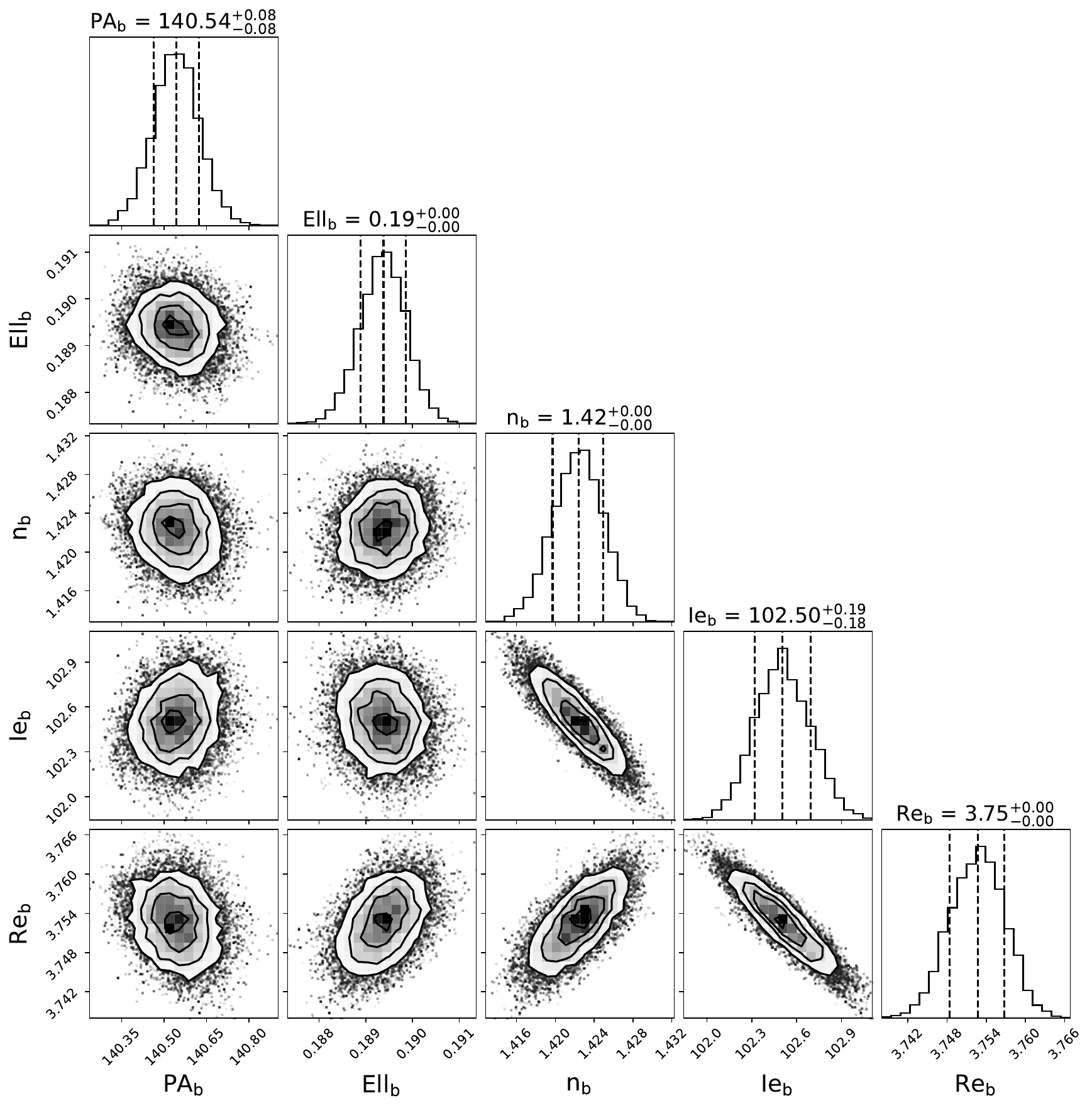} 
\caption{Results from the MCMC run on IC\,1438 employing the original S$^4$G image concerning the photometric bulge parameters in the last 5000 likelihood evaluations. Errors quoted correspond to 1$\sigma$.}
\label{fig:cornerbulge}
\end{figure*}

\begin{figure*}
\includegraphics[width=2\columnwidth]{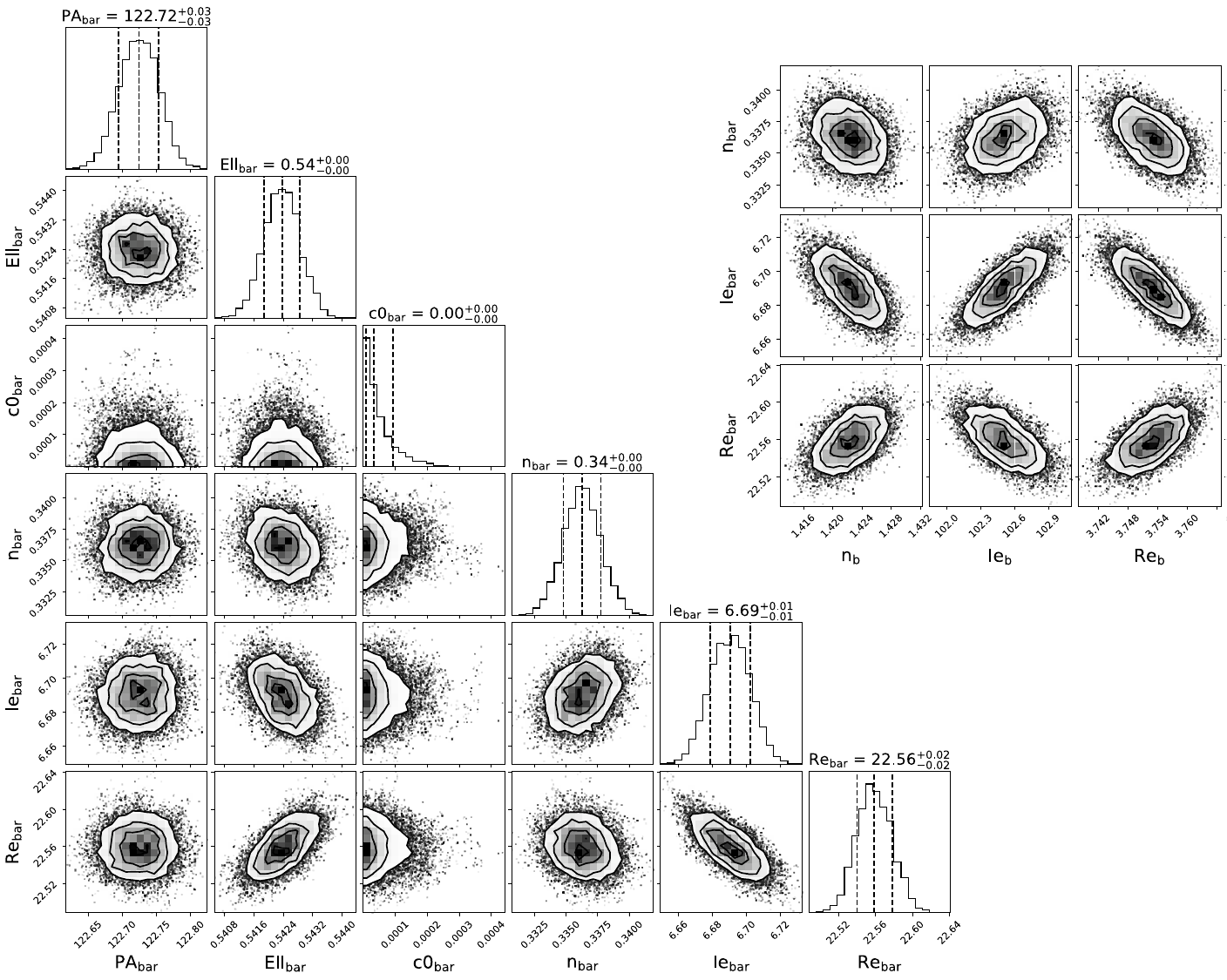} 
\caption{{\it Left:} same as Fig.\,\ref{fig:cornerbulge} but concerning the bar parameters. {\it Right:} relations between the posterior distributions of the main parameters of the photometric bulge and bar models.}
\label{fig:cornerbar}
\end{figure*}

\begin{figure*}
\includegraphics[width=1.8\columnwidth]{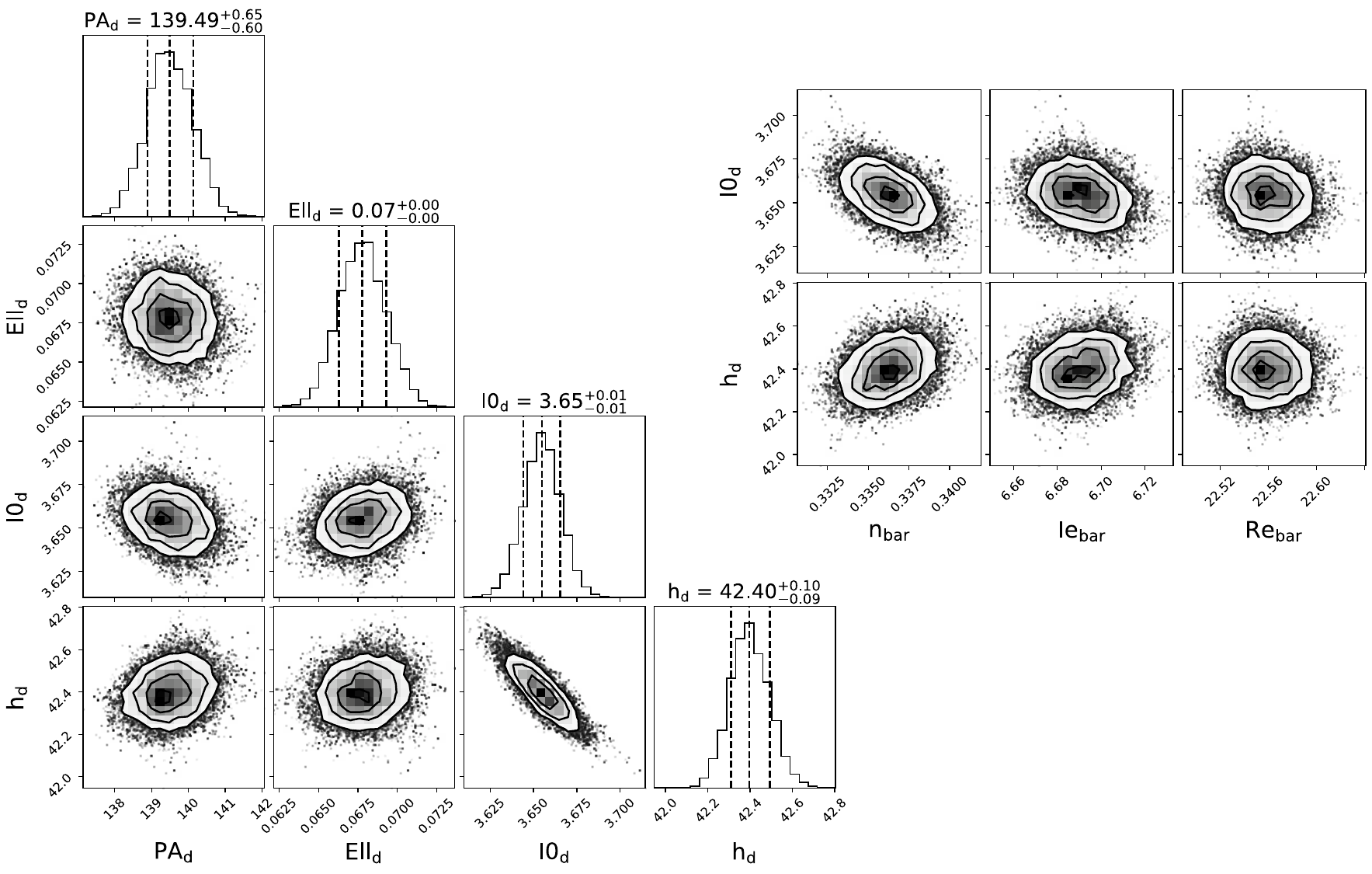} 
\caption{{\it Left:} same as Fig.\,\ref{fig:cornerbulge} but concerning the disc parameters. {\it Right:} relations between the posterior distributions of the main parameters of the disc and bar models.}
\label{fig:cornerdisc}
\end{figure*}

The final set of fits presented here employing the original S$^4$G galaxy images makes use of the Markov Chain Monte Carlo (MCMC) algorithm available within {\sc imfit}. The approach used by {\sc imfit} is an adaptation of the DE optimisation algorithm to MCMC, namely the DiffeRential Evolution Adaptive Metropolis (DREAM) algorithm of \citet[][see also \citealt{terBraak2006} and \citealt{Vrugt2016}]{Vrugt2009}. I used the default configuration of this implementation, which calculates one chain for each free parameter, produces 5000 generations for each chain in the ``burn-in'' phase, and produces up to 100000 generations for each chain or stops before if convergence is found. Most fits in fact reach this maximum number of generations and never achieve formal convergence. The fits use the Poisson Maximum-Likelihood-Ratio statistic for minimisation, which is similar to the Cash statistic but is always $\geq0$. The valid range for each fitted parameter is identical to that used in the DE fits. To compute the ``best fit'' result from the MCMC realisations, I take the mean value of each parameter using the last 5000 generations produced in the chain. These values end up being very similar to those found with the DE algorithm. This can be seen in Fig.\,\ref{fig:Compn_NMMCMC}, which compares the photometric bulge S\'ersic index determined from the MCMC realisations with those obtained with the NM fits (cf. Fig.\,\ref{fig:Compn_NMDE}, which makes the corresponding comparison between the DE and NM fits). In addition, Table \ref{tab:local} shows almost identical values for the comparisons between the DE and MCMC fits with the supervised NM fits.

However, the MCMC fits produce of course posterior distributions, which can be used to evaluate uncertainty ranges and correlations between model parameters. This, again, comes with a cost: the MCMC realisations take on average about 30 times longer than the DE fits\footnote{Using as a benchmark a fit involving $\approx$150k pixels and 17 free parameters, the LM algorithm (employing the $\chi^2$ statistic) takes 72s to converge. The NM algorithm (using the Cash statistic) takes 237s, and the DE algorithm (also using the Cash statistic)  takes 5025s. These benchmark runs were performed on a MacBook Pro 18 with an 8-core Apple M1 Pro chip.}. Figures \ref{fig:cornerbulge}, \ref{fig:cornerbar} and \ref{fig:cornerdisc} show such posterior distributions for the photometric bulge, bar and disc parameters of IC\,1438, respectively, for the last 5000 generations. While this galaxy is used here as an example, the other galaxies in the sample show generally similar results. The errors quoted in the figures correspond to 1$\sigma$, and one can see that, after tens of thousands of likelihood evaluations, the {\em statistical} uncertainties of most model parameters are negligible. It must be noted, however, that this does not take systematic uncertainties into account. More sophisticated models can lead to different optimised values for some of the model parameters, e.g., by taking into account the presence of AGN \citep{Gad08}, rings \citep{GaoHo17}, spiral arms \citep{Chugunov2024}, disc breaks \citep{KimGadShe14}, barlenses \citep{Li2025}, and dust \citep{GadBaeFal10,PasPopTuf13}.

Figures \ref{fig:cornerbulge} and \ref{fig:cornerbar} show the relations between the S\'ersic index, effective radius and intensity at the effective radius of the photometric bulge and bar model components, which are both modelled with a S\'ersic function. \citet{TruGraCao01} discussed the strong relations in the case of the photometric bulge, and concluded that those are not artificially introduced by the use of a S\'ersic function, but arise from relations between model-independent parameters. Interestingly, the relations concerning the bar parameters appear much less significant. In particular, the bar S\'ersic index seems to largely be independent of its effective radius and intensity at the effective radius. In the case of the disc, modelled with an exponential function, Fig.\,\ref{fig:cornerdisc} shows that the central intensity and scale length are strongly anti-correlated. Figure \ref{fig:cornerbar} also shows the relations between the main parameters of the photometric bulge and bar, while Fig.\,\ref{fig:cornerdisc} also shows the relations between the main parameters of the disc and bar. While some strong relations are found, the S\'ersic indices of the photometric bulge and bar are seen to be largely independent. No other significant relations were found. For example, I find no significant relations between the parameters of the photometric bulge and disc. A more in-depth discussion on these relations is left for a future study.

\subsection{The recovery of parameters other than the S\'ersic index}

\begin{figure*}
\includegraphics[width=2\columnwidth]{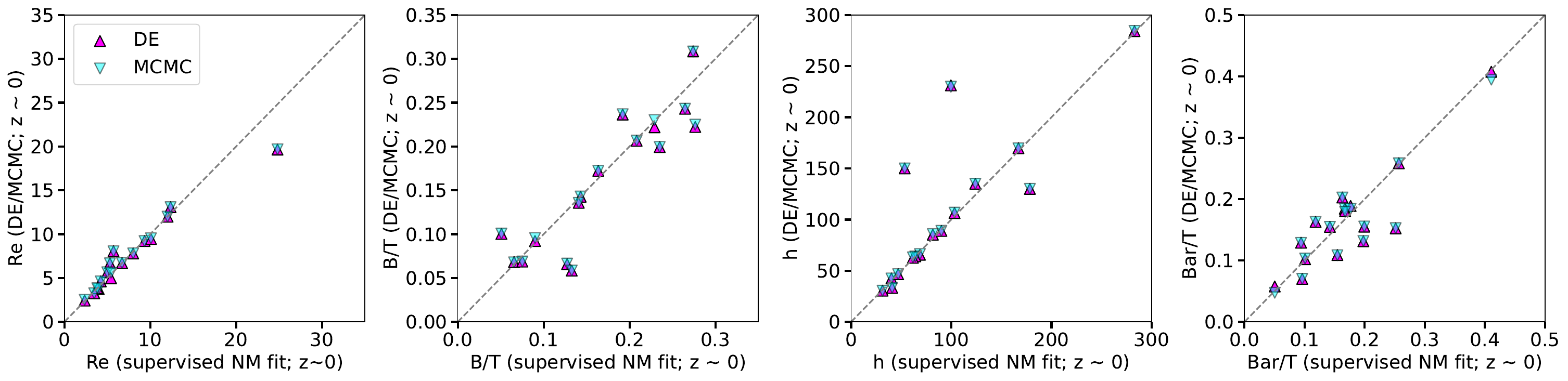}
\caption{Photometric bulge effective radius, photometric bulge-to-total luminosity ratio, disc scale length, and the bar-to-total luminosity ratio as derived using the original S$^4$G images with the unsupervised DE and MCMC fits (as indicated), compared to the corresponding results from the supervised NM fits. $r_e$ and $h$ are in pixels.}
\label{fig:Compparm0}
\end{figure*}

While the main focus of this paper is to assess how robustly nuclear discs can be identified through the S\'ersic index in photometric decompositions, it is also very instructive to evaluate how well other structural parameters fitted in such decompositions are recovered in the tests of different algorithms performed here. I will focus on four other structural parameters, namely, the photometric bulge effective radius $r_e$, the bulge-to-total luminosity ratio B/T, the disc scale length $h$ and the bar-to-total luminosity ratio Bar/T. In the case of the decompositions employing the original S$^4$G images, this is shown in Fig.\,\ref{fig:Compparm0} for the unsupervised DE and MCMC fits compared with the reference decompositions, i.e., the supervised NM fits. The figure shows that the recovery of these additional structural parameters is generally very good, and virtually identical for the unsupervised DE and MCMC fits, as was the case of the S\'ersic index of the photometric bulge (discussed above). Of notice are a few outliers in the recovery of $h$ -- but with excellent agreement otherwise -- and somewhat larger scatter in the luminosity ratios.

\section{Galaxy Image Decompositions Beyond the Nearby Universe}
\label{sec:beyond}

To test how accurately one can retrieve the structural parameters of photometric bulges, bars and discs for galaxies beyond the nearby Universe, the decompositions were repeated with the images artificially redshifted to $z\approx0.05$ and $z\approx0.1$. I reiterate here again that these redshifted images assume the same PSF FWHM as that of the original S$^4$G images, and thus, of course, the results presented here are not representative of decompositions that can be done using images with higher spatial resolution, for example from HST or JWST at these low redshifts. Nonetheless, as shown below, these tests give a very good indication of how a poorer physical spatial resolution can affect the results from such decompositions. In addition, other effects, such as band-shifting and cosmological surface brightness dimming -- which is very substantial for higher redshifts -- are not accounted for in these tests.

With the enlargement of the PSF FWHM in the redshifted images, there is no need for an oversampled PSF in these fits. Most of the NM fits are now to a large extent unsupervised and performed only once, but they employ for each galaxy the last set of initial guesses employed for the best corresponding NM fit described above. Accordingly, the DE and MCMC fits of the redshifted images also use the same lower and upper limits of each fitted parameter in the corresponding DE and MCMC fits discussed above. However, in some cases, the initial guesses for the NM fits at $z\approx0.05$ had to be adjusted so that a better fit could be achieved, and in these cases, this would be propagated to the fits at $z\approx0.1$. 

Furthermore, in some cases, due to the poorer spatial resolution, a few decompositions employing the images redshifted to $z\approx0.1$ do not include a bar or a photometric bulge component. This is the case for NGC\,3351 and NGC\,4981, for which there is no sign of a bar at $z\approx0.1$ (see Fig.\,\ref{fig:input}), and NGC 4303, whose fitted photometric bulge is so small at $z\approx0.1$ that the bulge-to-total ratio of the models fitted is below 0.1\%. Another interesting case is that of NGC\,4984, whose bar becomes quite rounder at $z\approx0.1$ (see Fig.\,\ref{fig:input}), which is reflected in the fitted models, which yield values for the bar ellipticity below 0.2. Finally, the ellipticity of the disc of NGC\,5728 had to be kept fixed in the fits employing the image redshifted to $z\approx0.1$, lest it would become overly eccentric. These occurrences will be discussed further below, but I note here already that no such issues occurred with the fits employing the images redshifted to $z\approx0.05$, except for the following: in the best DE fit for NGC\,7140, the models for the bar and disc components get swapped, in the sense that the exponential component is a factor of two more eccentric than the generalised S\'ersic component. Curiously, this does not happen in any other fit of NGC\,7140 or any other galaxy in this study, but this already shows that care must be exercised with fits where the physical spatial resolution is relatively poor, even if sophisticated optimisation algorithms are employed.

\begin{figure*}
\includegraphics[trim=0.2cm 0.2cm 0.2cm 0.2cm,clip=true,width=2\columnwidth]{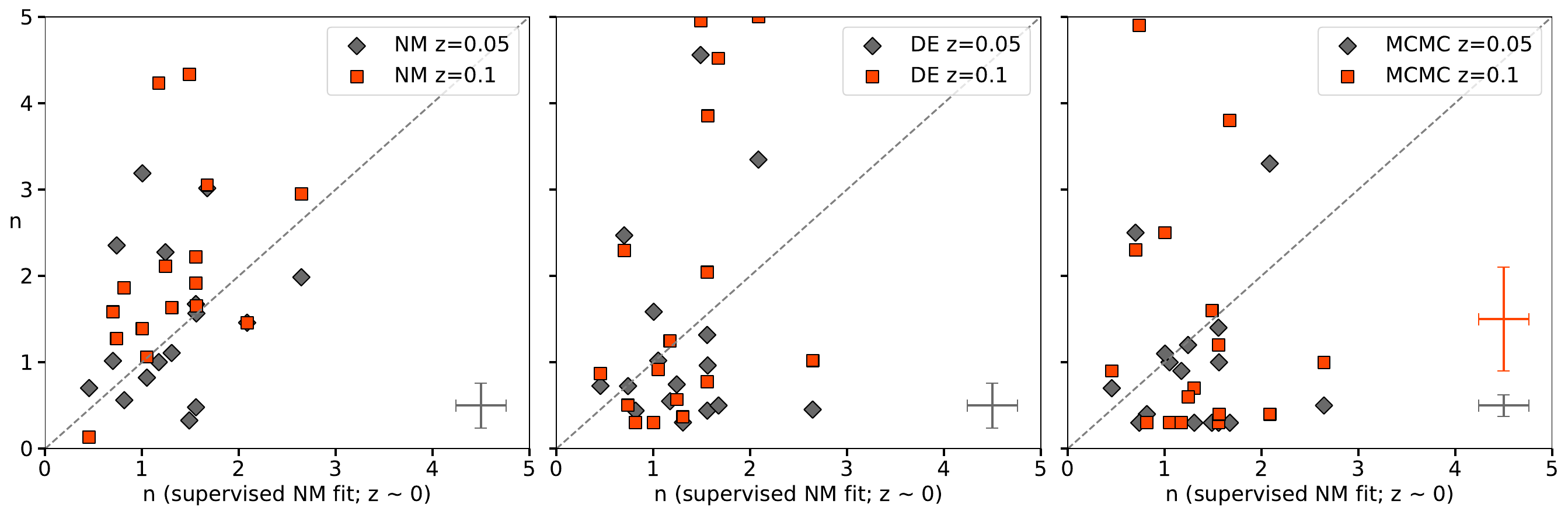}
\caption{Comparison between the photometric bulge S\'ersic index value derived from the two sets of redshifted images, as indicated, and the value derived with the Nelder-Mead supervised fits on the original S$^4$G images. The left panel shows the values derived when the redshifted images are fitted with the Nelder-Mead algorithm, while the central panel shows the values corresponding to the Differential Evolution algorithm, and the right panel shows the values obtained with the Markov Chain Monte Carlo algorithm. Also plotted on the bottom right corner of the left and central panels are fiducial errors bars derived by \citet{KimGadShe14} using {\sc budda}. The relative average error on the bulge S\'ersic index found by Kim et al. corresponds to 13\%, which, for the galaxies studied here, implies a mean 1$\sigma$ error of 0.26. In the right panel, the mean error bars are derived with the MCMC algorithm and are different for the two sets of redshifted images (0.12 for $z\approx0.05$ and 0.60 for $z\approx0.1$).}
\label{fig:Compn_highz}
\end{figure*}

Figure \ref{fig:Compn_highz} and Table \ref{tab:beyond} summarise the main results from the decompositions with the redshifted images in what concern the S\'ersic index $n$ of the photometric bulge component. It is clear that the scatter in the measurements of $n$ is much larger in the redshifted images when compared to our reference values obtained with the supervised NM fits of the original S$^4$G images (i.e., at $z\sim0$), and this is more so for the DE and MCMC fits (although at $z\approx0.05$ the NM and MCMC fits have similar scatter; compare the standard deviation values in Tables \ref{tab:beyond} and \ref{tab:local}). It is not surprising the the scatter is lower with the NM fits, since they employ as initial guesses the best fit results of the NM fits at $z\sim0$. Overall, the scatter in the fits employing the redshifted images, as parameterised with the standard deviation, is a factor $2-6$ larger than the scatter between different studies using the original (or similar) images. In addition, the NM fits of the images redshifted to both $z\approx0.05$ and $z\approx0.1$, as well as the DE fits of the images redshifted to $z\approx0.1$, yield systematically larger values for $n$. Altogether, this severely impacts the use of the S\'ersic index of the photometric bulge to distinguish classical bulges and nuclear discs. In the next section, I discuss how the results from the fits using the redshifted images, when compared to the results from the fits of the original S$^4$G images, demonstrate the difficulties in establishing the structural properties of galaxies beyond the nearby Universe, if the images employed do not provide enough physical spatial resolution.

\begin{figure*}
\includegraphics[width=2\columnwidth]{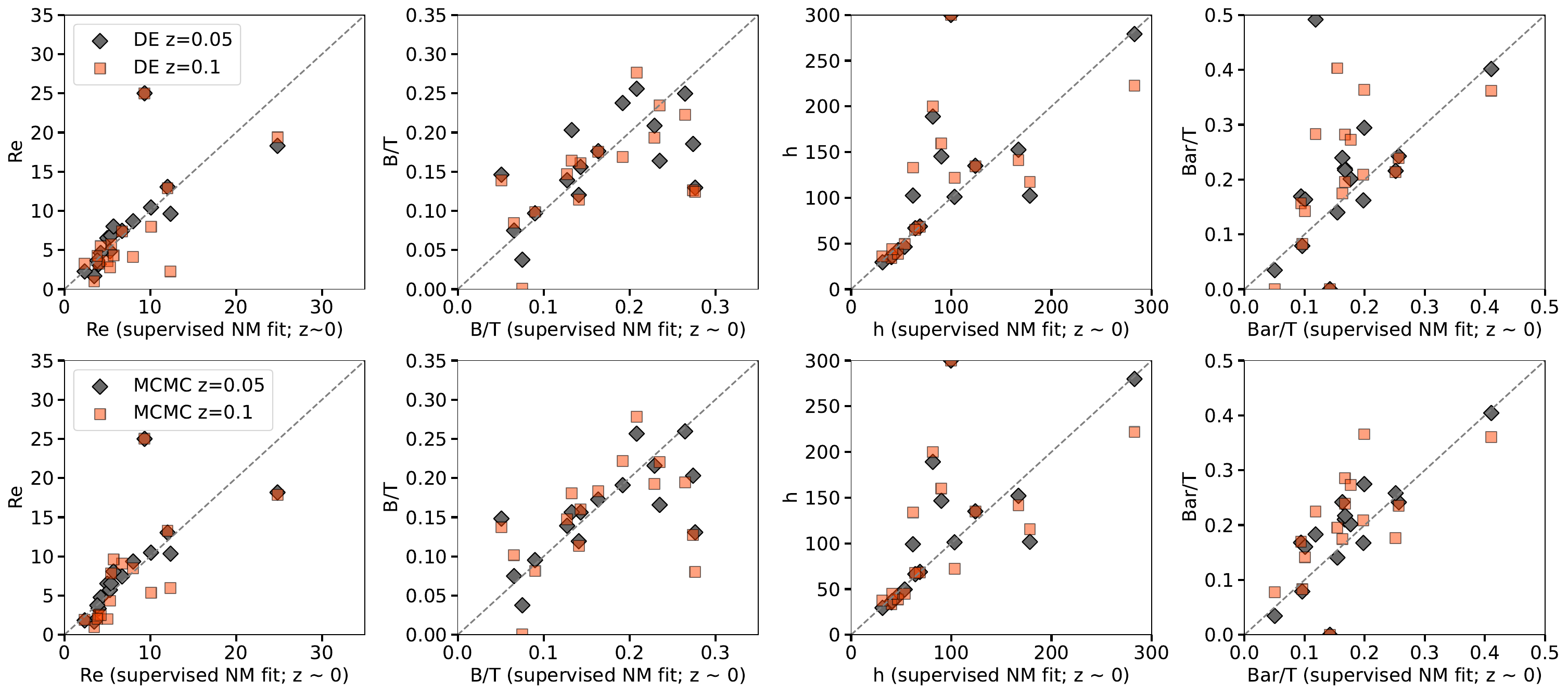}
\caption{Photometric bulge effective radius, photometric bulge-to-total luminosity ratio, disc scale length, and the bar-to-total luminosity ratio as derived using the artificially redshifted images with the unsupervised DE (upper row) and MCMC (bottom row) fits, compared to the corresponding results from the supervised NM fits employing the original S$^4$G images. $r_e$ and $h$ are in pixels.}
\label{fig:Compparm}
\end{figure*}

As done in Fig.\,\ref{fig:Compparm0} for the unsupervised DE and MCMC fits compared to the supervised NM fits, all using the original S$^4$G images, Fig.\,\ref{fig:Compparm} compares the values of the photometric bulge effective radius, photometric bulge-to-total luminosity ratio, disc scale length, and the bar-to-total luminosity ratio obtained with the artificially redshifted images. The trends seen in this figure indicate that the recovery of these parameters is to some extent better than the recovery of the photometric bulge S\'ersic index (cf. Fig.\,\ref{fig:Compn_highz}). However, there is still significant scatter, as shown in Table \ref{tab:Compparm}. The table shows that the scatter in the derivation of $n$ can go beyond 100\%, whereas the scatter in the other structural parameters is in the range of 30\% to 70\%.

\begin{table}
\centering
\caption{Same as Table \ref{tab:local} but for the supervised NM and unsupervised DE and MCMC fits employing the redshifted images, as indicated, compared to the supervised NM fits of the original S$^4$G images.}
\label{tab:beyond}
\begin{tabular}{lcccc}
\hline
$z\approx0.05$ & & & \\
Parameter & NM & DE & MCMC \\
\hline
$\left<\Delta_n\right>$ & 0.15 & -0.06 & -0.34 \\
SD ($\Delta n$) & 0.95 & 1.27 & 0.97 \\
\hline
$z\approx0.1$ & & & \\
Parameter & NM & DE & MCMC \\
\hline
$\left<\Delta_n\right>$ & 0.72 & 0.53 & 0.02 \\
SD ($\Delta n$) & 1.03 & 1.59 & 1.60 \\
\hline
\end{tabular}
\end{table}

\begin{table}
\centering
\caption{Relative scatter in the recovery of structural parameters with the unsupervised DE and MCMC fits employing the redshifted images, as indicated. The scatter is computed as the standard deviation of the corresponding fits of the 16 galaxies divided by the mean values of the corresponding parameter as obtained with the supervised NM fits of the original S$^4$G images.}
\label{tab:Compparm}
\begin{tabular}{lcccc}
\hline
Param. & DE$_{\mathrm{(z\approx0.05)}}$ & MCMC$_{\mathrm{(z\approx0.05)}}$  & DE$_{\mathrm{(z\approx0.1)}}$  & MCMC$_{\mathrm{(z\approx0.1)}}$  \\
\hline
$n$     & 0.96 & 0.73 & 1.21 & 1.21 \\
$r_e$ & 0.59 & 0.58 & 0.68 & 0.68 \\
B/T     & 0.36 & 0.33 & 0.40 & 0.46 \\
$h$    & 0.64 & 0.64 & 0.69 & 0.71 \\
Bar/T & 0.63 & 0.33 & 0.58 & 0.46 \\
\hline
\end{tabular}
\end{table}

\section{Discussion}
\label{sec:discuss}

\begin{table}
\centering
\caption{The number of photometric bulges with the S\'ersic index $n$ above 2 (top) and 2.26 (middle), i.e., 2 plus 0.26, which corresponds to the fiducial error in the photometric bulge S\'ersic index for the sample used here (and with the original S$^4$G images), as derived by \citet[that is, 13\%]{KimGadShe14}. These numbers, shown for each set of fits in this paper, would thus roughly be the number of classical bulges found. At the very bottom, I also show the number of photometric bulges with $n>2.60$, but only for the fits employing the images redshifted to $z\approx0.1$, since 0.60 is the mean 1$\sigma$ error found with the MCMC fits of these images.}
\label{tab:class}
\begin{tabular}{lccc}
\hline
Number of photometric bulges with $n>2$ & \omit & \omit & \omit \\
\hline
Sample & NM & DE & MCMC \\
\hline
$z\sim0$   & 2 & 2 & 2 \\
$z\approx0.05$  & 4 & 3 & 2 \\
$z\approx0.1$    & 6 & 6 & 4 \\
\hline
Number of photometric bulges with $n>2.26$ & \omit & \omit & \omit \\
\hline
Sample & NM & DE & MCMC \\
\hline
$z\sim0$   & 1 & 0 & 0 \\
$z\approx0.05$  & 4 & 3 & 2 \\
$z\approx0.1$    & 4 & 5 & 4 \\
\hline
Number of photometric bulges with $n>2.60$ & \omit & \omit & \omit \\
$z\approx0.1$    & 4 & 4 & 2 \\
\hline
\end{tabular}
\end{table}

\subsection{Classifying Photometric Bulges as Classical Bulges or Nuclear Discs}

As mentioned in the Introduction, it is common practice to use the S\'ersic index $n$ to identify whether the major central stellar component in a disc galaxy (when it exists) is a classical bulge ($n>2$) or a nuclear disc ($n<2$). However, several studies have pointed out that this practice is limited in its accuracy \citep[e.g.,][]{KorKen04,Gad09b,FisDro16,Kormendy2016}. These studies thus caution against using only the S\'ersic index to make such an assessment, and show that a more robust approach involves combining a number of different criteria alongside the S\'ersic index. These include: (i), assessing the presence of nuclear sub-structures such as nuclear bars, nuclear rings and nuclear spiral arms, as discussed above in the context of residual images; (ii), measuring parameters related to the stellar kinematics, essentially verifying if the central component is kinematically hot or cold; (iii), verifying the existence of star formation or young stellar populations (often present in nuclear discs, but less likely in classical bulges); and (iv), identifying outliers in scaling relations such as the \citet{FabJac76} and \citet{Kor77} relations, which are likely to be nuclear discs. Nonetheless, it is often the case that only the photometric bulge S\'ersic index is readily available, particularly in studies involving a large number of galaxies beyond the nearby Universe. Therefore, I will employ here the S\'ersic index only, but this is to be considered as an illustration of the effects of degraded spatial resolution, rather than an endorsement of the method.

Using then the simple-minded approach of employing the photometric bulge S\'ersic index $n$ to classify the central component of disc galaxies as either a classical bulge or a nuclear disc, Table \ref{tab:class} shows the number of classified classical bulges for the different fits with the original and redshifted images using three different thresholds for $n$. I remind the reader that all 16 galaxies in this study are shown to have rapidly-rotating nuclear discs, and none show evidence for a massive classical bulge (although NGC\,1291 is a strong candidate to host a `composite bulge', i.e., it may host a small, subdominant classical bulge at the centre of the nuclear disc). The numbers presented in Table \ref{tab:class} show that, when using the threshold to separate classical bulges and nuclear discs at $n=2$, all decompositions using the original images indicate the presence of two classical bulges in the sample. However, this does not take into account the statistical error on $n$. If we consider the fiducial statistical uncertainty for $n$ found by \citet{KimGadShe14} of 13\% (which, for the sample studied here, corresponds to a mean uncertainty of 0.26), then the threshold should be at $n=2.26$ (for a 1$\sigma$ confidence level). Using this later threshold then the NM fits indicate one classical bulge (in NGC\,1291), whereas the DE and MCMC fits indicate none, which is in accordance with the kinematical assessment.

Considering the threshold at $n=2$, with the images redshifted to $z\approx0.05$, the number of classical bulges increases with the NM and DE fits by factors of 2 and 1.5, respectively, but remains stable with the MCMC fits. With the same threshold, but with the images redshifted to $z\approx0.1$, the number of classical bulges increases with the NM and DE fits by a factor of 3 (when compared to $z\sim0$), and with the MCMC fits by a factor of 2.

Considering now the threshold at $n=2.26$, with the images redshifted to $z\approx0.05$, the number of classical bulges again increases substantially but less so for the MCMC fits. The indicated number of classical bulges in this case is the same as with the threshold at $n=2$ and again with the images redshifted to $z\approx0.05$, which means that the larger values of $n$ are all above the fiducial 1$\sigma$ confidence level. With the threshold at $n=2.26$, but with the images redshifted to $z\approx0.1$, the trend of an increase in the number of classified classical bulges continues, although this is stabilised with the NM fits -- and this number is less for the NM and DE fits than with the threshold at $n=2$.

One can also consider the mean uncertainties in $n$ estimated with the posterior distributions derived with the MCMC fits. However, for the fits using the original images, this is negligible (as can be seen for example in Fig.\,\ref{fig:cornerbulge}), and for the images redshifted to $z\approx0.05$ this is only 0.12 (see Fig.\,\ref{fig:Compn_highz}), which is less than the fiducial uncertainty from \citet{KimGadShe14}. Nevertheless, for the images redshifted to $z\approx0.1$, the MCMC-derived mean 1$\sigma$ uncertainty in $n$ is 0.60. I therefore show in Table \ref{tab:class} the number of classical bulges indicated with a threshold at $n=2.60$, but only for the fits using the images redshifted to $z\approx0.1$. Comparing these results with those from the fits using the original images, and with a threshold at $n=2.26$, the number of classified classical bulges jumps from 1 to 4 with the NM fits, 0 to 4 with the DE fits, and 0 to 2 with the MCMC fits.

\begin{figure}
\includegraphics[trim=0.2cm 0.2cm 0.2cm 0.2cm,clip=true,width=\columnwidth]{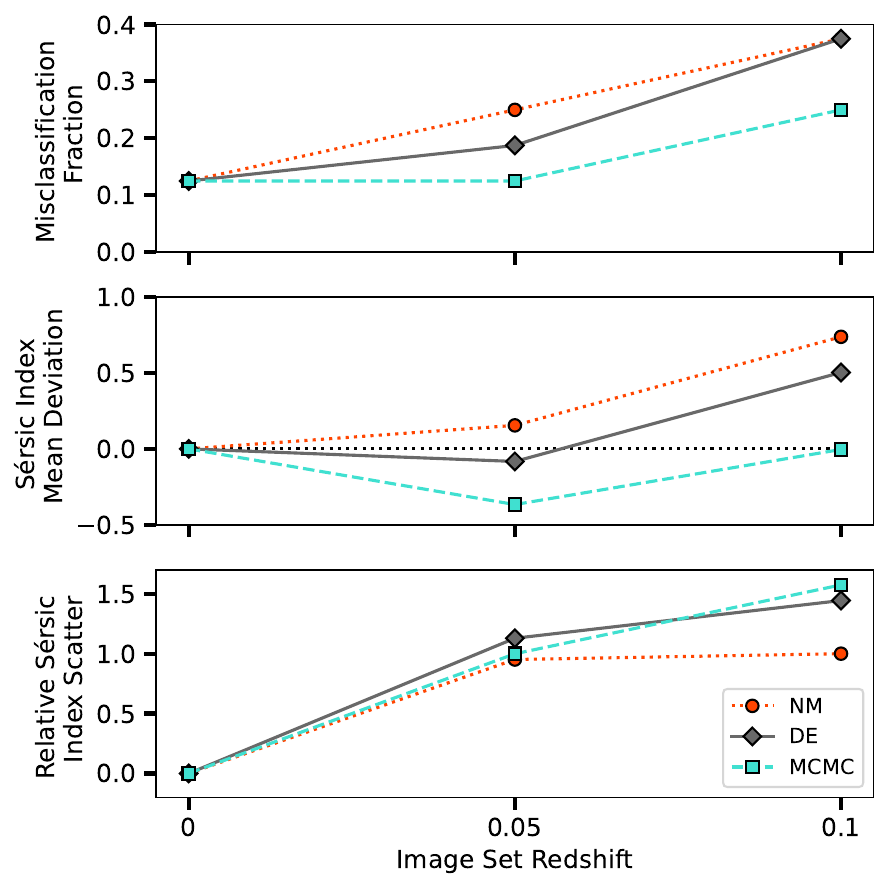}
\caption{{\it Top:} The fraction of photometric bulges misclassified as classical bulges when a S\'ersic index threshold of $n=2$ is considered, for the three image sets and three optimisation algorithms employed here, as indicated. {\it Middle:} As above, but now showing the mean deviation of $n$ with respect to the decompositions using the original S$^4$G images, separately for each optimisation algorithm. {\it Bottom:} As above, but now showing the {\em scatter} in $n$ with respect to the estimates using the original S$^4$G images, separately, again, for each optimisation algorithm. Clearly, the increased scatter in the derivation of $n$, when the spatial resolution deteriorates, results in an increase of the fraction of photometric bulges wrongly classified as classical bulges -- reaching nearly 40\% for the supervised NM and unsupervised DE optimisations. To a lesser (but still significant) extent, this also affects the unsupervised MCMC optimisation, even if, in this case, the mean deviation of $n$ is near zero, simply because the scatter is large.}
\label{fig:Misclass}
\end{figure}

Figure\,\ref{fig:Misclass} shows the fraction of photometric bulges misclassified as classical bulges (with the threshold at $n=2$) with the decompositions using the original S$^4$G images, as well as the images artificially redshifted to $z=0.05$ and $z=0.1$, and for the three optimisation algorithms. The figure also shows the mean deviation in the measurements of $n$ -- and the corresponding scatter -- with respect to the estimates using the original S$^4$G images. One can see that, as the spatial resolution deteriorates, the scatter increases substantially, leading to a significant increase in the fraction of photometric bulges wrongly classified as classical bulges. This fraction reaches up to nearly 40\% in the case of the supervised NM and unsupervised DE fits. The issue also significantly affects the unsupervised MCMC fits, although to a lesser extent. While, in this case, the mean deviation of $n$ is near zero, the scatter is large enough to lead to the misclassification of up to 25\% of the sample.

Altogether, these results suggest that there is a systematic bias in the fraction of disc galaxies hosting classical bulges -- when derived through photometric decompositions -- such that it can be significantly overestimated {\em if the images employed do not have high enough spatial resolution}. They also suggest that using MCMC fits and the corresponding uncertainties can to some extent alleviate this effect. Evidently, it is the {\em physical} spatial resolution that is the relevant parameter. In the original S$^4$G images of the galaxies studied here this is typically 170\,pc, which is enough to resolve the nuclear discs in these galaxies. In the images redshifted to $z\approx0.05$, this is 1.7\,kpc, and in the images redshifted to $z\approx0.1$, 3.4\,kpc. These redshifts are thus only an indication of the corresponding spatial resolution, but, of course, this will depend on the facility and instrumental setup used. Table \ref{tab:resols} shows the redshifts at which the physical spatial resolution corresponds to various values (including 1.7\,kpc and 3.4\,kpc) for various representative facilities. This can be used to assess what is the upper limit in redshift -- for a given instrumental setup -- above which the systematic bias discussed here starts affecting the results.

This analysis suggests that the effects of poor resolution may have affected the higher redshift end in \citet{Gad09b}, for which the average PSF FWHM is $1.5''$ and the sample is within the range $0.02\leq z \leq 0.07$, with Table \ref{tab:resols} indicating that such effects start taking place at least at $z\geq0.06$. The same applies to the results in  \citet{AllDriGra06} and \citet{DriAllLis07}, employing the Millennium Galaxy Catalog sample at $0.013\leq z \leq 0.18$ \citep{Driver2005} with a typical PSF FWHM of approximately $1.2''$ \citep[][similar to DECaLS]{Liske2003}. In this case, Table \ref{tab:resols} indicates that the systematic bias on the bulge S\'ersic index are in effect at least at $z\geq0.07$. Finally, the situation is the same also concerning the results from \citet{SimMenPat11}, which have to do with a sample in the range $0\leq z \leq 0.3$ and average PSF FWHM of $1.4''$, and so the effects of poor resolution should start affecting the results at least for $z\gtrsim0.06-0.07$. The systematic bias towards higher values of the bulge S\'ersic index may thus explain the discrepancy concerning the larger fraction of classical bulges in these studies, when compared to studies employing more nearby galaxies, as mentioned in the Introduction (although cosmic variance cannot yet be ruled out). It should be noted that here I find that imposing that the bulge effective radius be larger than half the FWHM does {\em not} avoid the bias in the bulge S\'ersic index, at least not when both parameters are still in the same order of magnitude.

Fortunately, however, with HST and JWST, the physical spatial resolution never reaches 1.7\,kpc (see table \ref{tab:resols}). With HST Wide Field Camera 3 (WFC3) at $1.6\mu m$ and JWST Near Infrared Camera (NIRCam) with the F444W filter, the worst spatial resolution is 1.3\,kpc at $z\approx1.6$. With JWST NIRcam with the F200W filter, this is 0.6\,kpc. At $z=0.1$, HST WFC3 at $1.6\mu m$ and JWST NIRCam with the F200W filter resolve structures as small as approximately 300\,pc and 150\,pc, respectively. With Euclid, nonetheless, with the $H_E$ filter, the resolution reaches 1.7\,kpc already at $z\approx0.20$, although with the $I_E$ filter the worst resolution is 1.5\,kpc at $z\approx1.6$.

Nevertheless, it must be noted that the known radii of nuclear discs range from $\approx100$\,pc to $\approx1$\,kpc \citep{Gadotti2020,deSa-Freitas2025}, and thus with a physical spatial resolution of 1\,kpc or more, most (if not all) nuclear discs are unresolved. Table \ref{tab:resols} shows the redshifts at which the physical spatial resolution corresponds to $\approx100$\,pc and $\approx1$\,kpc for various facilities. It shows that even HST WFC3 at $1.6\mu m$ and JWST NIRcam with the F444W filter miss virtually all nuclear discs already at $z\gtrsim0.59$. JWST NIRcam with the F200W filter does much better but still misses most nuclear discs (those with radii below 600\,pc) at $z\approx1.6$. Only MICADO at the Extremely Large Telescope will be able to resolve all nuclear discs with sizes above 100\,pc at all redshifts. This is thus a very important caveat in the interpretation of results concerning photometric bulges at distances where the typical physical spatial resolution is beyond $\sim1$\,kpc or even less. And it is a testimony to the blurred vision of galaxies we still have today, with a limited picture of the central kilo-parsec in more distant galaxies, which is now starting to become sharper only with the most advanced telescopes and instruments.

Moreover, it is important to have in mind that this physical spatial resolution problem also affects spectroscopic data, such as in integral field spectroscopy surveys like CALIFA and SAMI, with a typical spatial resolution of $\approx1$\,kpc and $\approx1.6$\,kpc, respectively \citep[see][]{Walcher2014,Bryant2015}. On the other hand, the typical spatial resolution in the ATLAS$^{\rm{3D}}$ and TIMER surveys is, respectively $\approx175$\,pc and $\approx100$\,pc \citep{CapEmsKra11,GadSanFal19}.

Finally, I note that all nuclear discs studied here are in (weakly or strongly) barred galaxies, and there have been claims that some unbarred galaxies may also host nuclear discs or nuclear rings \citep[see][]{KorKen04,Gad09b,FisDro10,ComKnaBec10}. While it is unclear if those galaxies still appear to be unbarred and hosting a nuclear disc with further scrutiny with better data, some of such nuclear discs may have formed through similar processes as those formed through bar-driven gas inflow. Any non-axisymmetric component in the main disc (such as spiral arms or oval distortions) can cause gas inflows; bars are just more efficient at it. Nuclear discs formed via external gas accretion due to interactions with other galaxies appear to be rarer \citep[see discussion in][]{Schultheis2025}, but in both cases the properties of the nuclear discs are not expected to be fundamentally different from those of bar-built nuclear discs, and thus the results presented here are likely to apply to such nuclear discs as well, regardless of the formation process.

\subsection{Classical bulges embedded in nuclear discs}

In our current understanding of the formation of classical bulges and nuclear discs, it is straightforward to devise scenarios in which a galaxy could have at its centre both a classical bulge and a nuclear disc. For example, one could speculate that early mergers could form the classical bulge, and the formation of a bar would lead to the building of the nuclear disc, either within the classical bulge or surrounding it. These possibilities were discussed by, e.g., \citet{Ath05b}, \citet{Gad09b}, and \citet{FisDro10}, and studied in detail in \citet[][see also \citealt{deLFalVaz13,deLSanMen19}]{ErwSagFab15}.

All such `composite bulges' in \citet{ErwSagFab15} are embedded at the centre of the nuclear disc and identified as structures rounder than the main galaxy disc, and with stellar kinematics dominated by velocity dispersion rather than rotation. However, because of their limited sizes (some have effective radii of only a few tens of parsecs), spatial resolution is even more critical in the identification of such small classical bulges than it is for the identification of nuclear discs. In fact, in this study, the strongest candidate to host a small classical bulge embedded in its nuclear disc is NGC\,1291 \citep[see][]{deLSanMen19}. But the S\'ersic index of the photometric bulge derived here in the decomposition of this galaxy, using the DE algorithm, is only 2.2, and therefore still consistent with being less than two within the uncertainties, and thus near exponential. This is at least in part due to the presence of the more extended nuclear disc, which is of course also included in the model of the photometric bulge. In fact, \citet{FisDro10} showed that the S\'ersic index of the photometric bulge in the case of `composite bulges' can depend strongly on the relative masses of the small classical bulge and its surrounding nuclear disc, as well as on the S\'ersic index of the small classical bulge alone. It is therefore unclear how to use the S\'ersic index to identify `composite bulges', unless there is enough spatial resolution to indicate the presence of multiple central components.

In this paper, I am concerned with the problem of correctly distinguishing {\em massive} classical bulges and nuclear discs. However, establishing a census of the small classical bulges embedded in nuclear discs -- which so far appears to require information on the central stellar kinematics with very high spatial resolution -- is another important open problem.

\subsection{Improving the structural analysis of galaxies with image decompositions}

In this subsection, I discuss the lessons learned from the analyses above in the context of progressing 2D decompositions of galaxy images to study the structural components of galaxies and what they can tell us about the processes involved in the formation and evolution of galaxies. The first major aspect that can be improved concerns the accuracy of the models employed to describe the light distribution of each structural component. As mentioned above in Sects. \ref{subsec:de} and \ref{subsec:mcmc}, the residual images in Fig.\,\ref{fig:modres} reveal the presence of substructures that are generally not included in the models, such as rings and spiral arms. With the usual simplistic models for, e.g., the disc and the bar, even with the powerful model optimisation algorithms employed here, these substructures are not captured by the best-fit models. The inclusion of these components in the models have been shown to increase the accuracy of the results \citep[e.g.,][]{GaoHo17,Chugunov2024}. The residual images also reveal that the inner parts of discs in barred galaxies are often over subtracted by the models. This is understood as the natural evolution of barred discs, whereby bars capture stars in the disc within the bar radius, trapping them in the resonant orbits that make up the bar \citep[see][]{Lynden-Bell1972,BinTre87,Ath13}, creating regions that resemble crescents surrounding the bar (but inside the bar radius), which are to some degree devoid of stars and recent star formation \citep[see, e.g.,][]{GaddeS03,Gad08,KimGadAth16}. These regions have also been called the `star formation desert' \citep[e.g.,][]{Donohoe-Keyes2019} and the `dark gaps' \citep[e.g.,][]{But17,Aguerri2023,Ghosh2024,Kim2025}. These bar-driven processes result in changes in the luminosity profile of the disc, which therefore cannot be described anymore with a single exponential \citep[see also][]{ErwPohBec08}. \citet{KimGadShe14} showed that accounting for such disc breaks also result in more accurate models for all structural components in a disc galaxy \citep[see also][for a different approach to characterise this problem]{Breda2020}. Other relevant effects to account for are AGN emission \citep{Gad08} and the effects of dust \citep{GadBaeFal10,PasPopTuf13}. Finally, bars are also often more complex than what can be captured by a single 2D boxy light distribution. For example, \citet{NeuGadWis19} employed up to three S\'ersic models following generalised ellipses to account for the observed complexity in the structure of bars, which is a result of their rich orbital scaffolding \citep[see also][]{Li2025}.

Such improvements to the models lead to a reduction in systematic effects and improved accuracy. Currently, these more complex models are typically used in dedicated studies involving relatively small samples, but we should aim at having these more realistic models being used routinely. However, with the significant increase in free parameters to fit, this will likely require the use of powerful optimisation algorithms to minimise any possibilities of degeneracies between some structural parameters, as well as to accurately capture the statistical uncertainties.

The analyses above demonstrate the severe effects of poor spatial resolution and the importance of using an oversampled PSF (in this study, with a five-fold oversampling). Figure \ref{fig:nexcess} suggests that the S\'ersic index of the photometric bulge may be overestimated without the use of an oversampled PSF, which may have secondary effects in the recovered structural parameters of other galactic components as well. Furthermore, as discussed in the previous subsection, a poor physical spatial resolution can severely hamper the accurate recovery of the S\'ersic index of the photometric bulge (Figs.\,\ref{fig:Compn_highz} and \ref{fig:Misclass}, and Table \ref{tab:Compparm}), imposing a systematic bias towards higher values of $n$, and it increases the scatter in the recovery of other structural parameters (Figs.\,\ref{fig:Compparm} and \ref{fig:Misclass}, and Table \ref{tab:Compparm}). It is therefore key that the PSF be carefully accounted for and considered in the interpretation of the results and corresponding uncertainties. Furthermore, with a physical spatial resolution of 3.4\,kpc, in some cases, the presence of a bar or even of a photometric bulge, is made difficult to recover. To put this result into context, for SDSS and DECaLS $r$-band images, this resolution corresponds to redshifts of $z\approx0.12$ and $z\approx0.16$, respectively, and therefore very much within the local Universe. For Euclid $H_E$ images this would be $z\approx0.59$, so still in what one could call a low-redshift regime. This means that galaxies such as NGC\,3351 and NGC\,4981 would not be classified as barred in these instances. For shorter or weaker bars, not even JWST is free from resolution effects wiping out bars. Le Conte et al. (in prep.) show that the F444W filter on NIRCam misses bars that are detected with the F200W filter (which has a factor 2 better resolution; see also \citealt{Liang2024,Guo2025}), and in both filters one can see the effects of the PSF in making bars appear rounder, just as it did here with NGC\,4984 with the S$^4$G image artificially redshift to $z\approx0.1$, corresponding to a physical spatial resolution of 3.4\,kpc.

The results in Sects. \ref{subsec:de} and \ref{subsec:mcmc} reveal that by employing the DE optimisation algorithm, an adaptation of which is also used in the MCMC fits, one can recover the values of structural parameters with unsupervised fits just as well as when one produces {\em supervised} fits employing the NM algorithm (which is faster but still prone to be misled by local statistical minima in the parameter space, although in this context it is better than the LM algortihm; see Figs. \ref{fig:Compn_NMDE}, \ref{fig:Compn_NMMCMC} and \ref{fig:Compparm0}). Crucially, not only the DE and MCMC fits are unsupervised, but they do not require initial guesses for all free parameters. Being unsupervised, and only requiring lower and upper limits for each free parameter, {\em the DE algorithm thus significantly removes the subjective character of these decompositions}. While the choice for lower and upper limits for a given parameter is still to a small degree subjective, the analyses presented here show that fairly wide ranges can be employed and still lead to an excellent agreement with the supervised NM fits, or to even superior fits. For example, the range used for the photometric bulge S\'ersic index, $n$, in all DE and MCMC fits, is $0.3\leq n \leq 5.0$ (see Sect.\,\ref{subsec:de} for the ranges employed in the other parameters). Removing the subjective component in 2D photometric decompositions of galaxy images is a major step forward and leads to more accurate fits (within the limits of the models employed). In fact, when considering the fiducial uncertainty in $n$, the unsupervised DE fits indicate no classical bulges in the sample, in agreement with the kinematical assessment of the central regions of these galaxies, better than the supervised NM fits, which find one classical bulge (see Table \ref{tab:class}; again, with the simplistic approach of classifying photometric bulges employing solely the S\'ersic index). The DE algorithm -- combined with MCMC techniques, particularly for more complex models or when parameter ranges are not straightforwardly defined -- should indeed become the standard for any photometric decomposition and, in particular, studies employing very large samples, where visual inspection of each fit is impractical.

Finally, the decision on which models to employ and, more importantly, the number of structural components to include in the fit, can also be guided by DE or MCMC fits, e.g., in a scheme that tries a number of fits with different characteristics (i.e., different models and number of components), and uses a statistical analysis to assess which fit is the more statistically justified. This could include, e.g., the use of the \citet{Akaike1974} information criterion (AIC; see \citealt{Silva-Lima2025} for a practical application of the AIC criterion in the context of spectroscopic models).

 \section{Summary and Concluding Remarks}
 \label{sec:conc}

In this paper, I have performed a number of 2D photometric fits employing $3.6\mu m$ images of nearby galaxies to produce models reflecting the structural properties of these galaxies. A previous spectroscopic analysis has determined that the central region of these galaxies is dominated by the presence of a rapidly-rotating nuclear disc (often associated with a photometric component with S\'ersic index $n$ less than two), and that these galaxies do not host any massive, kinematically hot central spheroid (i.e., a classical bulge, often associated with a photometric component with $n>2$). This study aimed at answering two main questions:

\begin{enumerate}

\item Can the Differential Evolution (DE) algorithm recover models without the need for iterations guided by subjective choices (i.e., without human supervision) as accurately as supervised fits employing less powerful (but faster) algorithms?

\item Are the photometric fits able to correctly indicate the presence of nuclear discs (and absence of classical bulges) in different regimes of physical spatial resolution?

\end{enumerate}

Different fits using different optimisation algorithms were thus employed to determine the corresponding best-fit models: supervised fits employing the Nelder-Mead (NM) algorithm (fast but more prone to be trapped in local minima), which were individually visually inspected and fine-tuned until a satisfactory fit was performed, unsupervised DE fits, and unsupervised MCMC fits also employing an adaptation of the DE algorithm. In addition, the fits were performed on the original images (with an average physical spatial resolution of 170\,pc), as well as two sets of artificially redshifted images, in which the physical spatial resolution is 1.7\,kpc and 3.4\,kpc, respectively.

The main results from these analyses can be summarised as follows:

\begin{itemize}

\item Unsupervised fits employing the DE algorithm recover structural models in excellent agreement with the supervised NM fits. In fact, an assessment of the residual images suggests that the DE fits are slightly better optimised. This implies that, by using the DE algorithm, one is able to obtain more accurate and more objective measurements of the structural parameters of galaxies.

\item The DE fits return values for the photometric bulge S\'ersic index $n$ that are consistent with $n<2$ within the uncertainties, and thus correctly indicate that the galaxies in the sample host nuclear discs, and not massive classical bulges (as previously determined with a spectroscopic analysis).

\item When the physical spatial resolution in the images is approximately 1.7\,kpc, the recovery of structural parameters is affected by a significant increase in the scatter. This is particularly so for the photometric bulge S\'ersic index, and is further exacerbated when the physical spatial resolution is approximately 3.4\,kpc.

\item The results from the fits of the artificially redshifted images show that there is a systematic bias that leads to an overestimation of the photometric bulge S\'ersic index. This may explain why the fraction of classical bulges in disc galaxies, found with samples at relatively high redshift, is larger than the fraction found in studies of more local galaxies, where the physical spatial resolution is suitable to recover central structures with radii of the order of a few hundred parsecs. This effect can also be a problem even for data taken with facilities such as Euclid, HST and JWST for high enough redshifts, and highlights that the utmost care should be taken when interpreting results on the central structure of disc galaxies, in the context of the physical spatial resolution of the data, be it photometric or spectroscopic.

\end{itemize}

Accounting for the bias presented above brings closer the fraction of classical bulges in studies employing samples in and beyond the nearby Universe, pointing out to a low fraction of such structures. An immediate consequence of this is the reaffirmation of what is presently a challenge to our current cosmological model, with a merger-driven picture of galaxy formation. Tackling this problem requires an accurate estimation of the fraction of classical bulges in the nearby Universe, employing both photometric, but perhaps more importantly, spectroscopic techniques, to directly measure stellar kinematics, as well as developing a more complete theoretical understanding of the role of mergers in the formation of central stellar structures in disc galaxies.

\section*{Acknowledgements}

I dedicate this work to the memory of my friend and collaborator, Dr. Peter Erwin, who left us unexpectedly and terribly early. Without his diligent effort in producing such a powerful tool as {\sc imfit} and making it publicly available, and without his help on using {\tt imfit-mcmc} on multiple threads, this paper would have not been possible. He provided the community with a wealth of knowledge on stellar structures in galaxies, and his passing is a significant loss to the field. I thank Francesca Fragkoudi for fruitful discussions that initiated this work and for encouraging me to pursue and complete this study. I am grateful to the anonymous reviewer who provided very helpful and thoughtful comments. The author is supported by STFC grants ST/T000244/1 and ST/X001075/1. This work used the DiRAC@Durham facility managed by the Institute for Computational Cosmology on behalf of the STFC DiRAC HPC Facility (www.dirac.ac.uk). The equipment was funded by BEIS capital funding via STFC capital grants ST/K00042X/1, ST/P002293/1, ST/R002371/1 and ST/S002502/1, Durham University and STFC operations grant ST/R000832/1. DiRAC is part of the National e-Infrastructure.

\section*{Data Availability}

All original $3.6\mu$m S$^4$G images used in this study are available at \href{https://irsa.ipac.caltech.edu/data/SPITZER/S4G/}{https://irsa.ipac.caltech.edu/data/SPITZER/S4G/}.



\bibliographystyle{mnras}
\bibliography{../../../../../papers/gadotti_refs} 

\appendix
\section{Radial Profiles from Ellipse Fits}
\label{app:ellipse}

\begin{figure*}
\includegraphics[trim=0.2cm 0.2cm 0.2cm 0.2cm,clip=true,width=1.8\columnwidth]{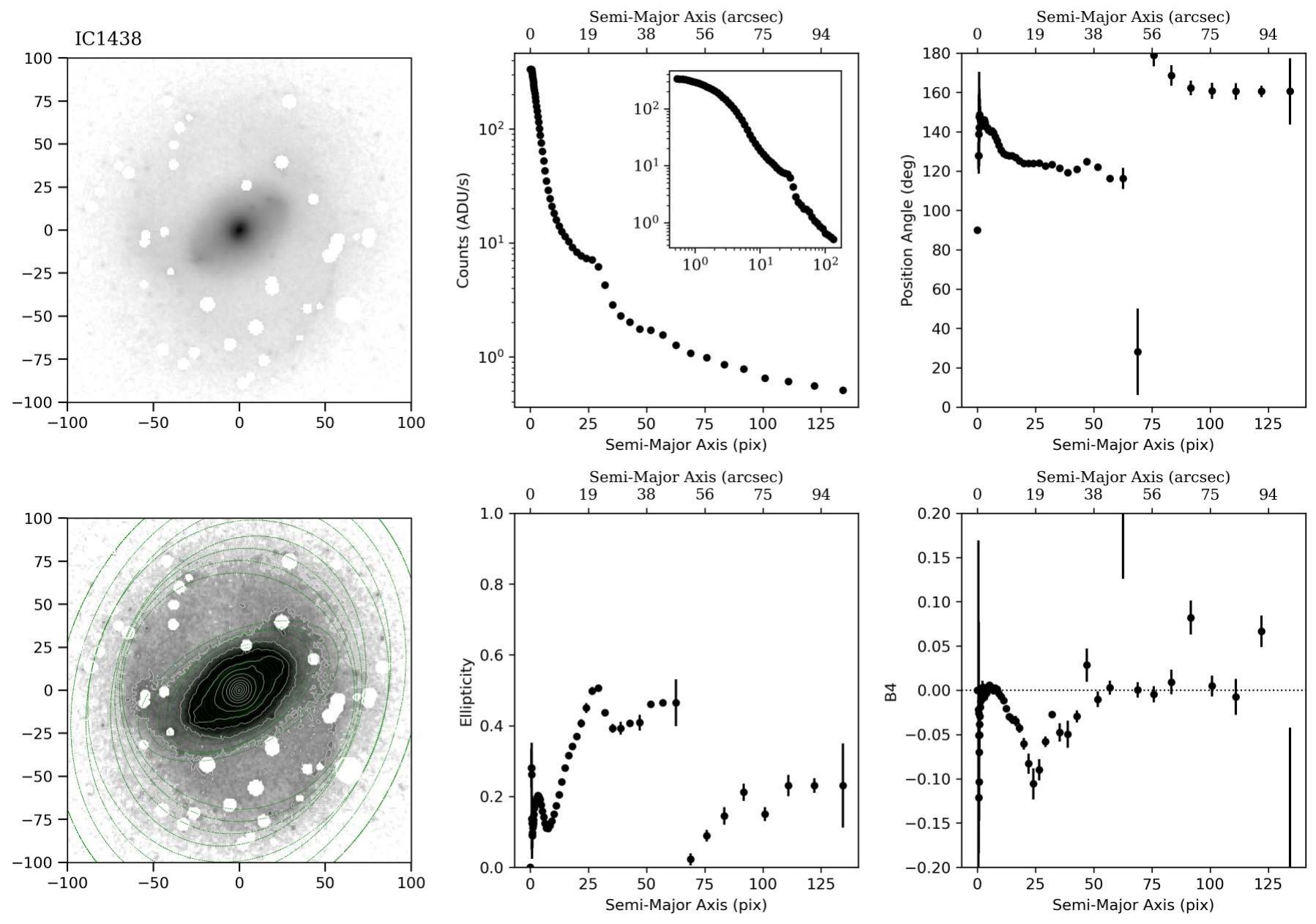}
\caption{\textit{Top left:} Original S$^4$G galaxy image (with coordinates in pixels); \textit{Bottom left:} Same image but with fainter features enhanced, and with isophotal contours overlaid in white and the ellipse fits to the isophotes overlaid in green. The radial profiles shown correspond the the isophotes' mean intensity (also shown in a log-log projection in the inset), ellipticity, position angle, and the b4 Fourier parameter.}
\label{fig:ellfits}
\end{figure*}

In this appendix, I show in Fig.\,\ref{fig:ellfits} the results of fitting ellipses to the isophotes of the 16 galaxies in the sample using the original S$^4$G images.

\begin{figure*}
\includegraphics[trim=0.2cm 0.2cm 0.2cm 0.2cm,clip=true,width=1.8\columnwidth]{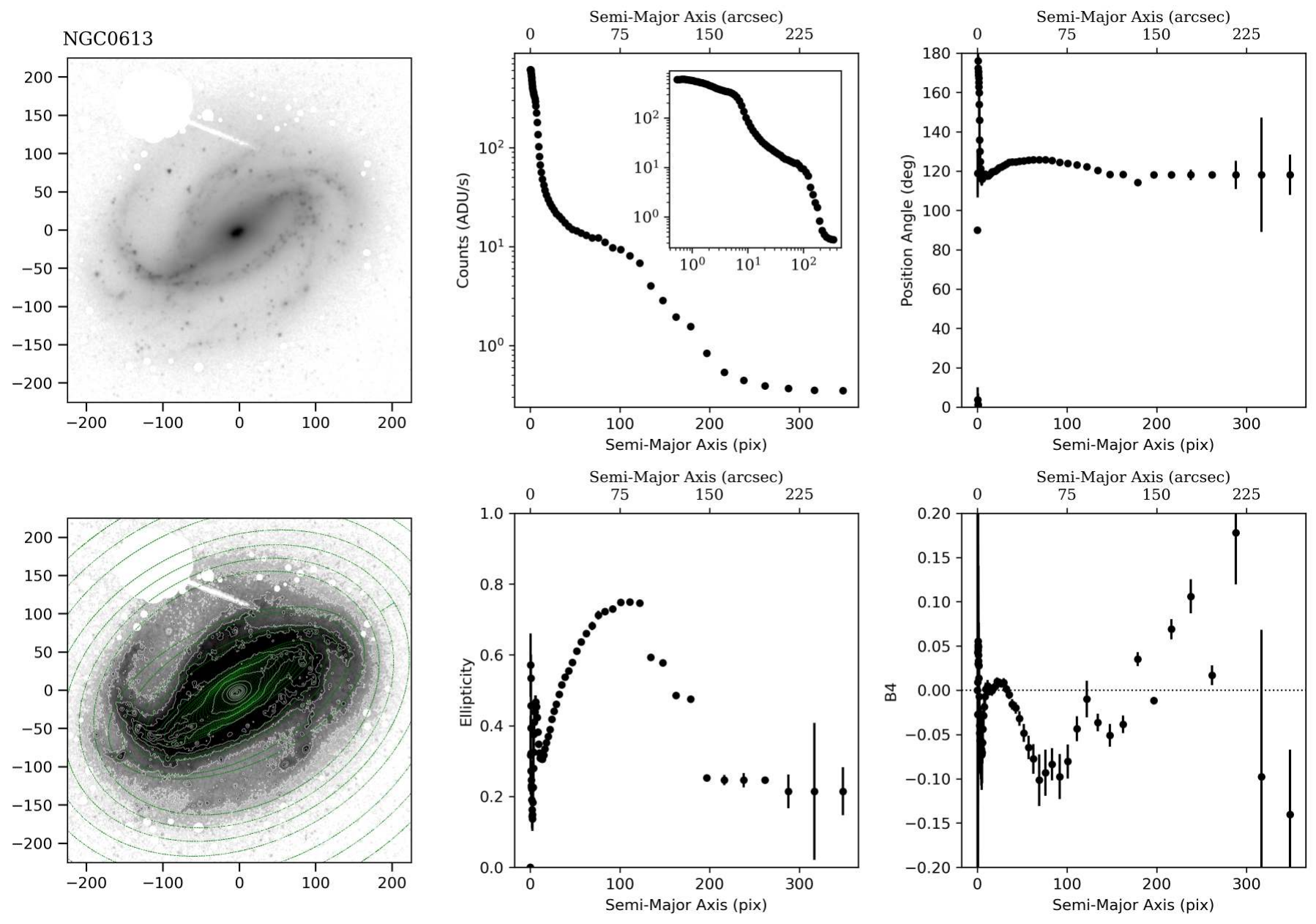}
\addtocounter{figure}{-1}
\caption{continued.}
\end{figure*}

\begin{figure*}
\includegraphics[trim=0.2cm 0.2cm 0.2cm 0.2cm,clip=true,width=1.8\columnwidth]{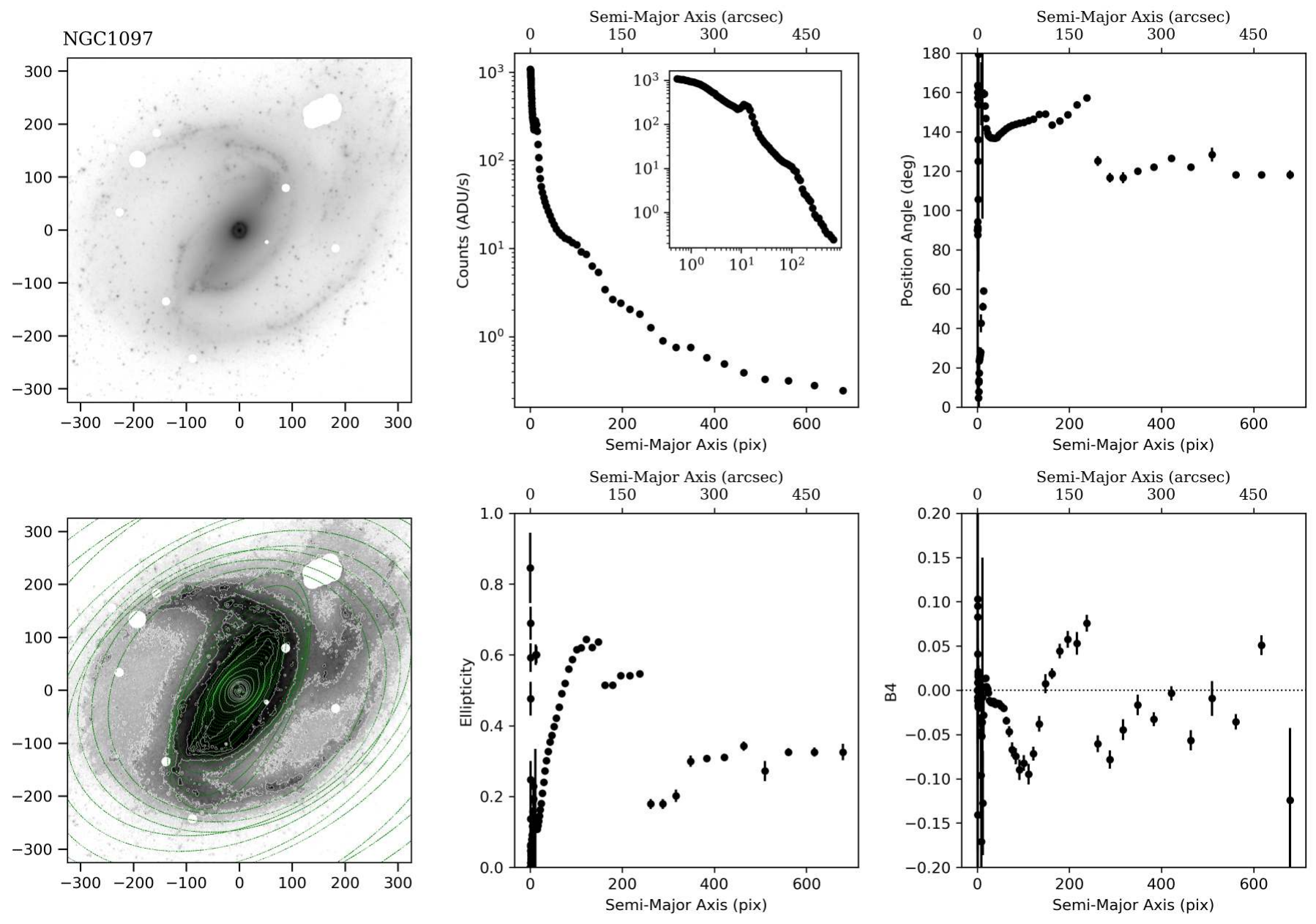}
\addtocounter{figure}{-1}
\caption{continued.}
\end{figure*}

\begin{figure*}
\includegraphics[trim=0.2cm 0.2cm 0.2cm 0.2cm,clip=true,width=1.8\columnwidth]{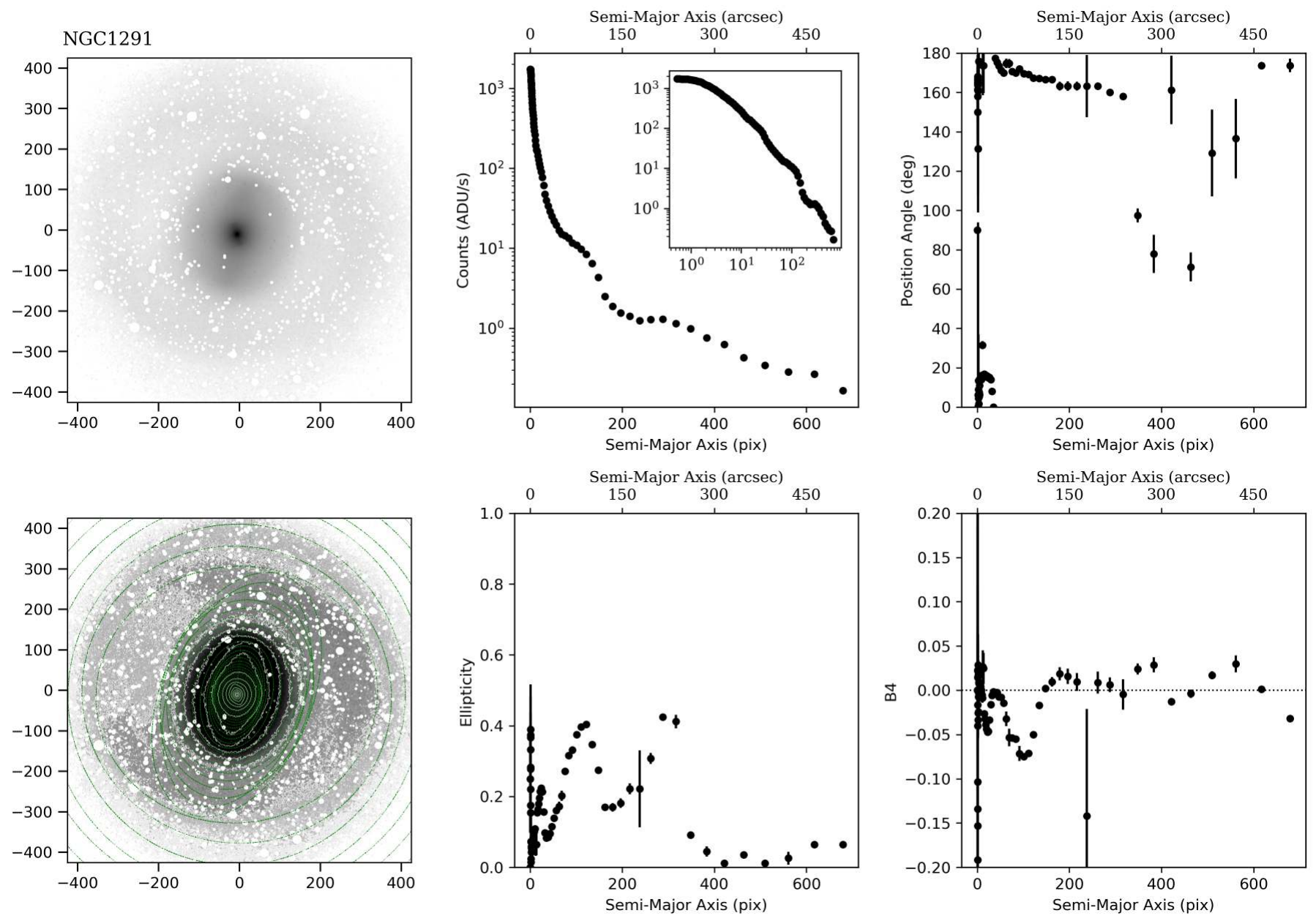}
\addtocounter{figure}{-1}
\caption{continued.}
\end{figure*}

\begin{figure*}
\includegraphics[trim=0.2cm 0.2cm 0.2cm 0.2cm,clip=true,width=1.8\columnwidth]{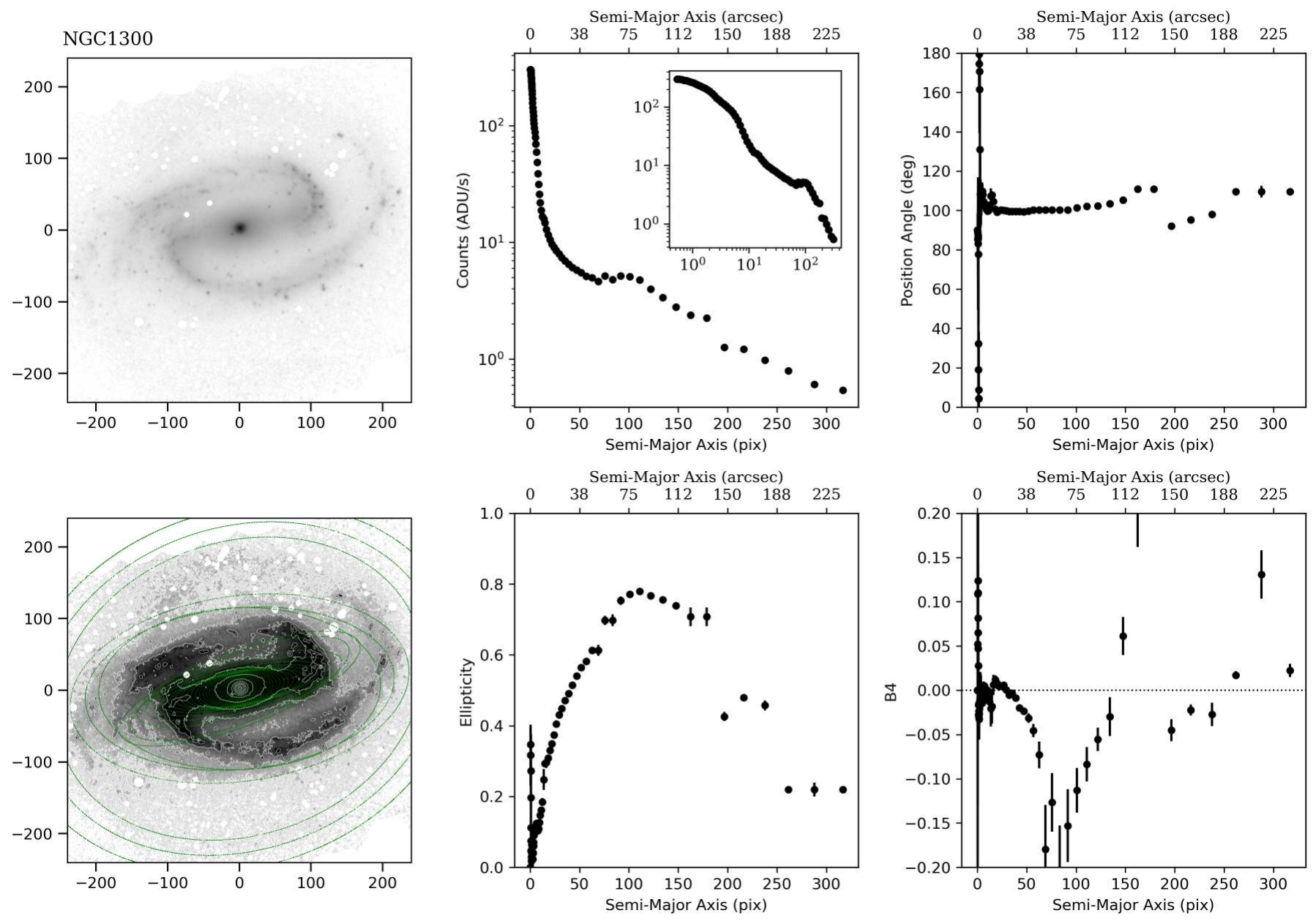}
\addtocounter{figure}{-1}
\caption{continued.}
\end{figure*}

\begin{figure*}
\includegraphics[trim=0.2cm 0.2cm 0.2cm 0.2cm,clip=true,width=1.8\columnwidth]{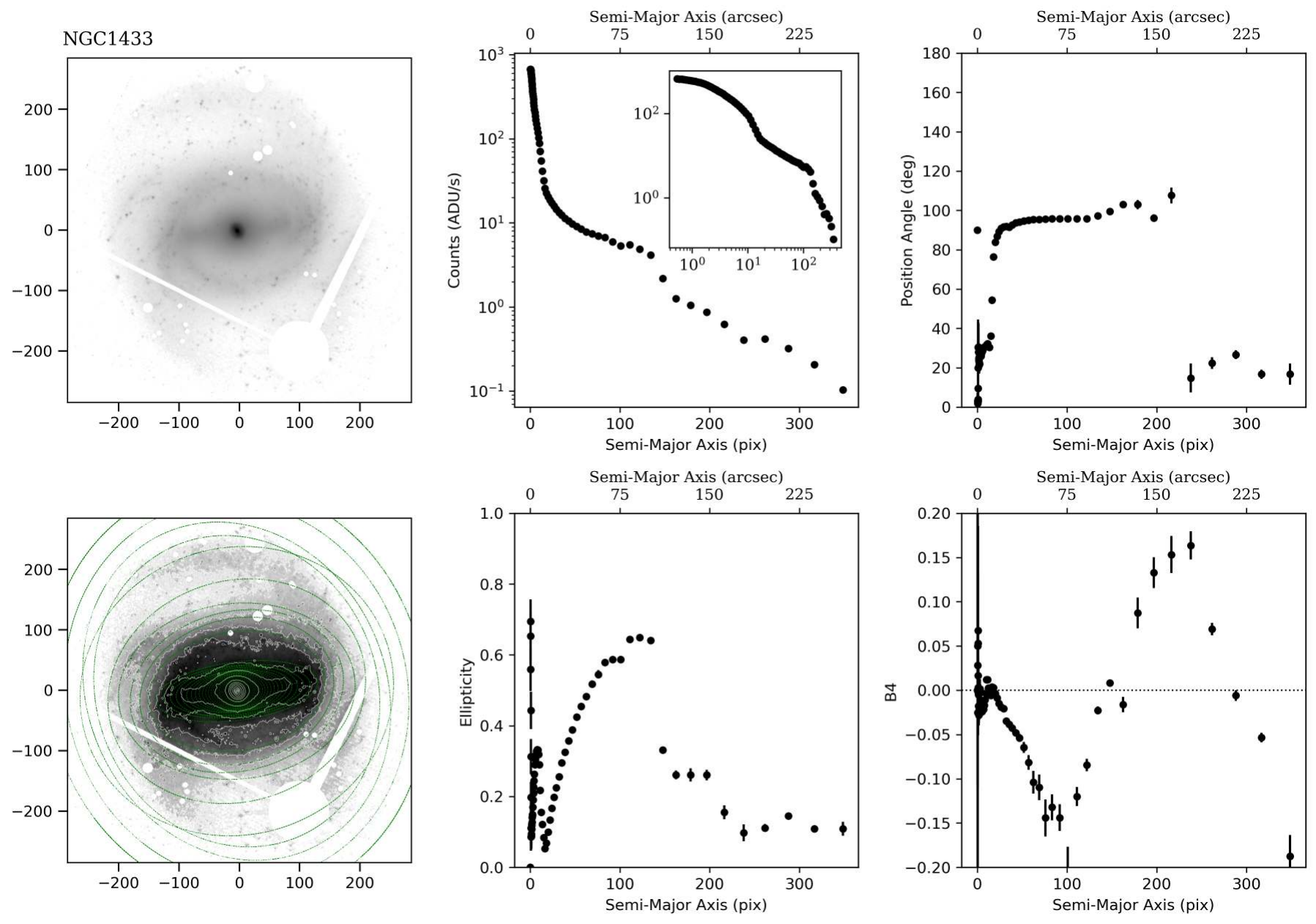}
\addtocounter{figure}{-1}
\caption{continued.}
\end{figure*}

\begin{figure*}
\includegraphics[trim=0.2cm 0.2cm 0.2cm 0.2cm,clip=true,width=1.8\columnwidth]{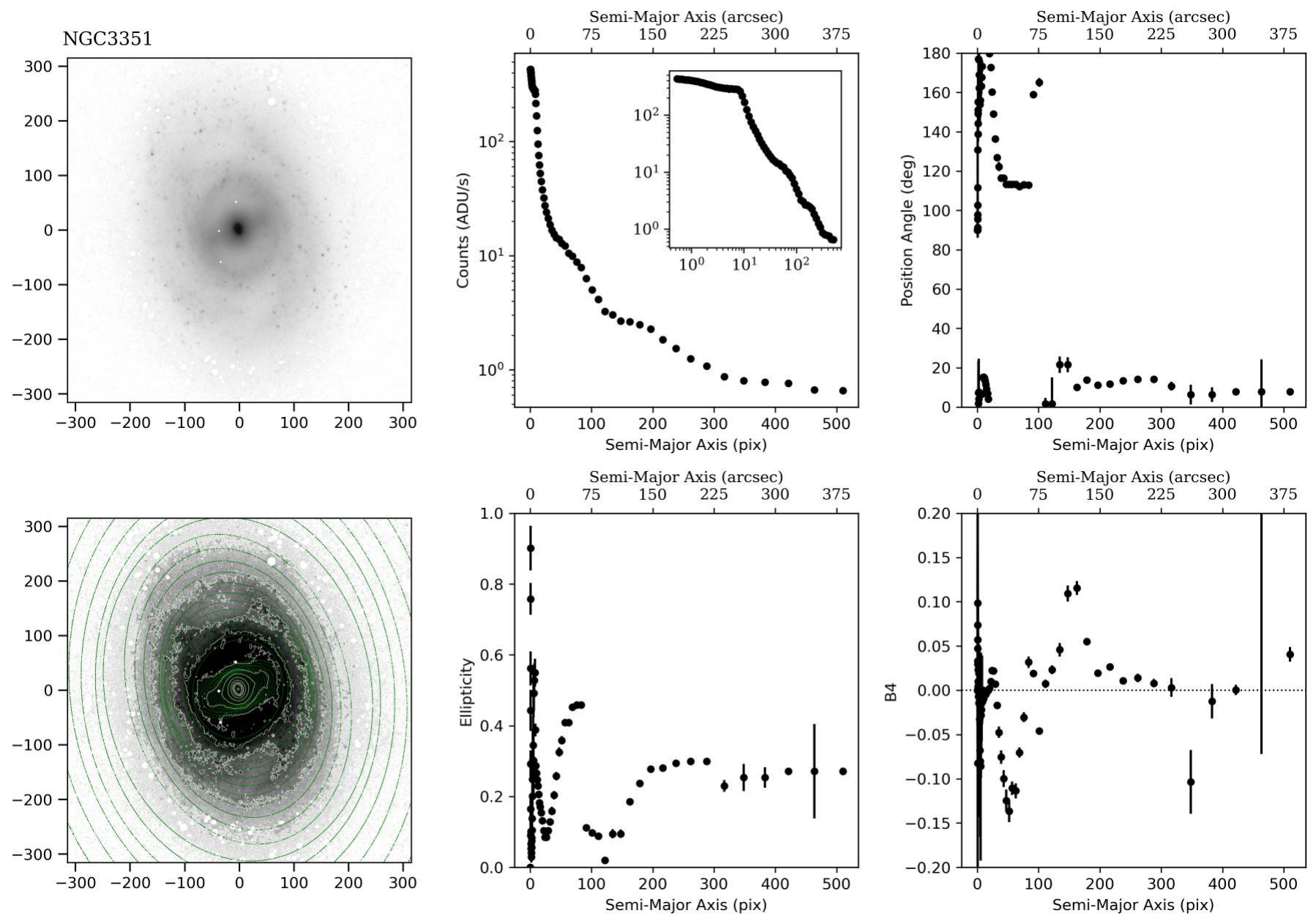}
\addtocounter{figure}{-1}
\caption{continued.}
\end{figure*}

\begin{figure*}
\includegraphics[trim=0.2cm 0.2cm 0.2cm 0.2cm,clip=true,width=1.8\columnwidth]{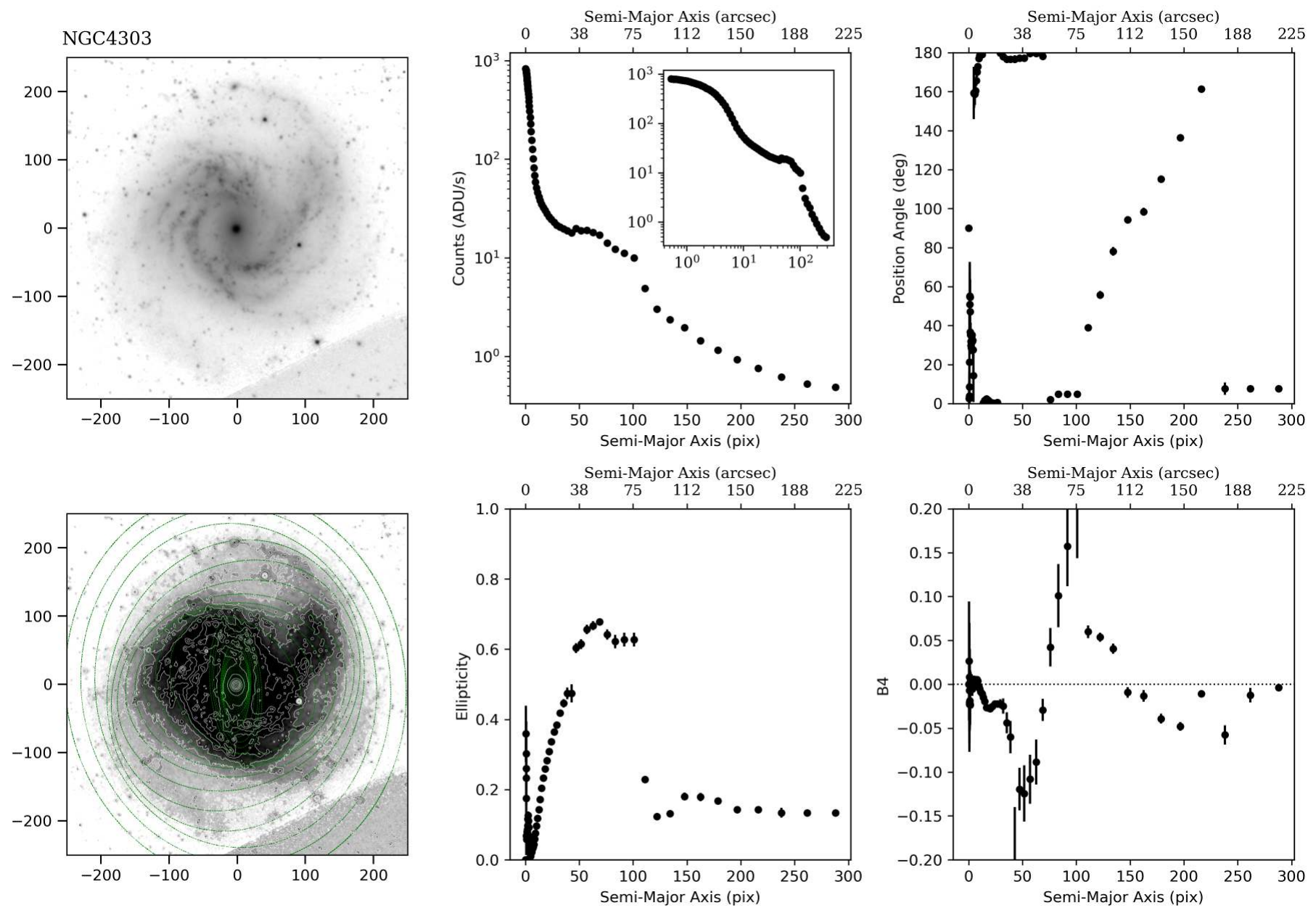}
\addtocounter{figure}{-1}
\caption{continued.}
\end{figure*}

\begin{figure*}
\includegraphics[trim=0.2cm 0.2cm 0.2cm 0.2cm,clip=true,width=1.8\columnwidth]{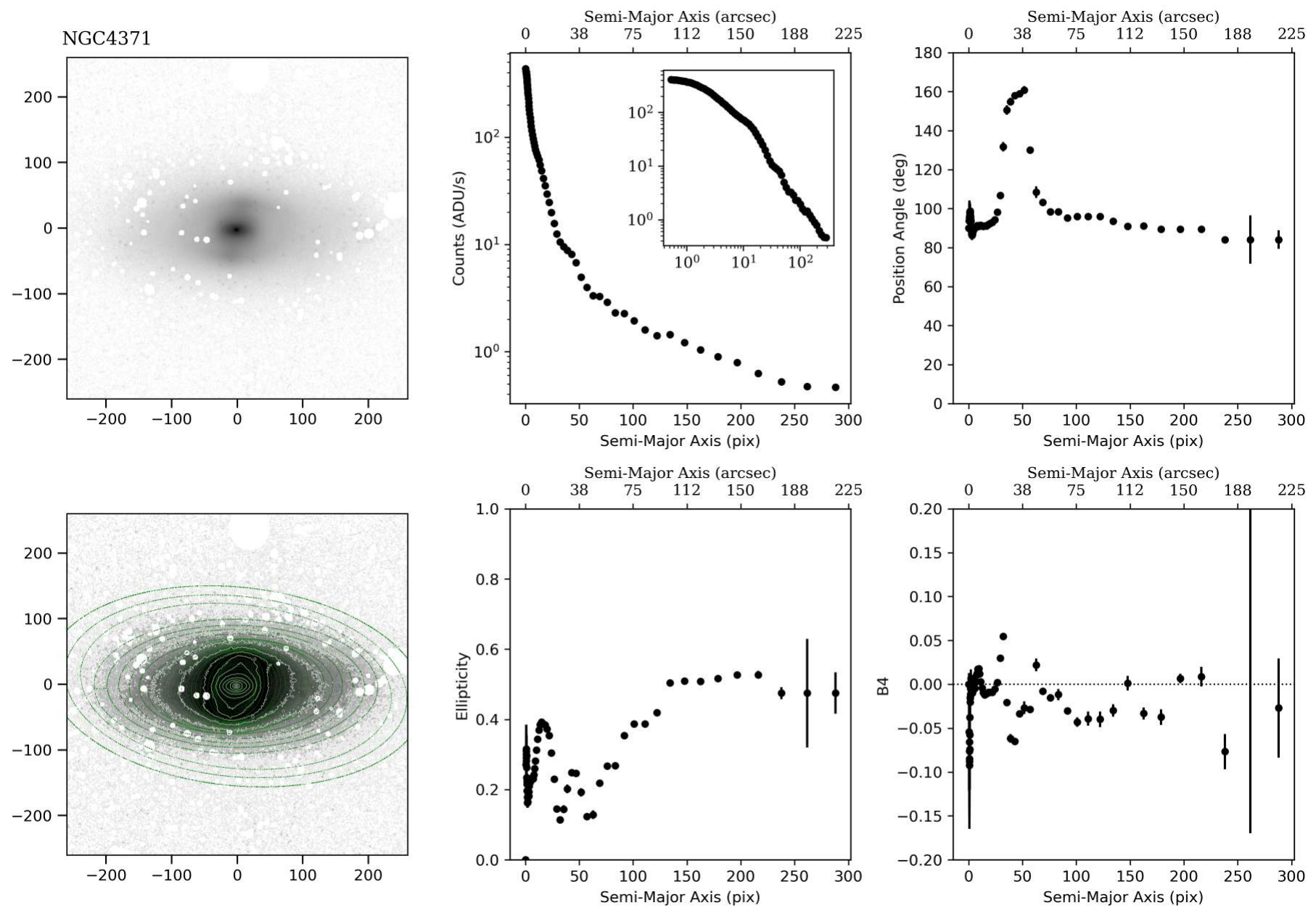}
\addtocounter{figure}{-1}
\caption{continued.}
\end{figure*}

\begin{figure*}
\includegraphics[trim=0.2cm 0.2cm 0.2cm 0.2cm,clip=true,width=1.8\columnwidth]{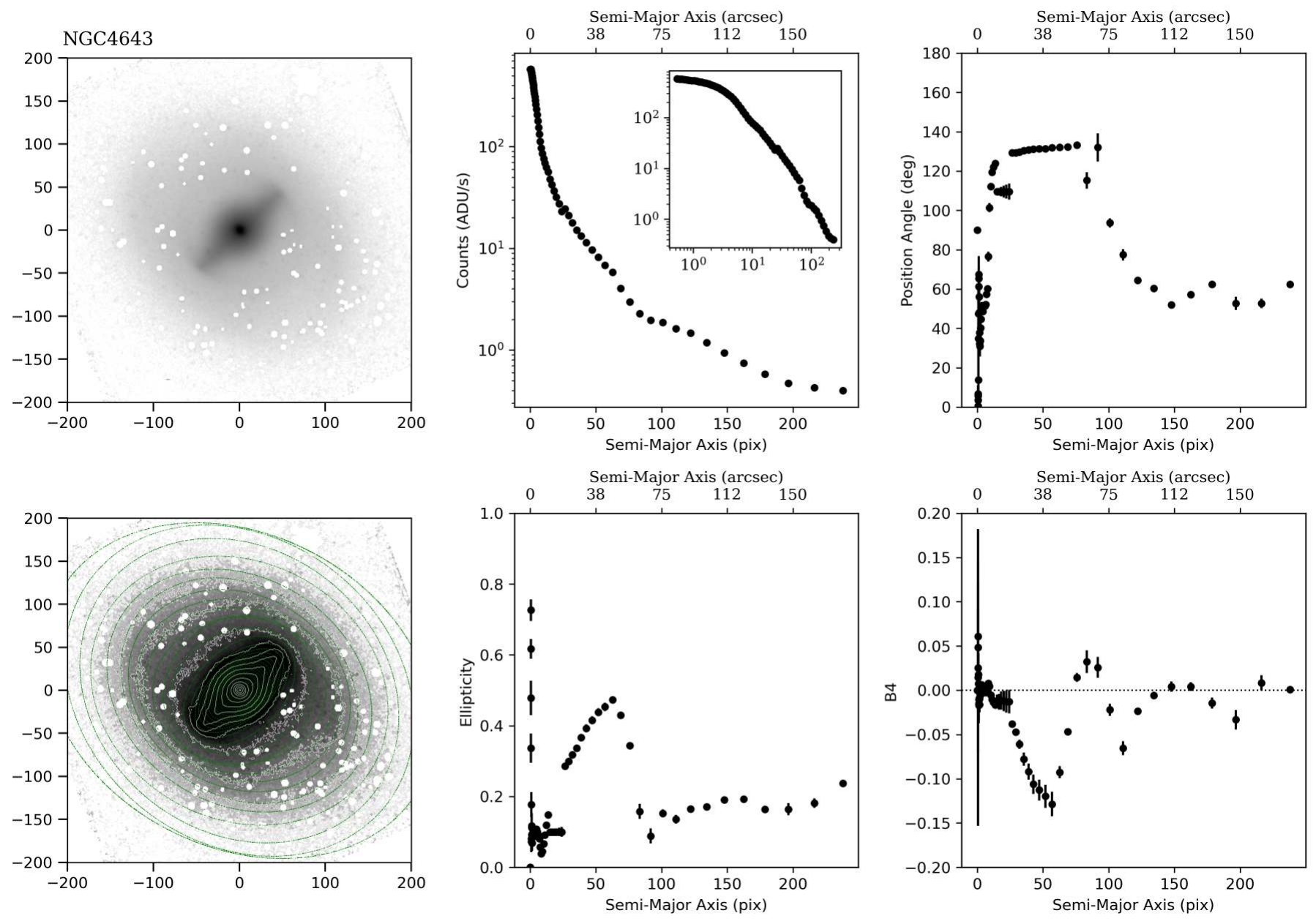}
\addtocounter{figure}{-1}
\caption{continued.}
\end{figure*}

\begin{figure*}
\includegraphics[trim=0.2cm 0.2cm 0.2cm 0.2cm,clip=true,width=1.8\columnwidth]{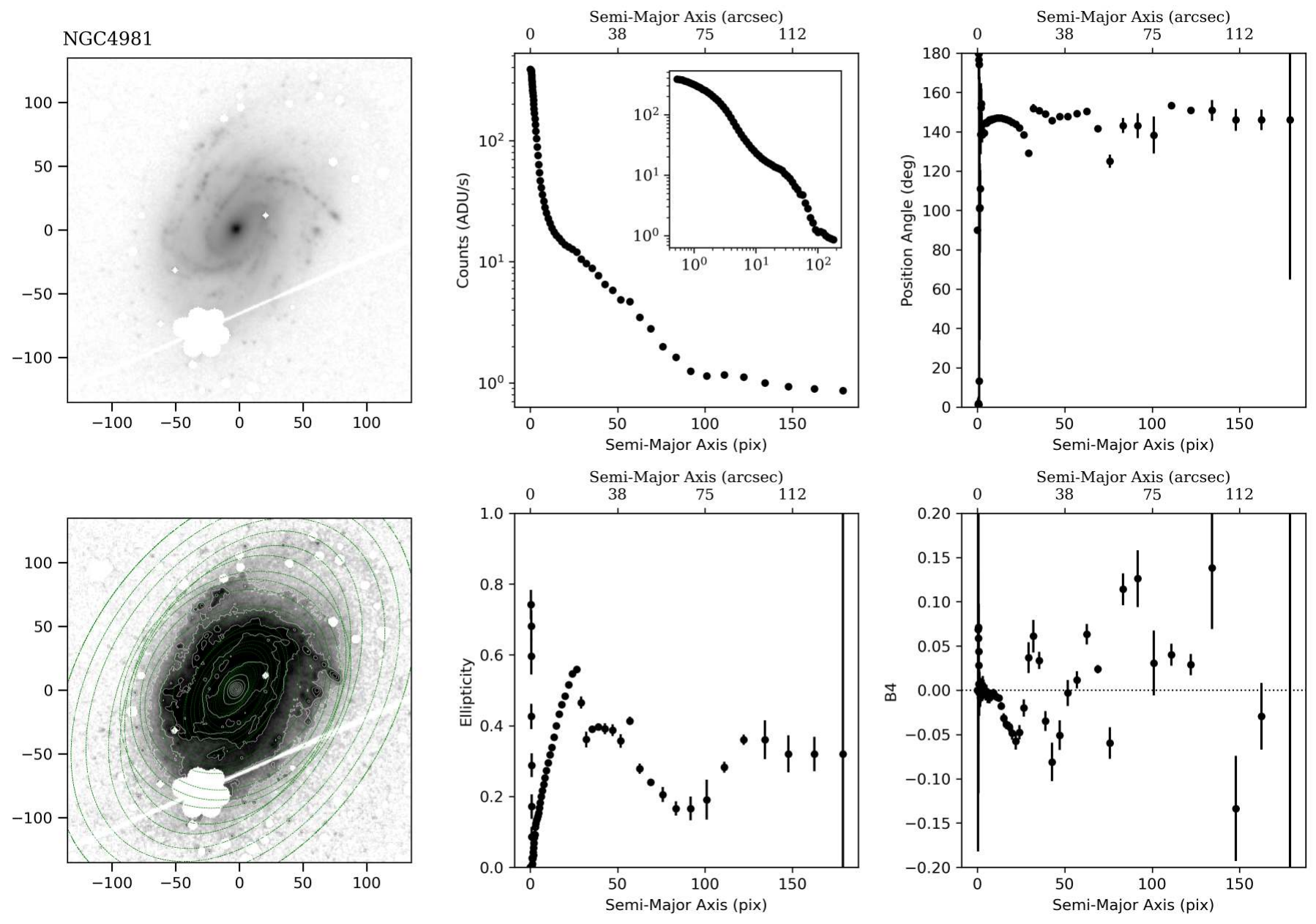}
\addtocounter{figure}{-1}
\caption{continued.}
\end{figure*}

\begin{figure*}
\includegraphics[trim=0.2cm 0.2cm 0.2cm 0.2cm,clip=true,width=1.8\columnwidth]{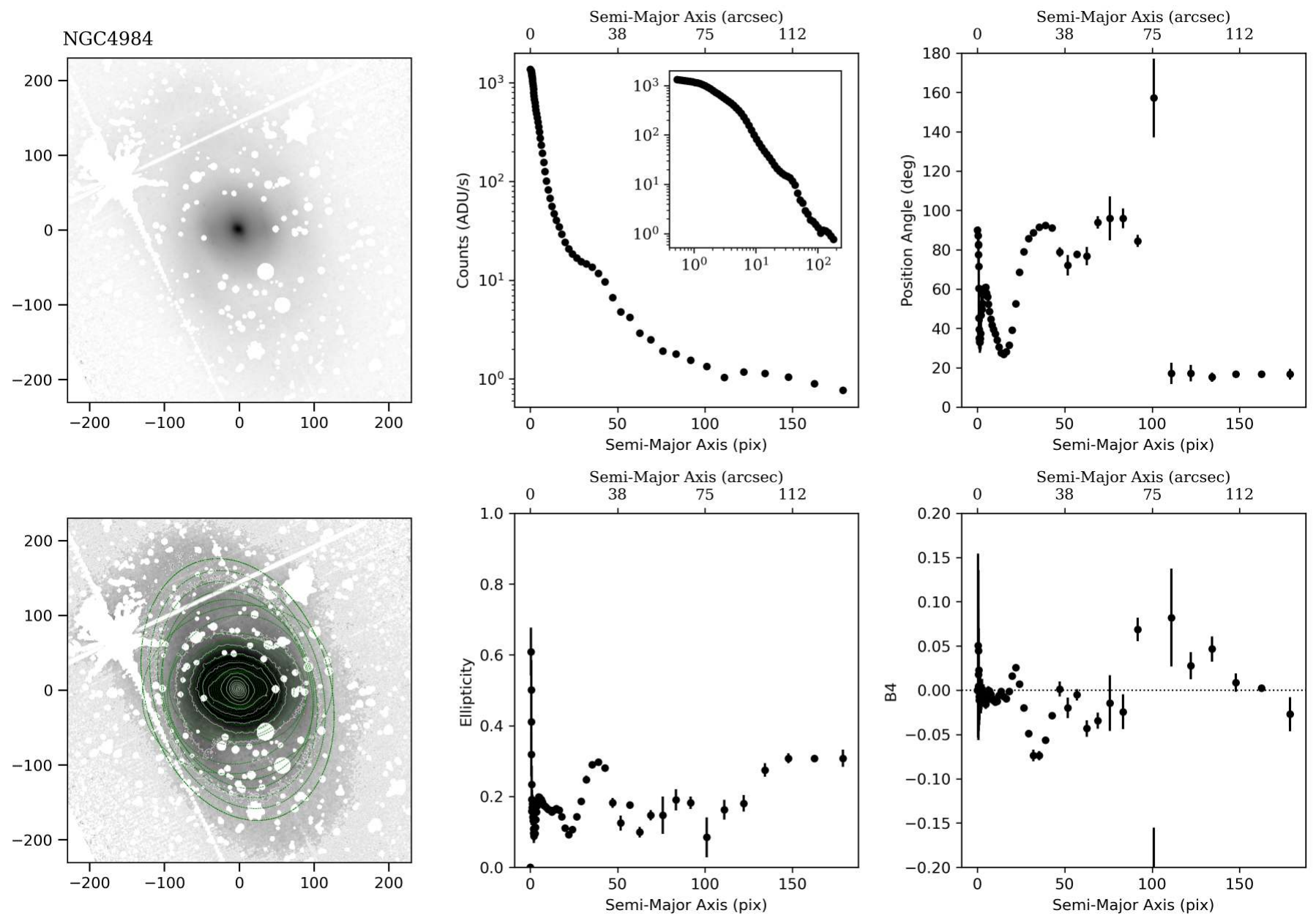}
\addtocounter{figure}{-1}
\caption{continued.}
\end{figure*}

\begin{figure*}
\includegraphics[trim=0.2cm 0.2cm 0.2cm 0.2cm,clip=true,width=1.8\columnwidth]{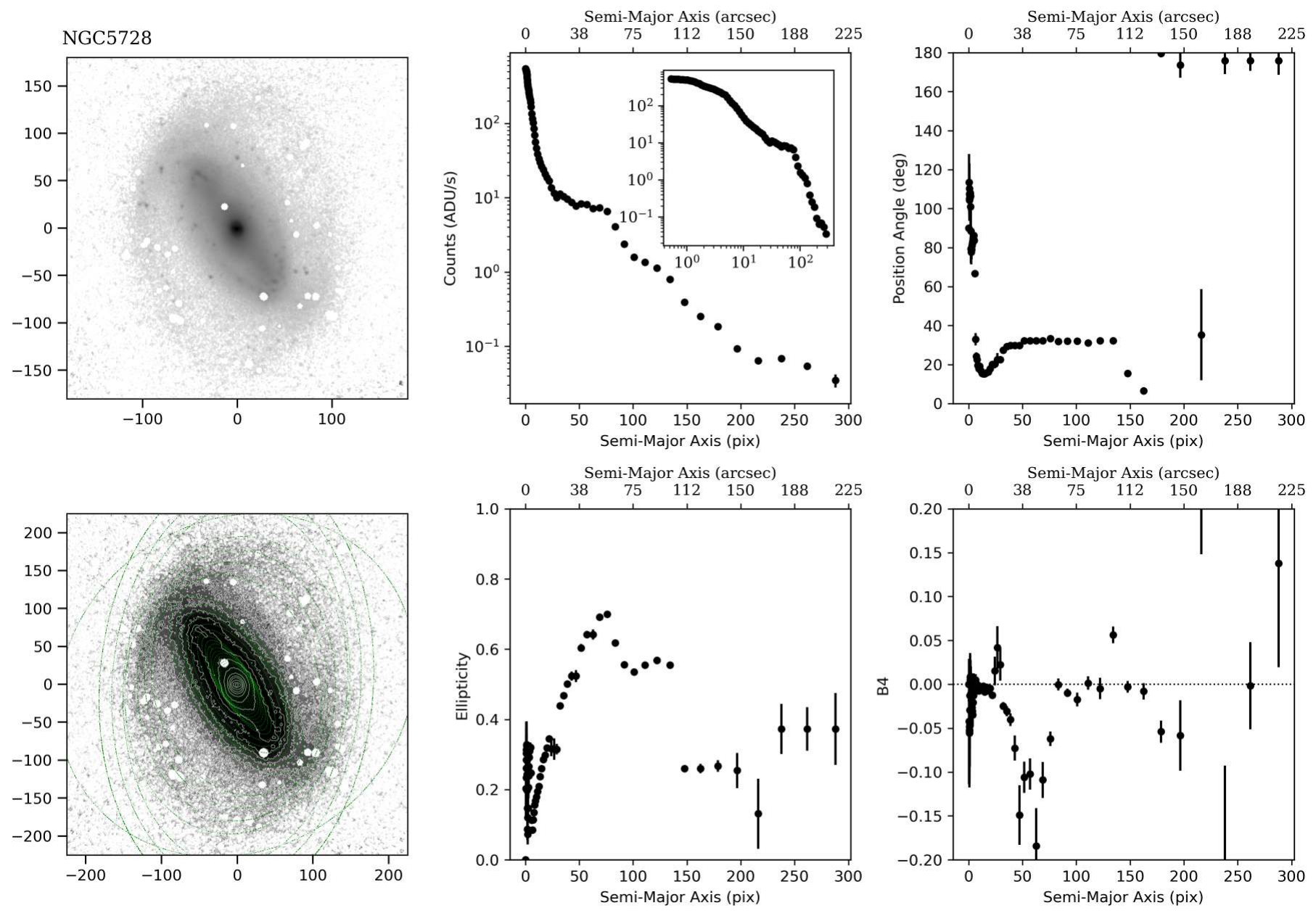}
\addtocounter{figure}{-1}
\caption{continued.}
\end{figure*}

\begin{figure*}
\includegraphics[trim=0.2cm 0.2cm 0.2cm 0.2cm,clip=true,width=1.8\columnwidth]{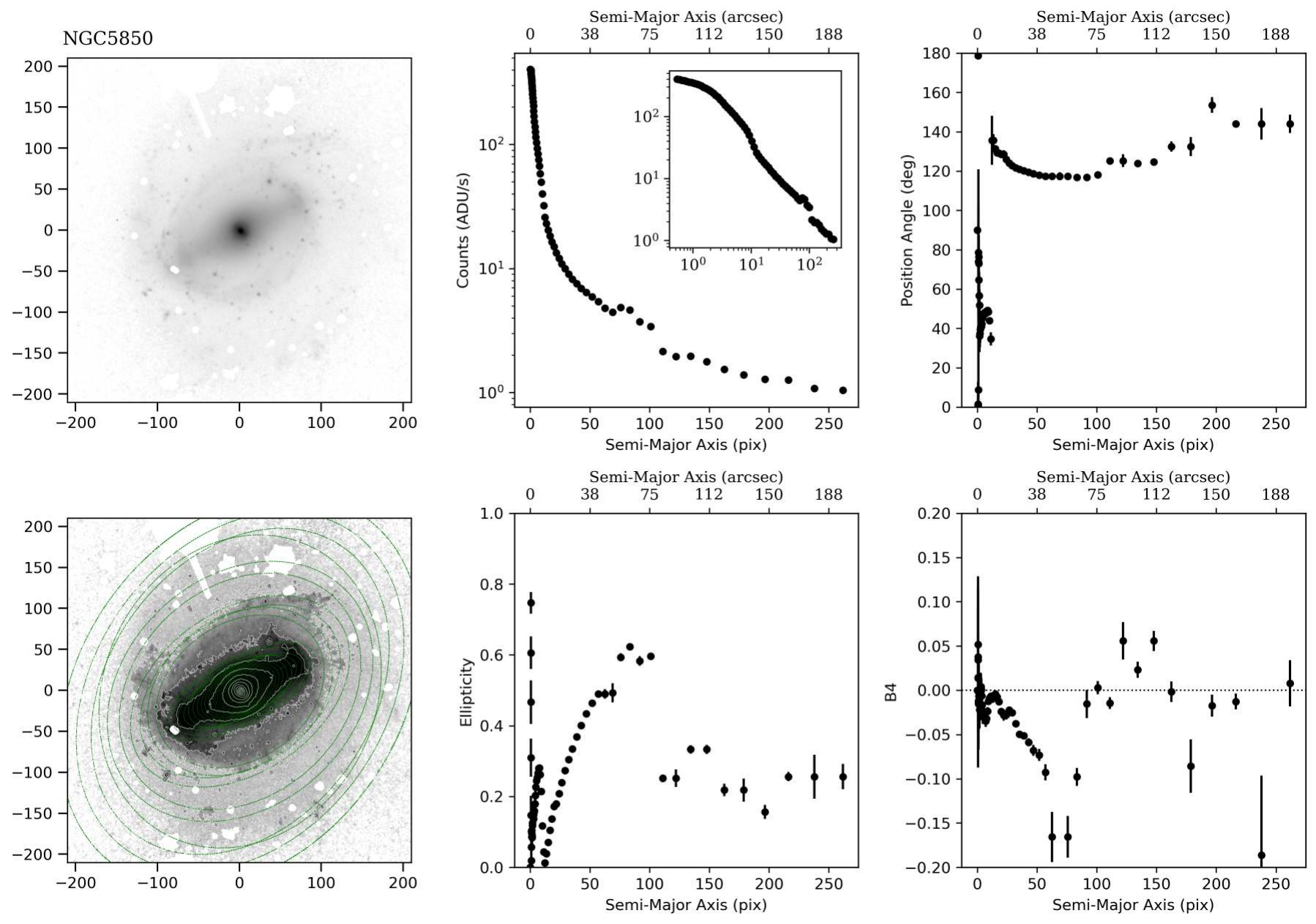}
\addtocounter{figure}{-1}
\caption{continued.}
\end{figure*}

\begin{figure*}
\includegraphics[trim=0.2cm 0.2cm 0.2cm 0.2cm,clip=true,width=1.8\columnwidth]{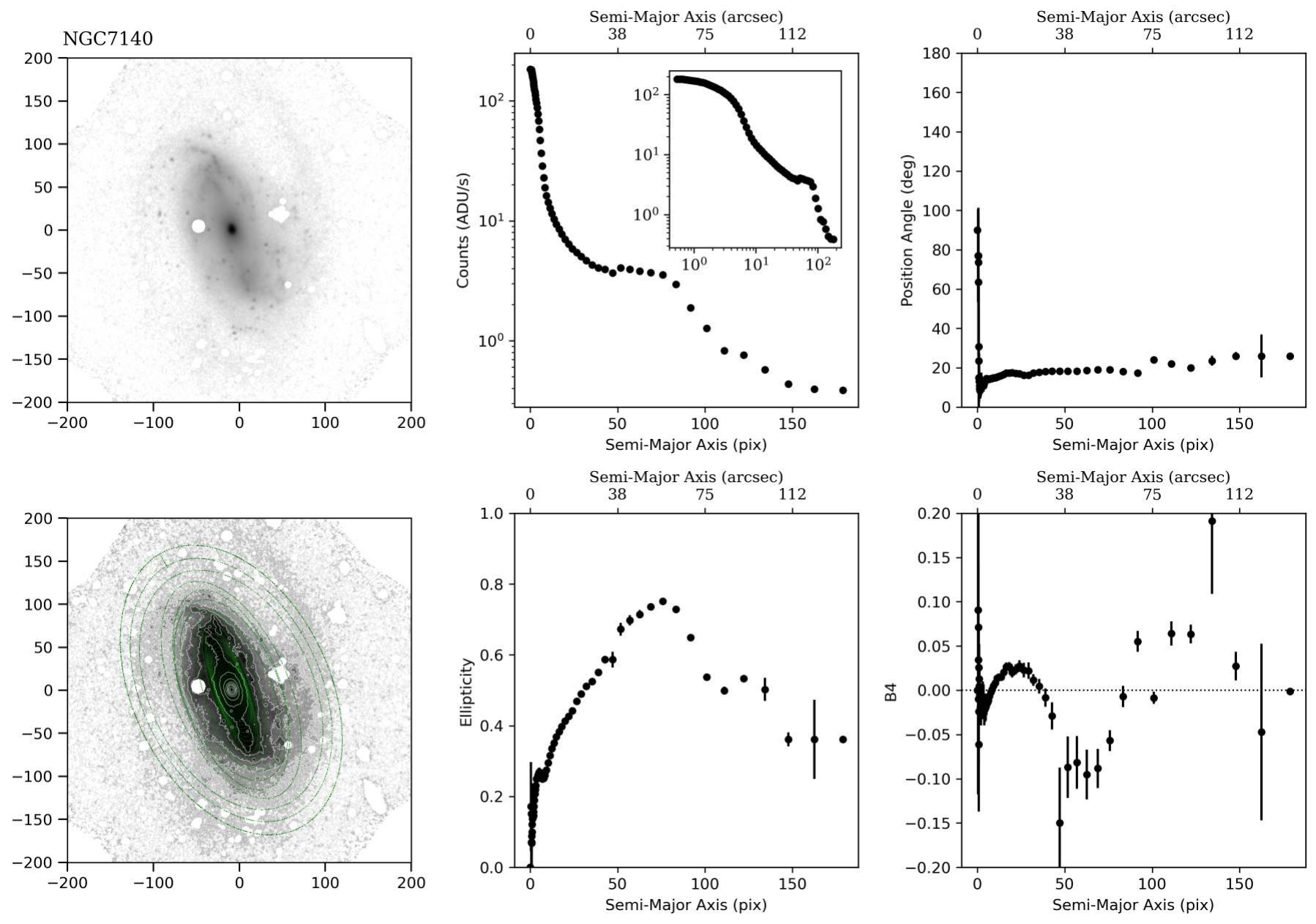}
\addtocounter{figure}{-1}
\caption{continued.}
\end{figure*}

\begin{figure*}
\includegraphics[trim=0.2cm 0.2cm 0.2cm 0.2cm,clip=true,width=1.8\columnwidth]{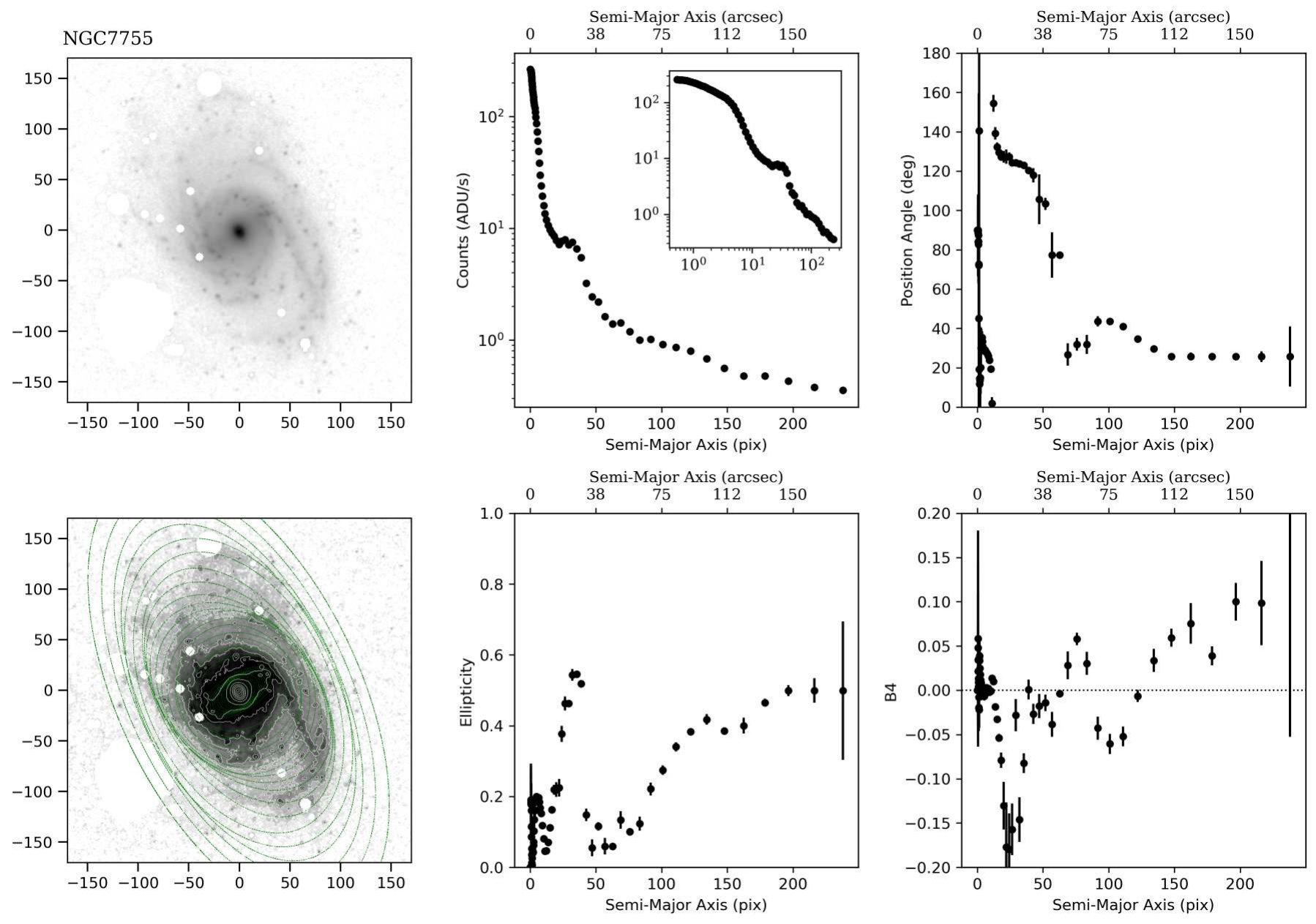}
\addtocounter{figure}{-1}
\caption{continued.}
\end{figure*}

\section{Structural Parameters}
\label{app:params}

In this appendix, I show in Table \ref{tab:strparams} the derived structural parameters of the photometric bulge, disc and bar obtained with the DE fits using the original S$^4$G images for the 16 galaxies in the sample.

\begin{table*}
\centering
\caption{Structural parameters of the photometric bulge, disc and bar obtained with the DE fits using the original S$^4$G images. Column (1) indicates the galaxy considered and column (2) shows the effective intensity of the photometric bulge, while columns (3), (4) and (5) show, respectively, the photometric bulge effective radius, S\'ersic index and ellipticity. Columns (6), (7) and (8) show, respectively, the bar effective intensity, effective radius and S\'ersic index, while columns (9) and (10) show the bar ellipticity and boxiness, respectively. The disc central intensity, scale length and ellipticity are shown in columns (11), (12) and (13), respectively. Finally, the bulge-to-total, bar-to-total and disc-to-total 3.6$\mu m$ luminosity ratios are show in columns (14), (15) and (16), respectively. Radii are in arcseconds, and intensities are in counts per second (on $0.75''$ pixels),  which can be converted to MJy/sr when multiplying by the {\tt FLUXCONV} header keyword in each S$^4$G image. Ellipticities are not deprojected.}
\label{tab:strparams}
\begin{tabular}{lccccccccccccccc}
\hline
Galaxy & $I_e$ & $r_e$ & $n$ & $\epsilon_{\mathrm b}$ & $I_{e,{\mathrm{bar}}}$ & $r_{e,{\mathrm{bar}}}$ & $n_{\mathrm{bar}}$ & $\epsilon_{\mathrm{bar}}$ & $c_0$ & $I_0$ & $h$ & $\epsilon_{\mathrm d}$  & B/T & Bar/T & D/T \\
(1) & (2) & (3) & (4) & (5) & (6) & (7) & (8) & (9) & (10) & (11) & (12) & (13) & (14) & (15) & (16) \\
\hline
IC 1438  &  102.40  &  2.8  &  1.4  &  0.19  &  6.69  &  16.9  &  0.3  &  0.54  &  0.0  &  3.65  &  31.8  &  0.07  &  0.24  &  0.18  &  0.57 \\
NGC 613  &  199.92  &  5.0  &  1.1  &  0.41  &  8.09  &  67.8  &  0.3  &  0.84  &  1.0  &  18.89  &  49.2  &  0.40  &  0.14  &  0.19  &  0.67 \\
NGC 1097  &  213.54  &  9.0  &  0.5  &  0.10  &  10.96  &  65.8  &  0.7  &  0.65  &  0.0  &  4.45  &  127.3  &  0.22  &  0.21  &  0.26  &  0.54 \\
NGC 1291  &  93.35  &  14.7  &  2.2  &  0.15  &  7.48  &  70.7  &  0.4  &  0.38  &  0.0  &  3.63  &  213.1  &  0.05  &  0.20  &  0.13  &  0.67 \\
NGC 1300  &  65.39  &  4.2  &  1.4  &  0.19  &  2.96  &  69.2  &  0.2  &  0.81  &  2.0  &  5.04  &  79.8  &  0.20  &  0.07  &  0.11  &  0.82 \\
NGC 1433  &  99.31  &  5.9  &  1.5  &  0.21  &  3.85  &  67.3  &  0.6  &  0.71  &  2.0  &  6.70  &  66.7  &  0.00  &  0.14  &  0.20  &  0.66 \\
NGC 3351  &  151.53  &  6.9  &  0.7  &  0.28  &  7.05  &  49.3  &  1.0  &  0.40  &  0.1  &  4.57  &  173.3  &  0.27  &  0.07  &  0.15  &  0.78 \\
NGC 4303  &  349.51  &  2.4  &  0.7  &  0.00  &  5.79  &  37.5  &  0.5  &  0.71  &  0.0  &  34.44  &  35.0  &  0.00  &  0.07  &  0.07  &  0.86 \\
NGC 4371  &  43.85  &  9.8  &  1.7  &  0.38  &  6.60  &  26.5  &  0.3  &  0.49  &  1.1  &  9.06  &  48.1  &  0.47  &  0.31  &  0.16  &  0.54 \\
NGC 4643  &  71.15  &  7.1  &  2.0  &  0.11  &  15.41  &  28.5  &  0.6  &  0.65  &  0.0  &  6.52  &  64.0  &  0.17  &  0.22  &  0.18  &  0.60 \\
NGC 4981  &  172.33  &  1.8  &  1.2  &  0.21  &  2.11  &  22.5  &  1.1  &  0.76  &  0.8  &  22.90  &  22.9  &  0.29  &  0.09  &  0.06  &  0.85 \\
NGC 4984  &  164.06  &  5.0  &  2.2  &  0.20  &  8.11  &  29.1  &  0.5  &  0.40  &  0.0  &  3.13  &  101.3  &  0.28  &  0.24  &  0.15  &  0.61 \\
NGC 5728  &  124.05  &  3.7  &  1.1  &  0.00  &  6.76  &  43.8  &  0.5  &  0.67  &  0.1  &  3.58  &  47.0  &  0.27  &  0.22  &  0.41  &  0.37 \\
NGC 5850  &  38.53  &  6.0  &  2.0  &  0.00  &  3.09  &  48.8  &  0.6  &  0.69  &  0.4  &  3.03  &  97.5  &  0.00  &  0.10  &  0.10  &  0.80 \\
NGC 7140  &  71.86  &  3.5  &  1.2  &  0.35  &  2.98  &  48.4  &  0.5  &  0.67  &  0.0  &  1.58  &  112.5  &  0.23  &  0.06  &  0.16  &  0.78 \\
NGC 7755  &  122.01  &  2.9  &  0.7  &  0.23  &  4.47  &  21.7  &  0.2  &  0.66  &  2.0  &  10.11  &  25.0  &  0.17  &  0.17  &  0.13  &  0.70 \\
\hline
\end{tabular}
\end{table*}

\section{Physical Spatial Resolution with Different Facilities}
\label{app:resols}

In this appendix, Table \ref{tab:resols} shows the redshifts at which the given physical spatial resolutions are achieved with different representative facilities. The physical spatial resolutions chosen here are 100\,pc, 1\,kpc, 1.7\,kpc and 3.4\,kpc. The latter two values correspond to the artificially redshifted images shown in Fig.\,\ref{fig:input} and thus to the effects presented and discussed in Sects. \ref{sec:beyond} and \ref{sec:discuss}. The lowest value (100\,pc) was chosen as it is the typical order of magnitude of the smallest nuclear discs observed to date \citep{deSa-Freitas2023b}, while the intermediate value of 1\,kpc was chosen because it corresponds to the largest nuclear disc identified to date \citep{Gadotti2020}. These two values therefore indicate in Table \ref{tab:resols} the capability of a given instrumental setup to resolve nuclear discs. For example, for Euclid in the $H_E$ band, images will have a physical resolution poorer than 1\,kpc already beyond $z\approx0.11$. For HST at $1.6\mu m$, this corresponds to $z\approx0.59$, which means that, even with HST, most (if not all) nuclear discs are unresolved beyond $z\approx0.59$.

\begin{table*}
\setlength{\tabcolsep}{3pt}
\centering
\caption{Redshifts at which the given physical spatial resolutions are achieved with different instrumental setups, as indicated. At redshifts higher than these, the data will have poorer resolution. E.g., for Euclid in the $H_E$ band, images will have a physical resolution poorer than 1\,kpc already beyond $z\approx0.11$. For HST at $1.6\mu m$, this corresponds to $z\approx0.59$, which means that, even with HST, most (if not all) nuclear discs are unresolved beyond $z\approx0.59$. The values are approximated and were computed using the interface provided by \citet{Wright2006}, assuming a Hubble constant of ${\rm H_0}=67.8\,\rm{km}\,\rm{s}^{-1}\,\rm{Mpc}^{-1}$ and $\Omega_{\rm m}=0.308$ in a Universe with flat topology \citep[see][]{AdeAghArn15}. It should be noted that the parameter that sets the physical resolution is the galaxy distance, so for very low redshits, the redshifts quoted here assume negligible peculiar motions with respect to the Hubble flow. It should also be noted that cosmological effects are such that the physical spatial resolution starts increasing again beyond $z\approx1.6$.}
\label{tab:resols}
\begin{tabular}{lcccc}
\hline
Setup (angular resolution) & $z$ for 100-pc resolution & $z$ for 1-kpc resolution & $z$ for 1.7-kpc resolution & $z$ for 3.4-kpc resolution \\
\hline
S$^4$G $3.6\mu m$ ($1.7''$)                                                           & 0.003    & 0.03     & 0.05    & 0.11    \\
SDSS $r$ ($1.5''$)                                                                            & 0.003    & 0.03    & 0.06    & 0.12    \\
DECaLS $r$ ($1.2''$)                                                                        & 0.004    & 0.04    & 0.07    & 0.16    \\
Euclid $H_E$ ($0.5''$)                                                                      & 0.009    & 0.11    & 0.20    & 0.59    \\
Euclid $I_E$ ($0.17''$)                                                                      & 0.029    & 0.46    & worst resol.: 1.5\,kpc@$z\approx1.6$    & --    \\
HST WFC3 $1.6\mu m$ ($0.15''$)                                                    & 0.034    & 0.59    & worst resol.: 1.3\,kpc@$z\approx1.6$    & --    \\
 JWST NIRCam F444W ($0.15''$)                                                    & 0.034    & 0.59    & worst resol.: 1.3\,kpc@$z\approx1.6$    & --    \\
JWST NIRCam F200W ($0.07''$)                                                     & 0.071    & worst resol.: 0.6\,kpc@$z\approx1.6$    & --    & --    \\
MICADO $K$ ($0.01''$)                                                                    & worst resol.: 0.09\,kpc@$z\approx1.6$    & --    & --    & --    \\
\hline
\end{tabular}

{\raggedright Sources for quoted image qualities: \par
S$^4$G: \citet{KimGadShe14} \par
SDSS: \citet{Gad09b} \par
DECaLS: \citet{Dey2019} \par
Euclid: \cite{EuclidCollaboration2025} \par
HST: \href{https://hst-docs.stsci.edu/wfc3ihb/chapter-7-ir-imaging-with-wfc3/7-6-ir-optical-performance}{https://hst-docs.stsci.edu/wfc3ihb/chapter-7-ir-imaging-with-wfc3/7-6-ir-optical-performance} \par
JWST: \href{https://jwst-docs.stsci.edu/jwst-near-infrared-camera/nircam-performance/nircam-point-spread-functions}{https://jwst-docs.stsci.edu/jwst-near-infrared-camera/nircam-performance/nircam-point-spread-functions} \par
MICADO: \cite{Davies2010} \par
}
\end{table*}

\bsp	
\label{lastpage}
\end{document}